\documentclass[authoryear,12pt, a4]{elsarticle}
\usepackage[latin1]{inputenc}
\usepackage{graphics, graphicx}
\usepackage{color}
\usepackage{rotating}
\usepackage{amssymb}
\usepackage{lineno}
\usepackage{hyperref}

\usepackage{natbib}

\journal{Icarus}
\begin{document}
\begin{frontmatter}

\title{Saturn's icy satellites investigated by Cassini - VIMS. \\
V. Spectrophotometry}

\author[label1]{Gianrico Filacchione\corref{cor1}} 
\author[label1]{Mauro Ciarniello}
\author[label1]{Emiliano D'Aversa}
\author[label1]{Fabrizio Capaccioni}
\author[label2]{Roger N. Clark}
\author[label3]{Bonnie J. Buratti}
\author[label4]{Paul Helfenstein}
\author[label5]{Katrin Stephan}
\author[label6]{Christina Plainaki}

\address[label1]{INAF-IAPS, Istituto di Astrofisica e Planetologia Spaziali, Area di Ricerca di Tor Vergata, via del Fosso del Cavaliere, 100, 00133, Rome, Italy}
\address[label2]{PSI Planetary Science Institute, Tucson, AZ, USA}
\address[label3]{Jet Propulsion Laboratory, California Institute of Technology, 4800 Oak Groove Drive, Pasadena, CA 91109, USA}
\address[label4]{Cornell Center for Astrophysics and Planetary Science, Cornell University, 404 Space Sciences Building, Ithaca, New York 14853, USA}
\address[label5]{Institute for Planetary Exploration, DLR, Rutherfordsta{\ss}e 2, 12489, Berlin, Germany}
\address[label6]{ASI, Agenzia Spaziale Italiana, via del Politecnico, snc, 00133, Rome, Italy}
\cortext[cor1]{Corresponding author, email gianrico.filacchione@inaf.it}

\begin{abstract}
Albedo, spectral slopes, and water ice band depths maps for the five midsized saturnian satellites Mimas, Enceladus, Tethys, Dione, and Rhea have been derived from Cassini-Visual and Infrared Mapping Spectrometer (VIMS) data. The maps are systematically built from photometric corrected data by applying the Kaasalainen-Shkuratov model \citep{Kaasalainen2001, Shkuratov2011}. In this work a quadratic function is used to fit phase curves built by filtering observations taken with incidence angle $i\le70^\circ$, emission angle $e\le70^\circ$, phase angle $10^\circ \le g \le 120^\circ$, and Cassini-satellite distance $D \le 100.000$ km. This procedure is systematically repeated for a subset of 65 VIMS visible and near-infrared wavelengths for each satellite. The average photometric parameters are used to compare satellites' properties and to study their variability with illumination conditions changes. We derive equigonal albedo, extrapolated at g=0$^\circ$, not including the opposition effect, equal to 0.63$\pm$0.02 for Mimas, 0.89$\pm$0.03 for Enceladus, 0.74$\pm$0.03 for Tethys, 0.65$\pm$0.03 for Dione, 0.60$\pm$0.05 for Rhea at 0.55 $\mu$m. The knowledge of photometric spectral response allows to correct individual VIMS spectra used to build maps through geolocation. Maps are rendered at a fixed resolution corresponding to a $0.5^\circ \times 0.5^\circ$ bin on a longitude by latitude grid resulting in spatial resolutions of 1.7 km/bin for Mimas, 2.2 km/bin for Enceladus; 4.7 km/bin for Tethys; 4.5 km/bin for Dione; 6.7 km/bin for Rhea. These spectral maps allow establishing relationships with morphological features and with endogenic and exogenic processes capable to alter satellites' surface properties through several mechanisms. The hemispheric dichotomies in albedo and spectral indicators between leading and trailing hemispheres are common properties of all midsized satellites: the accumulation of fine E ring grains is responsible for the higher albedo measured across the leading hemispheres of Tethys, Dione, Rhea, and the trailing side of Mimas. Conversely, the dark and red-colored material visible across the trailing hemispheres of Tethys, Dione, and Rhea is associated with the implantation of cold plasma particles. The thermal anomaly lenses located on the leading equatorial regions of Mimas and Tethys have been resolved and mapped. VIMS data evidence that the distribution of water ice band depths on Mimas lens is biased by the presence of the large Herschel impact crater pointing to a different regolith size distribution with respect to Tethys. Maps show local changes of albedo and spectral indicators in correspondence of recent impact craters (Inktomi on Rhea, Creusa on Dione) and on Dione's wispy terrains are caused by the exposure of pristine water ice. Enceladus' tiger stripes, the active sources of plumes in the southern polar region, despite being partially resolved on VIMS maps allow measuring an exceedingly high band depths in comparison with the rest of the satellite's surface. Moreover, Enceladus' smooth terrains located on the leading hemisphere around $(lon,lat)=(90^\circ, 30^\circ)$ show peculiar properties (low infrared albedo, positive 0.35-0.55 $\mu$m slope and maximum band depth) possibly associated with the presence of a buried diapir in this area. The variability of average spectral albedos and indicators induced by phase angle is investigated and exploited to establish a comparison among satellites' properties as a function of orbital distance from Saturn.
\end{abstract}


\end{frontmatter}

\section{Introduction}
In this work, we are exploiting the full Cassini-Visual and Infrared Mapping Spectrometer (VIMS) \citep{Brown2004} dataset built from observations collected between 2004 and 2017 to derive visible and near-infrared albedo maps of the surfaces of the five inner midsize satellites of Saturn (Mimas, Enceladus, Tethys, Dione, and Rhea). The adopted methodology has been already applied to Dione \citep{Filacchione2018a}, Tethys \citep{Filacchione2018b}, and Rhea \citep{Filacchione2020} datasets but with respect to these previous works here we introduce further improvements and make a comparison of the results among satellites.
The exploitation of a complex hyperspectral dataset like the one collected by VIMS needs a systematic approach designed for disentangling photometric effects caused by illumination and viewing conditions from intrinsic properties of the surface due to composition and regolith properties. In this respect, VIMS data are extremely valuable because they were collected from very different observation geometries allowing to measure the response of the light reflected and scattered from a given point on the surface. Although the high redundancy of the data allows a better retrieval of the photometric response, it raises the complication to extract the overall information to be represented on a map. To resolve these issues, we have adopted a method detailed in the following capable to derive the average photometric response of each satellite and then we have applied it to every single spectrum. The resulting equigonal albedo and spectral indicator maps are made available as online data. 
 \par 
The study of surface morphology, composition, and physical properties of Saturn's satellites is one of the primary scientific goals of the remote sensing instruments aboard Cassini \citep{Matson2002}. Focusing our discussion on visible and infrared spectrophotometric results we can summarize here some of the main results achieved. 
Saturn's satellites are among the brightest objects of the solar system because they are composed of almost pure crystalline water ice, a material characterized by high reflectance in the visible and infrared spectral range: the geometric albedo of Enceladus is 1.0, Tethys 0.8, Mimas and Dione 0.7, Rhea 0.6 \citep{Jaumann2009}. The large heliocentric distance (9.02 AU at perihelion) and the high reflectance of the satellites keep the surface diurnal temperature well below 100 K \citep{Howett2010, Filacchione2016b}. In this range of temperatures, water ice can be in both amorphous or crystalline forms: the two forms are recognizable through IR spectroscopy. Laboratory measurements on samples kept at controlled temperatures \citep{Mastrapa2008} demonstrate that it is possible to discriminate the crystalline form from the amorphous one whose reflectance spectrum does not display the 1.65 $\mu$m absorption while presents a shifted, broadened and weaker 3.1 $\mu$m Fresnel peak.
The amorphous ice can form through direct condensation of vapor to solid at a temperature below $\approx$100 K. When amorphous ice is heated above T$\approx$150 K it becomes crystalline through an exothermal irreversible reaction. At cryogenic temperatures, the water ice spontaneous thermal metamorphism cannot occur blocking also the grains sintering and coalescence processes \citep{Gundlach2018}. The only possibility to reform amorphous ice is through the accumulation of high doses of high-energy irradiated particles that can damage the crystalline lattice structure making it chaotic \citep{Strazzulla1992}. This is the case of the surface of Jupiter's moon Europa which experiences higher temperatures and fluxes of magnetospheric particles. Notwithstanding the high abundance of water ice, Europa is warm enough \citep{Filacchione2019} to experience sintering during its lifetime, whereas the sinter time-scale of Saturn's moon Enceladus is longer than the age of our Solar System \citep{Schaible2017}.
Among the two possible states of water ice, e.g. amorphous or crystalline, VIMS data point out that the latter is the dominant species on Saturn's satellites. Despite the very diagnostic crystalline feature at 1.65 $\mu$m  falls within the spectral position of the junction between two of the VIMS-IR order sorting filters resulting in a deterioration of the instrument response, and the concurrent effects caused by temperature and scattering from sub-micron grains, \cite{Clark2012} made a thorough spectral analysis of the icy satellites' data finding out that the overall properties match with a dominance of the crystalline form. 
 \par 
One of the major findings of Cassini has been the observation of the high cryovolcanic activity occurring on Enceladus' South Pole Terrains (SPT) where a network of five parallel fractures (called sulci) are located \citep{Spencer2006, Nimmo2007, Spitale2007}. From these fractures, discrete jets (plumes) made of water vapor and ice grains with traces of carbon dioxide, carbon monoxide or molecular nitrogen, and methane are released \citep{Waite2006}. This activity is not continuous but is triggered by orbital position and tidal stresses resulting in more intense jets at the apocenter than at the pericenter \citep{Hedman2013, Ingersoll2020}. 
The plume ice grains have a size distribution centered in the 1-5 $\mu$m radii range \citep{Hedman2009, Ingersoll2011}. The ice grains launched with velocities exceeding the escape speed feed the wide Saturn's E ring whereas the ones with small velocities fall back on the surface of Enceladus, coating with a preferential pattern (Fig. 8 in \cite{Kempf2010}).
 This flux of fine water ice grains makes Enceladus's surface extremely bright due to the constant replenishment of fresh material. Spectral analysis of infrared absorption bands confirms that the water ice is mostly in crystalline form \citep{Brown2006, Filacchione2007, Filacchione2012} with small fractions of amorphous ice over the areas located between the tiger stripes as evidenced by a systematic deviation towards shorter wavelengths of the positions of the 1.5-2.0 $\mu$m band and 3.1 $\mu$m Fresnel peak \citep{Brown2006, Newman2008}.
 \par 
The other four inner midsize satellites (Mimas, Tethys, Dione, and Rhea) continue to show very high visible albedo and prominent water ice absorption bands in the infrared range but also a strong reddening in the 0.35-0.55 $\mu$m caused by the presence of contaminants acting as chromophores able to induce color variations across the surface. From a spectral perspective, these satellites show some common characteristics: 1) strong brightening and deep water ice absorption bands in correspondence of recent impact craters and ejecta, e.g. Inktomi on Rhea \citep{Stephan2012, Scipioni2014, Filacchione2020} or Creusa on Dione \citep{Stephan2010, Scipioni2013, Filacchione2018a}; 2) complex systems of "wispy terrains", or chasmata, made of fractures, graben, scarps, and troughs possibly due to extensional and shear stresses occurring on the trailing hemispheres of Dione and Rhea \citep{Roatsch2008}. Chasmata appear brighter than the neighboring terrains because fresh, high albedo material is displaced on the surface; 3) leading-trailing hemisphere dichotomy caused by the preferential bombardment of E ring particles and high energy magnetospheric electrons on the leading hemisphere and cold plasma on the trailing hemisphere \citep{Schenk2011}. These exogenous processes differently alter surface composition: 
a) E ring fine grains cause surface brightening through the deposition of fresh ice falling above the leading hemispheres of Tethys, Dione, and Rhea (orbiting outside Enceladus' orbit) and on the trailing hemisphere of Mimas (orbiting internal to Enceladus);    
b) high energy electrons cause surface weathering and sputtering \citep{Paranicas2012} resulting in the formation of dark, thermal anomaly terrains located inside equatorial lens areas on the leading hemispheres of Mimas and Tethys \citep{Howett2011, Howett2012};
c) cold plasma gives rise to darkening and reddening of the regolith through the implantation of ions taking place on trailing hemispheres \citep{Schenk2011, Paranicas2012, Hendrix2018}.
\par
With this study, we are aiming to compare the spectrophotometric properties of Saturn's icy satellites by adopting a common protocol to process the full VIMS dataset. This strategy allows us to minimize systematic effects in the processing of the dataset, to disentangle surface composition from illumination effects, and to achieve comparable results among satellites allowing us to better understand their surface composition and physical properties and to correlate this information with geographical location and surface morphology. This approach is applied not only to every single satellite but also to the whole population of midsized satellites, allowing us to explore the spectral variability as a function of their orbital position which in turn corresponds to different Saturn's magnetospheric environment. Moreover, by retrieving the spatial distribution of the spectral albedo and derived indicators, it becomes possible to understand how the surfaces are altered by exogenous processes (magnetospheric particles, cosmic and UV rays, E ring grains, meteoroid impacts) and to distinguish these effects from endogenous ones (tectonism, resurfacing of fresh material). 
\par
A discussion about previous studies and how spectral results are affected by photometric response is in section \ref{sct:background}. In section \ref{sct:vims_dataset} a description of the VIMS dataset and methodology to render maps is given. The application of the photometric correction method used to derive equigonal albedos on 65 spectral channels follows in section \ref{sct:photometry_correction}. Albedo and spectral indicators maps are presented and discussed in section \ref{sct:albedo_maps}. Section \ref{sct:comparative_analysis} addresses the comparison of the photometric parameters among satellites. Finally, in the Conclusions, a discussion of the main results reached in this work is given.

\section{Background}
\label{sct:background}
The "standardization" of images of a target acquired at different illumination and viewing geometries is the principal goal of any photometric model. By applying a photometric correction is, in fact, possible to scale the \emph{specific flux (I/F)} which is a relative measurement dependent on incidence, emission, and phase angles, in \emph{spectral albedo}, an absolute quantity related to the surface's intrinsic properties and independent from geometry conditions. 
The irradiance to solar flux (I/F) is given by:
\begin{equation}
\frac{I}{F}(s,\lambda,l)=\frac{R(s,b,l) \cdot \pi \cdot DN(s,b,l) \cdot D_{Sun}^2}{t \cdot SI(\lambda)}
\end{equation}
where $R(s,b,l)$ is VIMS instrumental transfer function for sample $s$, band $b$, and line $l$; $D_{Sun}$ is the heliocentric distance (in AU) of the target from the Sun and $SI(\lambda)$ is the solar irradiance measured at 1 AU \citep{Thekekara1973}. The $I/F(\lambda)$ is a function of the surface's composition and physical properties, illumination, and viewing geometries.
\par
A thorough analysis of planetary surfaces data returned by imaging instruments aboard spacecrafts needs therefore the application of a proper photometric correction because illumination and viewing conditions rapidly change due to mission operations and orbital constraints: this means that to compare images of the same area on different observations or to build global maps of a target it is necessary to convert the observed I/F in spectral albedos. An alternative way to proceed is to scale the observed I/F to a standard illumination condition (e.g. incidence angle i=30$^\circ$, emission angle e=0$^\circ$, phase angle g=30$^\circ$, see in example \cite{Hicks2011} for application to Moon data or \cite{Pieters1983} for laboratory measurements) by adopting a given model (M):

\begin{equation}
I/F_{corrected} = \frac{I/F(i,e,g)}{I/F_{M}(i,e,g)}I/F_{M}(30^\circ, 0^\circ, 30^\circ) 
\end{equation}

The two more-used photometric models applied to correct planetary images are the ones developed by \cite{Hapke1993a} and by Kaasalainen-Shkuratov \citep{Kaasalainen2001, Shkuratov2011}. 
Following the same approach already applied in previous works \citep{Filacchione2018a, Filacchione2018b, Filacchione2020} we adopt the Kaasalainen-Shkuratov (KS) model in which I/F is modeled using two factors accounting for local topography - through incidence and emission angles - and phase angle. The main advantage of this empirical method is the simplicity to handle the phase angle separately from the local topography. 

In the following, we apply the KS model to derive equigonal albedo maps at a null phase of the five inner midsize satellites of Saturn.
\subsection{Previous results}
The KS method has been applied to derive albedo maps of many solar system bodies, including Mercury \citep{Domingue2016}, asteroids 4 Vesta \citep{Schroeder2013}, 21 Lutetia \citep{Longobardo2016}, Itokawa \citep{Abe2006} and the Earth's Moon \citep{Kaidash2009, Velikodsky2011, Shkuratov2011}. Different authors \citep{Besse2013, Buratti2011b,  Hicks2011, Hillier1999} have modeled the surface phase function through polynomial series adjusted to better fit observational data. As a general rule, polynomials with exponential terms are necessary to best match non-linear trends observed in the opposition effect regime ($g < 10^\circ$). At higher phases ($g > 20^\circ$) n-th degree polynomials are sufficient to model the phase curves.
\par
So far, the color maps produced by \cite{Schenk2011} remain the best effort to map Saturn's icy moons from Cassini ISS images. In this work an empirical simplified Lunar-Lambertian photometric function \citep{McEwen1991} has been adopted to equalize ISS images and achieve seamless mosaics but without giving a quantitative estimate of the albedo. At UV wavelengths icy satellites' phase curves have been analyzed \citep{RoyerHendrix2014} and normal albedo maps of Mimas were obtained \citep{Hendrix2012}.  
Several studies have investigated the spectral and geologic properties of the icy moons by mapping VIMS data and using several spectral indicators like band parameters, colors, and slopes for Dione \citep{Jaumann2006, Clark2008, Stephan2010, Scipioni2013}, Rhea \citep{Stephan2012, Scipioni2014}, Tethys \citep{Jaumann2006, Stephan2016}, Enceladus \citep{Jaumann2006, Jaumann2008, Scipioni2017, Combe2019}. All these works do not have applied any specific photometric model but were relying on empirical corrections based on data normalization. Up to now, very few studies have provided quantitative analyses of the disk-resolved photometric properties of the icy satellites of Saturn.  \cite{Robidel2020} have produced albedo maps and water ice band properties maps of Enceladus at 8 infrared wavelengths by applying the KS model. A similar method has been used systematically at multiple visible and infrared wavelengths for Tethys \citep{Filacchione2018b}, Dione \citep{Filacchione2018a}, and Rhea \cite{Filacchione2020}.   

\subsection{How photometry affects spectral analysis}

Apart from changing the reflectance of the surface, illumination conditions influence also the measurements of the spectral indicators, e.g the visible spectral slopes and absorption bands, due to the dependence of scattering properties from phase angle. Observations and laboratory measurements on different targets and materials show that phase reddening can follow a monotonic linear or a quadratic law. On powdered basaltic samples \citep{Gradie1986}, on Vesta \citep{Longobardo2014}, and on hydrated minerals \citep{Pommerol2008} spectral slopes and absorption band depths systematically decrease at low and at high phases resulting in a quadratic trend. Conversely, on other low-albedo solar system objects, like comet 67P \citep{Ciarniello2015} or asteroid Bennu \citep{Fornasier2020}, has been observed a phase reddening effect with a monotonic phase dependence of the spectral slope. This monotonic trend is characterizing also Saturn's main rings where visible slopes and water ice bands depth are linearly increasing with phase angle \citep{Ciarniello2019}. The nature of the two-phase reddenings trends is investigated by \cite{Schroeder2014} who study the behavior of semi-transparent particles in laboratory-controlled conditions. This study evidences that the quadratic trend characterizes smooth surfaces preferentially made by forward-scattering particles while the monotonic trend prevails when the surface roughness increases at the scale of the single particles. 
\par
\begin{figure}[h!]
\centering
	\includegraphics[width=19cm]{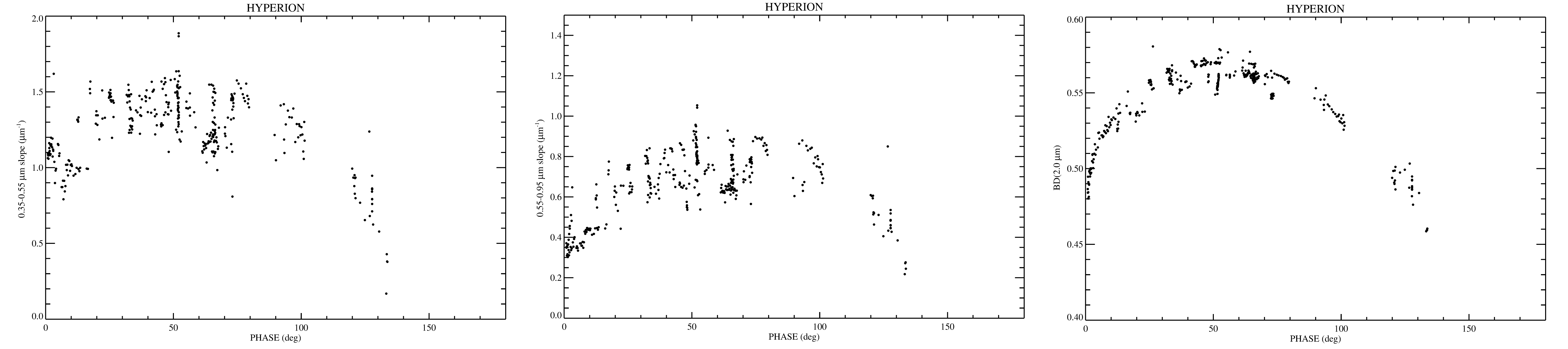}
    \caption{Variability of Hyperion's spectral indicators as a function of phase angle: 0.35-0.55 $\mu$m spectral slope (left panel), 0.55-0.95 $\mu$m spectral slope (center), and 2 $\mu$m water ice band depth (right). Full-disk observation data derived from \citep{Filacchione2012} dataset.}
    \label{fig:1}
\end{figure}

VIMS data of Saturn's icy satellites show that phase reddening has a quadratic trend: to illustrate this effect, we show the variations of the 0.35-0.55 $\mu$m and 0.55-0.95 $\mu$m spectral slopes, and the 2 $\mu$m water ice band depth for Hyperion as a function of the phase angle (Fig. \ref{fig:1}). Hyperion is particularly suitable to evidence this dependence because it shows uniform spectral properties when observed at hemispherical scale lacking the typical leading-trailing asymmetry characterizing the regular satellites of Saturn. 
\par   
These data refer to disk-integrated observations described in \cite{Filacchione2012}: the three spectral indicators show a clear quadratic dependence from phase having minimum values towards null and high ($g>100^\circ$) phases and maxima between $50^\circ \le g \le 70^\circ$. An increase of the order of $50\%$, $100\%$, and $18\%$ is observed when the solar phase rises from $g=0^\circ$ to $\approx60^\circ$ for 0.35-0.55 $\mu$m spectral slope, 0.55-0.95 $\mu$m spectral slopes, and 2 $\mu$m water ice band depth, respectively. After reaching the maximum values, a steady decrease from $g \approx60^\circ$ to $\approx130^\circ$ is visible on all curves.
 \par 
Such experimental results can be explained through the relative efficiency of the single and multiple scattering regimes as a function of the phase angle: the multiple scattering is more efficient at intermediate phases while the single scattering dominates at low and high phases  \citep{Hapke1993a}. This explains the trends observed on the spectral slopes and band depth data. In the opposition effect regime, e.g. when phase angle $g \rightarrow 0^\circ$, the coherent backscattering effect dominates causing a non-linear increase of the reflectance and at the same time a drastic decrease of the band depth \citep{Pitman2017}.

Albedo is also affecting the slope of the phase curve because multiple scattering is prevalent in bright materials while it is much smaller in dark materials: on the former materials, the photons are reflected many times among contiguous grains before being absorbed or reflected while in the latter photons are easily absorbed on the first interaction with the grains. As many studies have demonstrated, exists an anti-correlation between the reflectance of the target and the steepness of the phase function curve \citep{Longobardo2016, Longobardo2017, Longobardo2019}. Since we are deriving an average photometric response, on areas characterized by very high or low reflectance with respect to the average, we would expect photometric residuals. Apart from albedo, also surface roughness can induce photometric residuals \citep{Domingue2016}. This is the main reason for which we are not applying this method to Iapetus, where an average photometric response would be unable to properly correct at the same time the bright trailing and the dark leading hemispheres given the large variation of albedo between the two units. In similar cases, one would need a more complex treatment consisting of the separate characterization of the bright and dark terrains and in the following evaluation of the results in the transition areas between the two units.

In conclusion, to map the spatial distribution of a surface's albedo and spectral indicators is necessary to apply a proper photometric correction able to remove the dependency from illumination conditions. This is particularly important when merging and comparing together observations collected from varying geometries where it is necessary to decouple the illumination effects from intrinsic properties of the surface (like composition and grain size). We'll return to this point ahead in section \ref{sct:comparative_analysis} where a discussion about the variability of the spectral parameters as a function of phase angle is given for the midsized satellites.

\section{VIMS dataset}
\label{sct:vims_dataset}
A subset of the VIMS dataset has been filtered to select targeted observations of Mimas, Enceladus, Tethys, Dione, and Rhea satellites. The dataset contains observations acquired in a wide range of illumination/viewing geometries and spatial resolutions as occurred during the 13 years-long Cassini mission (2004-2017). 
VIMS spectra of the satellites are processed through a well-consolidated calibration pipeline \citep{Filacchione2012}: raw data are converted in radiance factor units, I/F (where I is the reflected intensity of light from the surface and $\pi$F is the plane-parallel incident solar flux) by applying the RC17 instrumental transfer function \citep{Clark2012}. The geographic location in latitude and longitude, the incidence, emission and phase angles, and the spacecraft distance of each pixel center and four corners have been computed through SPICE reconstructed kernels (\cite{Acton1996}; \cite{Acton2018}). 
\par
Lacking global digital shape models for the whole surfaces of the satellites, the geometric parameters are computed with respect to the ellipsoidal shape of each satellite and therefore the local morphology is not taken into account. In principle, this could introduce biases on rough terrain areas and craters' rims when applying the photometric correction because the local incidence and emission angles could be significantly different with respect to the ones computed from the ellipsoid shape, an effect well-described in \cite{Li2009}. It is also true that midsized satellites topography is quite relaxed and does not diverge too much from the ellipsoid shape: as an example, the maximum elevation variations of Ithaca Chasmata on Tethys are between -5 to +6 km with respect to the ellipsoid surface across a distance of more than 100 km \cite{Giese2007}. Since VIMS observations are acquired with multiple illumination and viewing geometries, and the method applied uses a statistical analysis of the signal falling on a given bin of the map (as discussed later), in case of high data redundancy, one would expect that this effect will become less relevant.
\par
By using geometry information, all pixels having center direction intercepting the satellites' surfaces have been mined from the dataset resulting in a subset suitable to compute photometric parameters. Pixels falling on the edge of the limb and partially filled by the target are filtered out. In Table \ref{tbl:dataset} are listed the total number of pixels (or spectra) for the five satellites. We report separate values for the VIS (at $\lambda=0.55\ \mu$m) and IR (at $\lambda=1.82\ \mu$m) datasets having the two VIMS channels the capability to operate independently with different operative modes resulting in different spatial resolutions. For the purposes of this work, the data are further filtered to optimize the computation of the photometric parameters with the following conditions: incidence angle $i\le70^\circ$; emission angle $e\le70^\circ$; phase angle $10^\circ \le g \le 120^\circ$; Cassini-satellite distance $D \le 100.000$ km; unsaturated signal. The rendering of the albedo and spectral indicators maps, discussed later and shown in Fig. \ref{fig:8}-\ref{fig:12}, has been performed with the same filtered parameters used to compute photometric correction but with the exception of the phase angle which has been reduced to the $10^\circ \le g \le 90^\circ$ range. This choice allows reducing seams among data collected at extreme phase angles.
 \par 
The filtering of incidence angle and emission angles is necessary to avoid grazing illumination geometries and oblique views in which the effects of the shadows and pixel distortion on the surface become critical for the reconstruction of maps. The condition on the incidence angle allows to filter out pixels occurring on the nightsides.
The condition on the distance $D \le 100.000$ km corresponds to data having spatial resolutions better than 50$\times$50  km/px for observations acquired by VIMS in nominal mode and 16.7$\times$16.7 km/px or 25$\times$50 km/px in high resolution by VIMS-VIS and IR channels, respectively \citep{Brown2004}. The saturation check is computed on raw data (dark current removed) in Digital Numbers (DN) by flagging out signals with DN$>3500$ for each of the 65 bands considered in this work. This condition allows to remove saturated (at 4096 DN) and out of linearity range signals. The resulting filtered dataset is reported in Table \ref{tbl:dataset}.   

\begin{table}[h!]
\begin{center}
\begin{tabular}{|c|r|r|r|r|}
\hline
\textbf{Satellite} & \textbf{VIS dataset} & \textbf{VIS filtered} & \textbf{IR dataset} & \textbf{IR filtered} \\
\hline
Mimas & 47267 & 13767 & 38941 & 11020 \\
Enceladus & 385252 & 159309 & 430298 & 189675 \\
Tethys & 280174 & 113148 & 303309 & 126505 \\
Dione & 421323 & 150701 & 470494 & 158125 \\
Rhea & 720724 & 173826 & 578200 & 178421 \\
\hline
\end{tabular}
\end{center}
\caption{Summary of the VIMS dataset. For each satellite are listed the total number of available spectra and the number of filtered ones at 0.549 $\mu$m (VIS) and 1.82 $\mu$m (IR) with conditions: incidence angle $i\le70^\circ$, emission angle $e\le70^\circ$, phase angle $10^\circ \le g \le 120^\circ$, Cassini-satellite distance $D \le 100.000$ km, unsaturated signal.}
\label{tbl:dataset}
\end{table}
 \par
\begin{figure}[h!]
\centering
	\includegraphics[width=18cm]{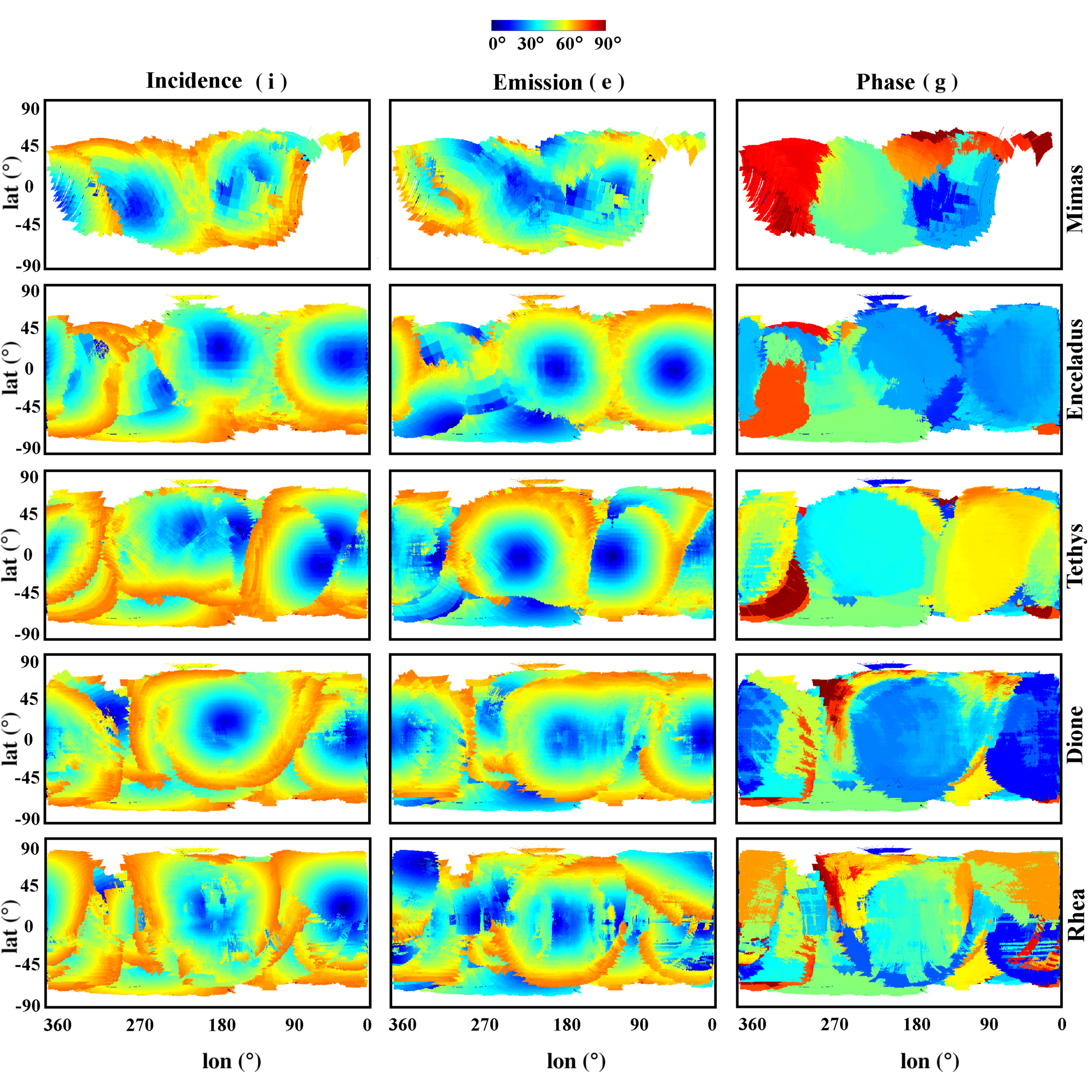}
    \caption{VIMS processed dataset: resulting median incidence (left column), emission (center) and phase (right) angles cylindrical maps for Mimas, Enceladus, Tethys, Dione and Rhea (from top to bottom). The range of variability of the angles is $i\le70^\circ$, $e\le70^\circ$ and $10^\circ \le g \le 90^\circ$. Note that for g we show the same range used to render the albedo maps shown in Fig. \ref{fig:8}-\ref{fig:11}}
    \label{fig:2}
\end{figure}

 
As a general approach in this work, all maps are rendered in simple cylindrical projections with fixed angular resolution on longitude and latitude of $0.5^\circ \times 0.5^\circ$ per bin. At the equator this angular resolution corresponds to a scale of 1.7 km/bin for Mimas, 2.2 km/bin for Enceladus; 4.7 km/bin for Tethys; 4.5 km/bin for Dione; 6.7 km/bin for Rhea. Following the same methodology discussed in \cite{Filacchione2016a} and \cite{Filacchione2016b}, the maps are built by projecting the value assigned to each pixel to an area defined by the geographical locations of the four corners and then sampled by the $0.5^\circ \times 0.5^\circ$ bins grid. The process is repeated for all pixels in the dataset keeping information of the distribution of points across each bin. Maps are finally rendered by showing the resulting median value. This method is used to render the distribution of the incidence, emission, and phase angles shown in Fig. \ref{fig:2} and equigonal albedo and spectral indicators maps discussed in section \ref{sct:albedo_maps}. These maps allow us appreciating also the overall coverage of the satellites' surfaces which, with the exception of Mimas and Rhea, show a continuous coverage in longitude while polar regions for latitudes $\ge \pm80^\circ$ are not covered due to the filtering on the incidence angle.

\clearpage

\section{Photometric correction method}
\label{sct:photometry_correction}
The formulation discussed in \cite{Shkuratov2011} is applied to derive the average photometric response of the icy satellites' surfaces. The radiance factor (I/F) spectra of each satellite, selected and filtered through the conditions discussed in the previous section, are associated with the illumination and viewing geometry of each illuminated pixel using its incidence (i), emission (e), and phase (g) angle as:  

\begin{equation}
\frac{I}{F}(\lambda,i,e,g)=D(i,e,g) \times F(\lambda, g)
\label{eq:1}
\end{equation}

and is expressed as the product of the disk function D(i,e,g) and the phase function F($\lambda$, g) modeling the equigonal albedo. The disk function D is modeled according to the Akimov model \citep{Shkuratov1999} as:

\begin{equation}
D(i,e,g)=D(\beta, \gamma, g)= \cos \left( \frac{g}{2} \right) \cos \left[ \frac{\pi}{\pi-g} \left( \gamma -\frac{g}{2} \right) \right] \frac{(\cos \beta)^{\frac{g}{\pi-g}}}{\cos \gamma}
\label{eq:2}
\end{equation}

being $\gamma$ and $\beta$ the photometric longitude and latitude, respectively:

\begin{equation} 
\gamma = arctan \left( \frac{cos(i)-cos(e)cos(g)}{cos(e)sin(g)} \right)
\label{eq:3}
\end{equation}

\begin{equation} 
\beta=arccos \left(\frac{cos(e)}{cos(\gamma)} \right)
\label{eq:4}
\end{equation}

The phase function $F(\lambda, g)$ is modeled in this work by means of a 2-nd degree polynomial fit. In this way eq. \ref{eq:1} becomes:

\begin{equation} 
\frac{\frac{I}{F}(\lambda)}{D(i,e,g)}=a_{0}+a_{1}g+a_{2}g^2
\label{eq:5}
\end{equation}

since D(g=0)=1, this equation allows computing the normal albedo $a_0$, defined as the best fit solution to observed I/F data extrapolated at null phase:

\begin{equation} 
\frac{I}{F}(\lambda, g=0)=a_0
\label{eq:6}
\end{equation}

Therefore, the normal albedo $a_0$ is equal to the equigonal albedo at g=0$^\circ$ derived from the polynomial fit. Noteworthy, this method does not model photometric behavior for observations taken in the opposition surge regime, e.g. observations corresponding to phase angles $\lesssim10^\circ$ in which non-linear effects (coherent backscattering, shadow-hiding opposition effect) prevail \citep{Hapke1993b, Ciarniello2014}. 
For this reason, the method applied is used to model observations taken in the phase angle range $10^\circ \le g \le 120^\circ$ where the phase curve trend is quadratic-like. The extrapolation of the phase curve at $g=0^\circ$ does not take into account the opposition effect and therefore our estimation is somewhat lower than the actual normal albedo as defined in \cite{Hapke1993a}.  
Given the non-linear nature of the phase curve for $0^\circ \le g \le 10^\circ$, it would be necessary to include observations taken in the opposition effect regime to infer the normal albedo. Unfortunately, given the Cassini mission orbital scenario, such observations are quite sparse, poorly resolved, and not sufficient for a reliable derivation of the albedo maps. For this reason, we prefer to use observations taken at intermediate phase angles where the photometric response can be derived more easily. For $g \ge 10^\circ$ both equigonal and normal albedo converge to the spectral albedo (see discussion in section \ref{sct:comparative_analysis}).

\begin{figure}[h!]
\centering
	\includegraphics[width=18cm]{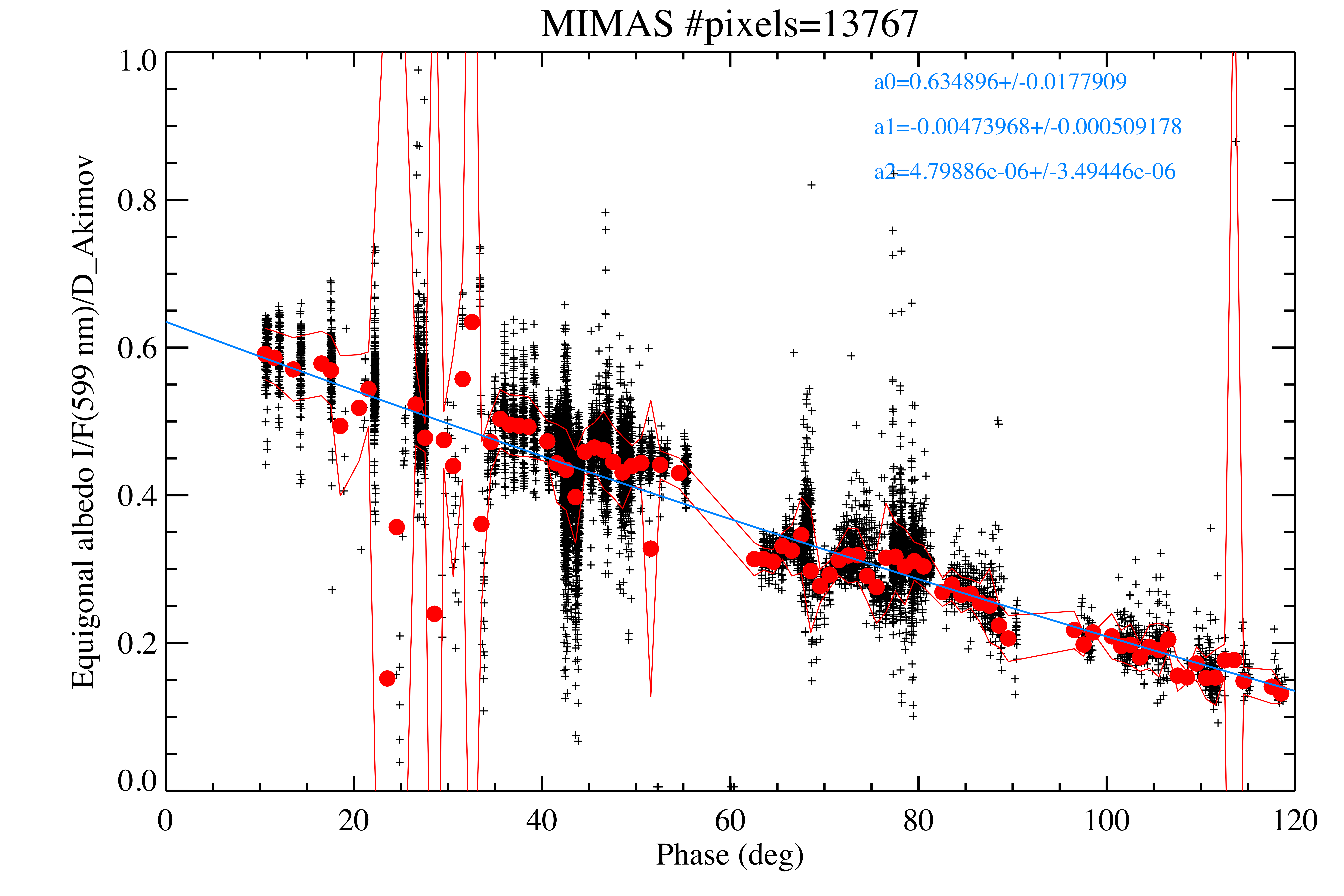}
    \caption{Mimas phase curve fit at $\lambda$=0.599 $\mu$m. Black crosses are the experimental data, red dots are the average values computed across 1 deg phase bins with associated standard deviation, blue curve is the best fit with parameters $a_0$ (adimensional), $a_1$ (in deg$^{-1}$), $a_2$ (in deg$^{-2}$) reported in the label. }
    \label{fig:3}
\end{figure}

\begin{figure}[h!]
\centering
	\includegraphics[width=18cm]{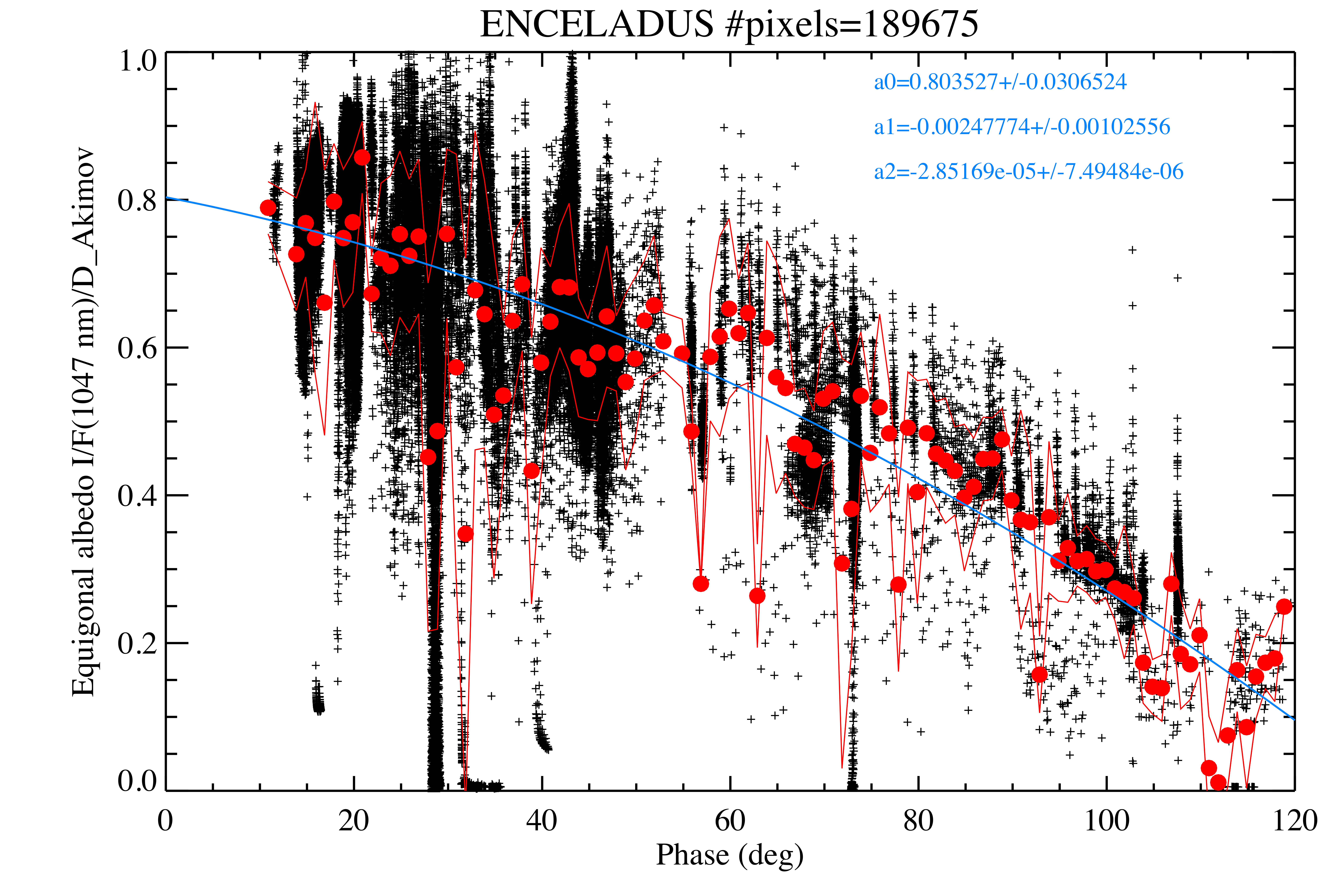}
    \caption{Enceladus phase curve fit at $\lambda$=1.047 $\mu$m. Same caption as in Fig. \ref{fig:3}.}
    \label{fig:4}
\end{figure}

\begin{figure}[h!]
\centering
	\includegraphics[width=18cm]{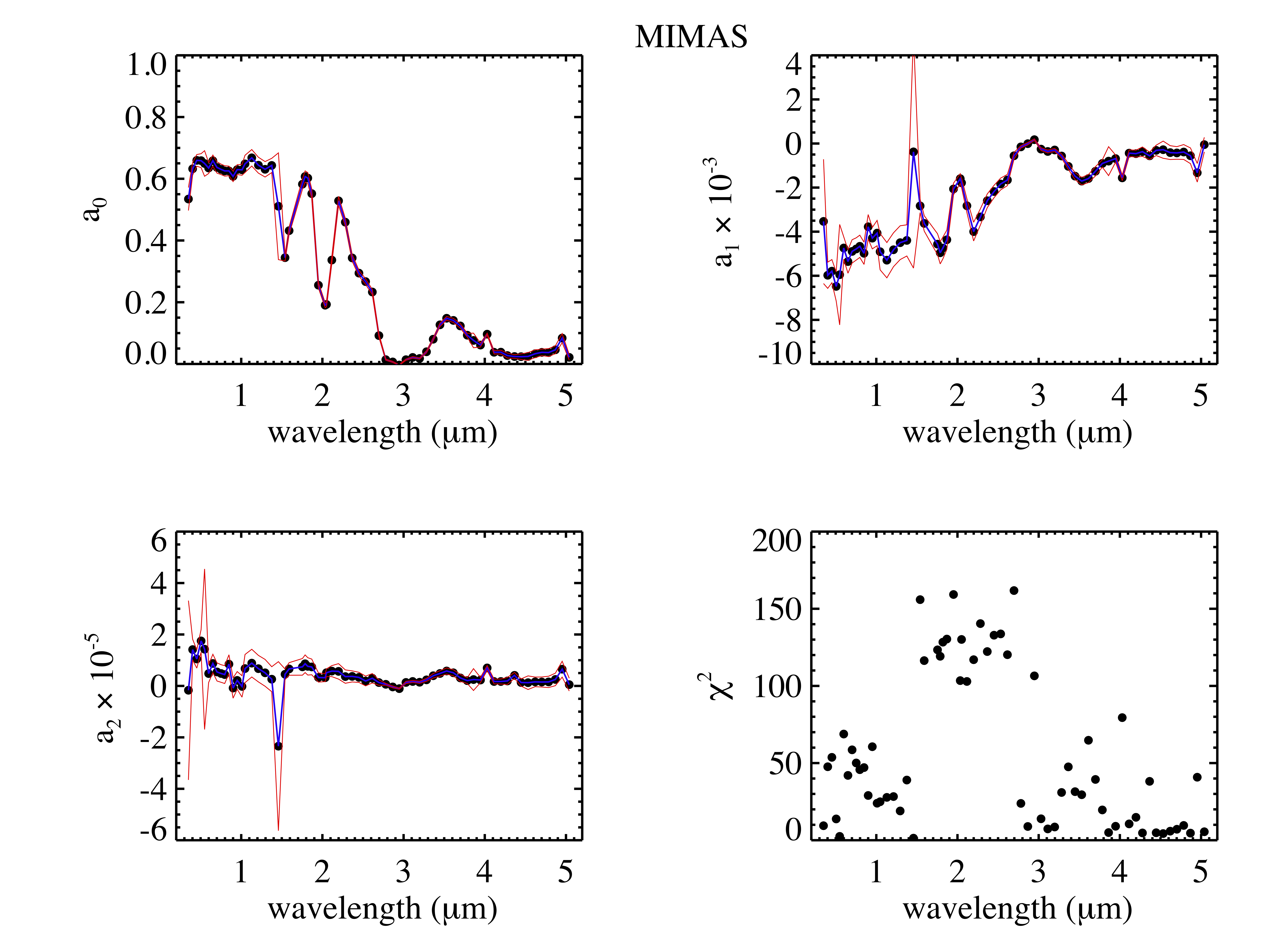}
    \caption{Mimas' photometric coefficients: equigonal albedo spectrum $a_0$ (top left panel), linear term $a_1$ (top right), quadratic term $a_2$ (bottom left), quality of the fit $\chi^2$ (bottom right). $a_0$ is given in adimensional units, $a_1$ in deg$^{-1}$ and $a_2$ in deg$^{-2}$.}
    \label{fig:5}
\end{figure}

\begin{figure}[h!]
\centering
	\includegraphics[width=18cm]{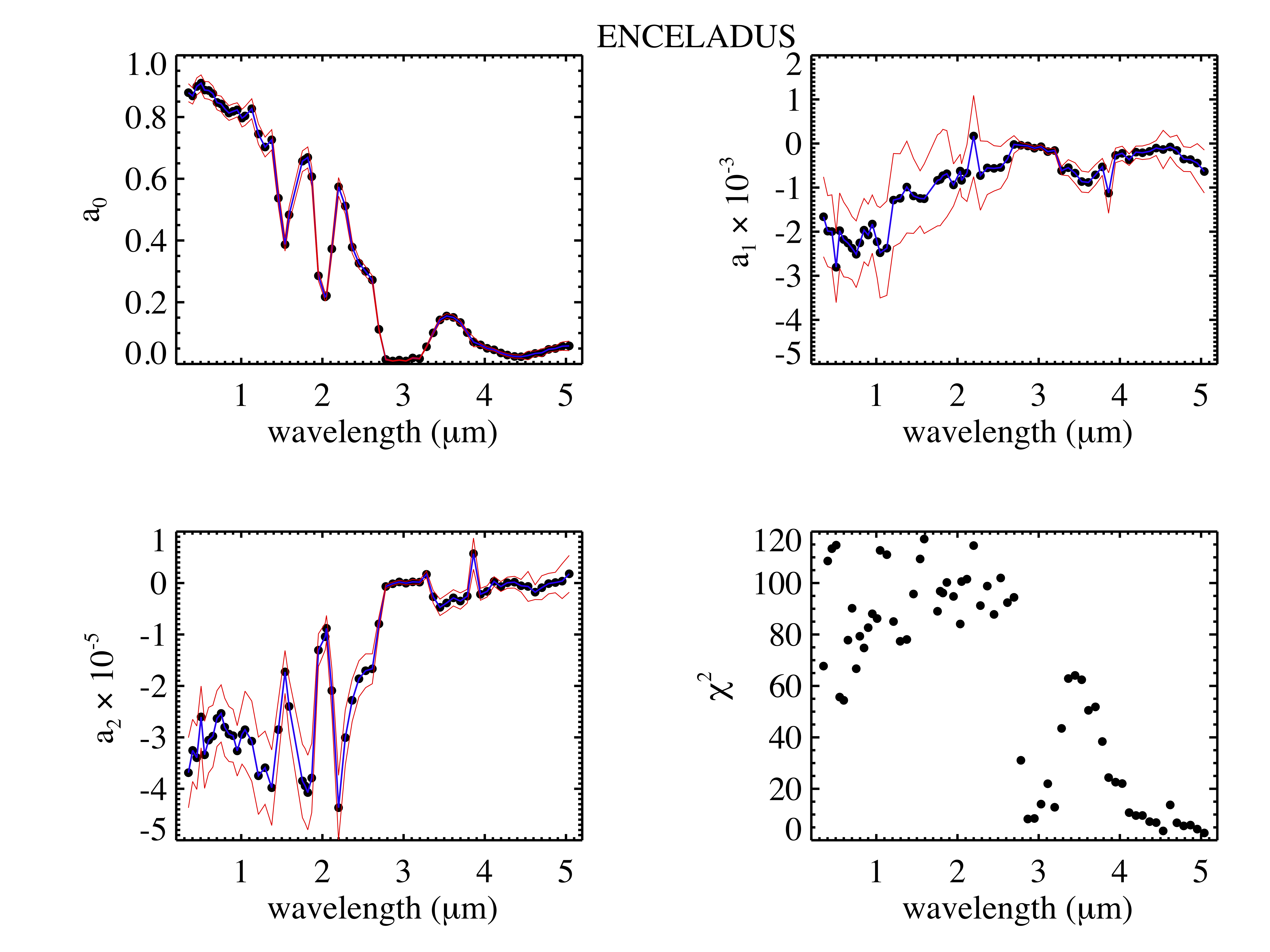}
    \caption{Enceladus' photometric coefficients. Same caption as in Fig. \ref{fig:5}.}
    \label{fig:6}
\end{figure}

\begin{figure}[h!]
\centering
	\includegraphics[width=18cm]{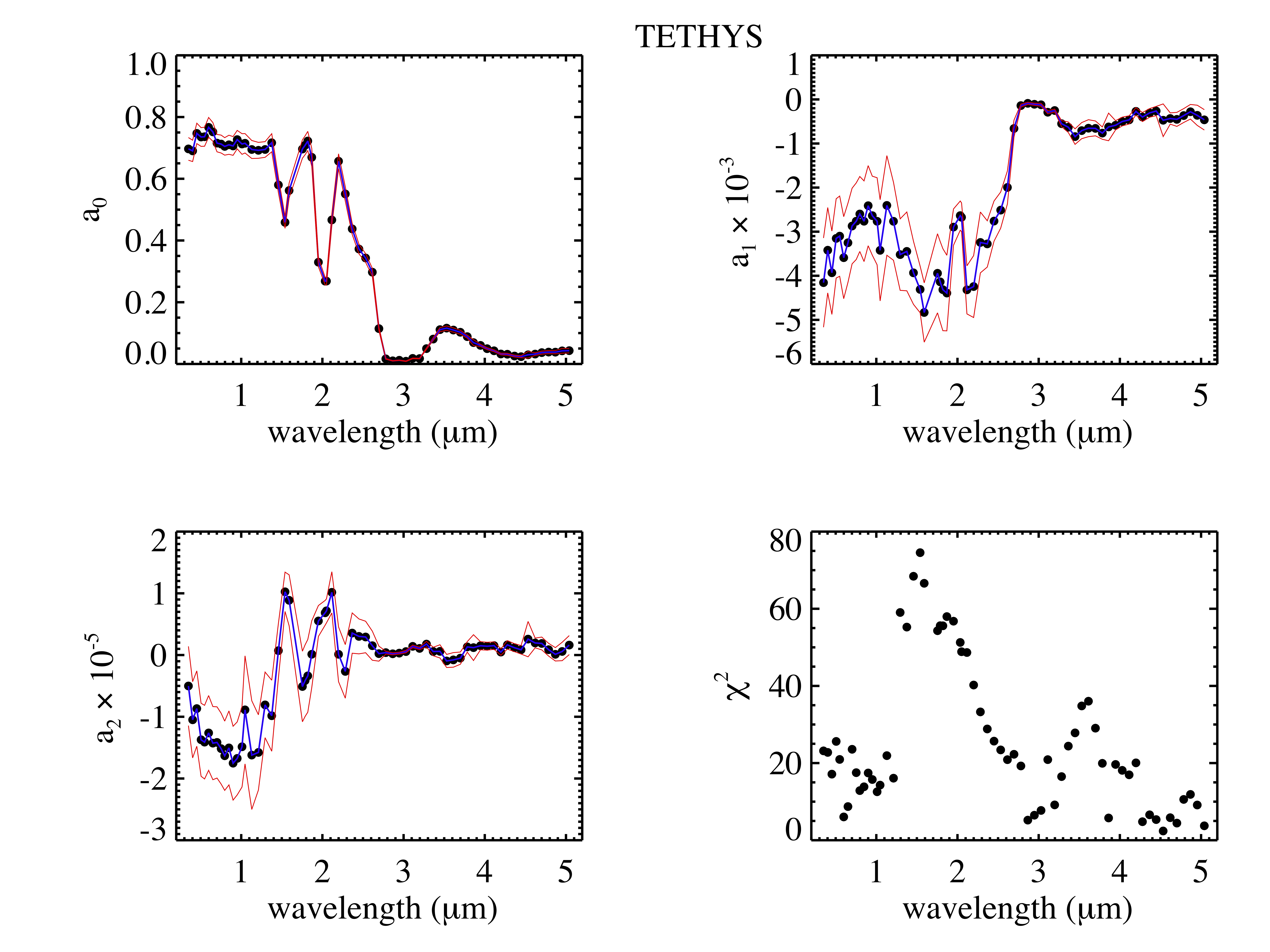}
    \caption{Tethys' photometric coefficients. Same caption as in Fig. \ref{fig:5}.}
    \label{fig:7}
\end{figure}

\begin{figure}[h!]
\centering
	\includegraphics[width=18cm]{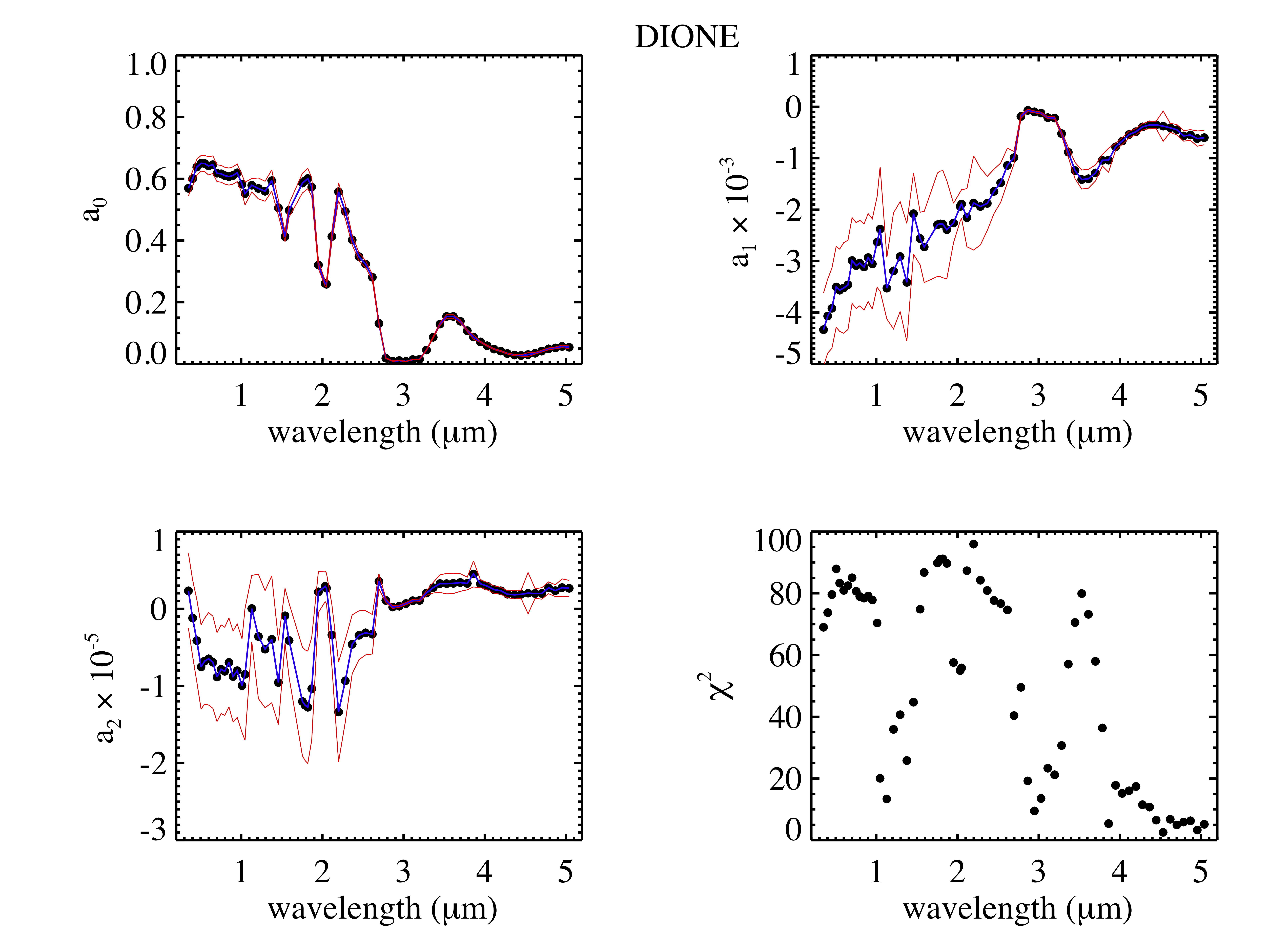}
    \caption{Dione's photometric coefficients. Same caption as in Fig. \ref{fig:5}.}
    \label{fig:8}
\end{figure}

\begin{figure}[h!]
\centering
	\includegraphics[width=18cm]{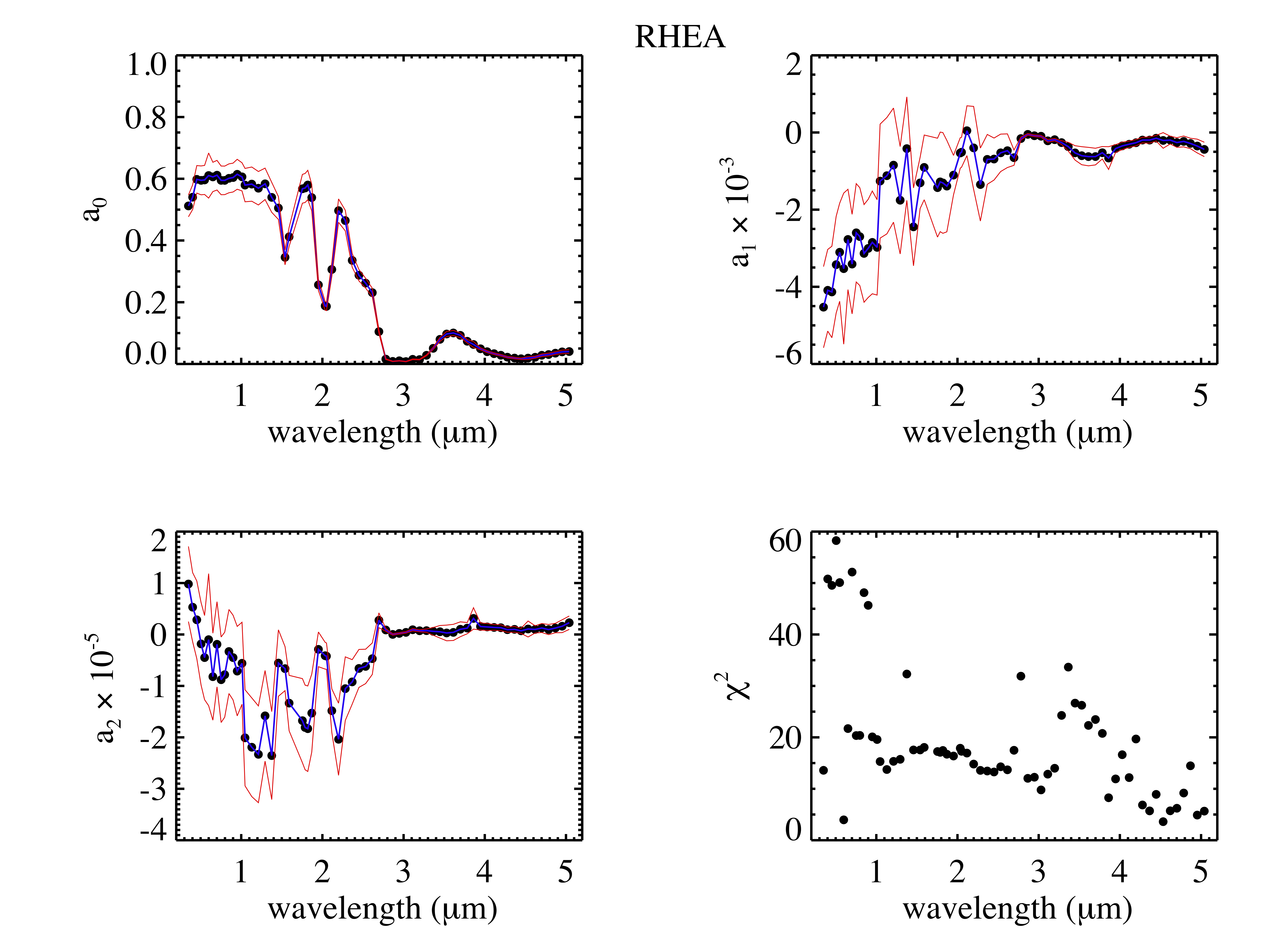}
    \caption{Rhea's photometric coefficients. Same caption as in Fig. \ref{fig:5}.}
    \label{fig:9}
\end{figure}

With respect to previous photometric analyses of Dione \citep{Filacchione2018a}, Tethys \citep{Filacchione2018b}, and Rhea \citep{Filacchione2020}, in this work we have further improved the overall photometric reduction process by implementing the following procedures:
\begin{itemize}
	\item the phase curves are derived independently on 65 VIMS spectral bands spanning between 0.35 and 5.04 $\mu$m for the entire datasets of Mimas, Enceladus, Tethys, Dione, and Rhea. The corresponding wavelengths are listed in Tables \ref{tbl:mimas_parameters}-\ref{tbl:rhea_parameters}. 

	With respect to previous works, the photometric correction has been extended to a larger number of spectral bands (65). The decision to process about one-fifth of the available VIMS bands (352) has been taken to maintain the computation time within reasonable limits. The 65 spectral bands over which the photometric correction is computed are not selected with a fixed sampling but are tuned to optimize spectral slopes and absorption band depths retrieval. 

	\item the fitting procedure has been improved by filtering out the contribution of pixels having an irregular photometric response, i.e pixels located on the limb of the satellite images where the geometry calculation is not accurate and the filling is uncertain. The phase curve data are averaged within bins of 1$^\circ$ in phase and a quadratic fit is computed on the average values plus/minus the associated standard deviation by minimizing the $\chi^2$. As an example, in Fig. \ref{fig:3}-\ref{fig:4} are shown the phase curve data (following eq. \ref{eq:5}) for Mimas at 0.599 $\mu$m and for Enceladus  at 1.047 $\mu$m, respectively. These are examples of fit results performed on scarce (13767 pixels for Mimas) and high populated datasets (189675 pixels for Enceladus). The black crosses indicate the single pixels data, red dots the average values within 1$^\circ$ phase bins, the red lines the associated standard deviation, and the blue line the resulting best fit curve. 
	
	\item the photometric parameters $a_0, a_1, a_2$ are now provided with their associated uncertainties and with the best fit $\chi^2$ value for each of the 65 spectral bands. These values are shown in Fig. \ref{fig:5}-\ref{fig:9} and tabulated in Tables \ref{tbl:mimas_parameters}-\ref{tbl:rhea_parameters}.
	
\end{itemize}

 \par 
By removing the illumination/viewing geometry effects, this method allows a direct comparison of the satellites' average spectrophotometric properties which will be discussed in detail in the next section \ref{sct:comparative_analysis}. 
The equigonal albedo ($a_0$) spectra of midsized satellites shown in the top left panels in Fig. \ref{fig:5}-\ref{fig:9} share similar water ice spectral features but with distinct properties for each satellite: 1) a red (positive) spectral slope in the 0.35-0.55 $\mu$m range, more intense on Mimas, Rhea and Dione, and weaker on Enceladus and Tethys; 2) a neutral slope in the 0.55-0.95 $\mu$m range on Rhea and progressively bluer (negative) on Mimas, Tethys, Dione, and Enceladus; 3) intense 1.5 and 2.0 $\mu$m bands due to water ice, with the latter band more intense than the former; 4) the faint water ice Fresnel peak at 3.1 $\mu$m is visible on all satellites within the broad 3.0 $\mu$m absorption band; 5) the water ice peak at about 3.6 $\mu$m is also visible on all satellites, with higher levels seen on Mimas and Rhea. Formal errors are of the order of a few percent and in general maximum values are affecting wavelengths $\lambda <$1.8 $\mu$m due to the larger variability of albedo across surfaces while they are remarkably lower at longer wavelengths where the albedo variations are smaller. At 0.549 $\mu$m the resulting equigonal albedos of the satellites are, in order of orbital distance from Saturn: 0.63$\pm$0.02 for Mimas, 0.89$\pm$0.03 for Enceladus, 0.74$\pm$0.03 for Tethys, 0.65$\pm$0.03 for Dione, and 0.60$\pm$0.05 for Rhea. These values are in agreement with the radial trend observed in the geometrical albedo ($p$) by \cite{Verbiscer2007} which have measured 0.96 on Mimas, 1.38 on Enceladus, 1.23 on Tethys, 1.00 on Dione, and 0.95 Rhea. On both analyses, the highest value is measured on the fresh and bright surface of Enceladus with a progressive reduction observed moving outwards to the outer edge of the E ring. We recall here that $p$ is, in general, higher than $a_0$ being defined as the ratio between a satellite's reflectance measured at null solar phase and that of a Lambertian disk observed at the same geometry: differently from $a_0$ as estimated in this work, the geometrical albedo, in fact, includes the opposition effect surge. 
 \par 
Linear term $a_1$ solutions are plotted in Fig. \ref{fig:5}-\ref{fig:9} top right panels. On all satellites the $a_1$ term is negative in the visible range with values of -5.95$\cdot 10^{-3}$ deg$^{-1}$ for Mimas, -1.98$\cdot 10^{-3}$ deg$^{-1}$ for Enceladus, -3.10$\cdot 10^{-3}$ deg$^{-1}$ for Tethys and Rhea, -3.57$\cdot 10^{-3}$ deg$^{-1}$ for Dione at $\lambda=0.549 \ \mu$m. The $a_1$ spectra converge towards null values in the infrared range with maximum values in correspondence of the strong water ice band at 3 $\mu$m. Similarly to $a_0$, the largest errors are associated with $\lambda <$1.8 $\mu$m.  
 \par 
Apart from Mimas, the  quadratic term $a_2$ (Fig. \ref{fig:5}-\ref{fig:9} bottom left panels) shows a trend similar to $a_1$ albeit with numerical values about two orders of magnitude smaller. On Mimas we note that the $a_2$ term is always positive and do not have any variations across the water ice bands wavelengths (apart for one point at 1.5 $\mu$m): 
this behavior is peculiar since it deviates from the rest of the other satellites' trends and it could be explained with a lower contribution of the multiple scattering which in turn makes the phase curve more linear on Mimas than on the rest of the satellites (see in example the Mimas' linear trend in Fig. \ref{fig:5} with respect to Enceladus in Fig. \ref{fig:6}).

Finally, the $\chi^2$ plots (bottom right panels) show the overall quality of the photometric fits at each wavelength.  In general, higher fluctuations of the photometric parameters are systematically associated with the visible range up to 2.5 $\mu$m and to the 1.5 and 2.0 $\mu$m absorption bands as a result of the intrinsic spectral variability of the surfaces. Beyond 2.5 $\mu$m a$_1$ and a$_2$ are almost constant around the null value.
Apart from these effects, we note also systematic instrumental effects around 1 $\mu$m, where the bridging between the two VIMS spectral channels happens, and in the 1.1-1.3 $\mu$m range where high signals levels, close to the saturation regime, preferentially occur due to the maximum instrument response at these wavelengths. 
\par 
With respect to the previous photometric analysis of Enceladus data by \cite{Robidel2020}, our analysis finds systematically lower equigonal albedo values on both continuum and band centers: at $\lambda \approx$ 1.36 and 2.0 $\mu$m equigonal albedo is 0.726 and 0.217 in this work (Table \ref{tbl:enceladus_parameters}) while \cite{Robidel2020} (Table 1) reports 0.771 and 0.242, respectively. This difference of about 5-10$\%$ is probably due to the different VIMS dataset processed and photometric fit method (linear vs. quadratic) and should be taken into account when comparing equigonal albedos from different authors.

\clearpage

\section{Equigonal Albedo and Spectral Indicators Maps for Saturn's icy satellites}
\label{sct:albedo_maps}

Albedo maps of the midsized satellites are systematically built for 65 visible and infrared wavelengths by applying the following procedure: 
\begin{itemize}
\item \emph{a)} VIMS I/F spectra acquired with incidence angle $i\le70^\circ$, emission angle $e\le70^\circ$, phase angle $10^\circ \le g \le 90^\circ$, Cassini-satellite distance $D \le 100.000$ km and unsaturated signal are photometrically converted in equigonal albedo $a_{0}(\lambda,n)$ from eq. \ref{eq:5}: 

\begin{equation} 
a_{0}(\lambda, n)=\frac{\frac{I}{F}(\lambda,n)}{D(i,e,g,n)}-a_{1}(\lambda)g(n)-a_{2}(\lambda)g(n)^2
\label{eq:6}
\end{equation}

where n is the n-th spectrum intercepting a given $0.5^\circ \times 0.5^\circ$ tile on the longitude/latitude grid
and $a_{1}(\lambda)$, $a_{2}(\lambda)$ are the average photometric parameters listed in Tables \ref{tbl:mimas_parameters}-\ref{tbl:rhea_parameters}. Geometry parameters (i,e,g) corresponding to the n-th spectrum are computed following the method discussed in section \ref{sct:vims_dataset}. Note that, with respect to the fits performed to retrieve the photometric parameters, here we reduce the variability of the phase angle to $g < 90^\circ$ on the albedo maps as a strategy to maximize spatial coverage while minimizing the occurrence of seams among data taken at extreme phases. However, despite this optimization residual seams persists on some satellite's maps at certain wavelengths;
\item \emph{b)} for each tile, the median value of the $a_{0}(\lambda, n)$ distribution is computed and assigned to a location (lat,lon) on the corresponding map (Fig. \ref{fig:10}-\ref{fig:14}). The projection of the pixel footprint makes use of the pixel center and four corners locations sampled on the reference grid. This is the same method implemented to build similar maps on previous works \citep{Filacchione2016a, Filacchione2016b, Filacchione2018b, Filacchione2018a, Filacchione2020}. The number of spectra available on each tile is shown in the corresponding redundancy map for each satellite (for sake of brevity we report redundancy only for the VIS channel at 0.55 $\mu$m; similar distributions are in general found at other wavelengths);
\item \emph{c)} in Fig. \ref{fig:10}-\ref{fig:14} are shown visible (0.55 $\mu$m) and infrared (1.82 $\mu$m) equigonal albedos and four spectral indicators: \emph{1)} 0.35-0.55 $\mu$m spectral slope; \emph{2)} 0.55-0.95 $\mu$m spectral slope; \emph{3)} water ice 1.5 $\mu$m band depth; \emph{4)} water ice 2.0 $\mu$m band depth. Both spectral slopes are computed as the angular coefficient of the line best fitting equigonal albedo data in the respective spectral range after normalization at 0.55 $\mu$m \citep{Filacchione2012}. Spectral slopes are very diagnostic of water ice grains distribution and the presence of contaminants and chromophores mixed in water ice particles: when the water ice grains increase from fine to coarse sizes, the 0.35-0.55 $\mu$m slope becomes redder while the 0.55-0.95 $\mu$m turns out bluer. For a given water ice grain distribution the presence of a small amount of contaminants and chromophores ($< 1 \%$ in both intraparticle or intramolecular mixing) can introduce a significant reddening on the 0.35-0.55 $\mu$m slope \citep{Clark2008, Clark2012}. Conversely, the 0.55-0.95 $\mu$m slope is influenced by the presence of dark contaminants in intimate and areal mixing. 
Both water ice band depths are computed as $1-a_c/a_b$ \citep{Clark1999} where $a_c$ and $a_b$ are the albedos corresponding to the local continuum (interpolated at the wavelength of the band minimum) and to the band minimum. Also, band depths, like spectral slopes, are dependent on the water ice grain distribution and the abundance of contaminants \citep{Filacchione2012}.   
\end{itemize}

\subsection{Mimas}
\label{sct:mimas}
\begin{figure}[h!]
\centering
	\includegraphics[width=18cm]{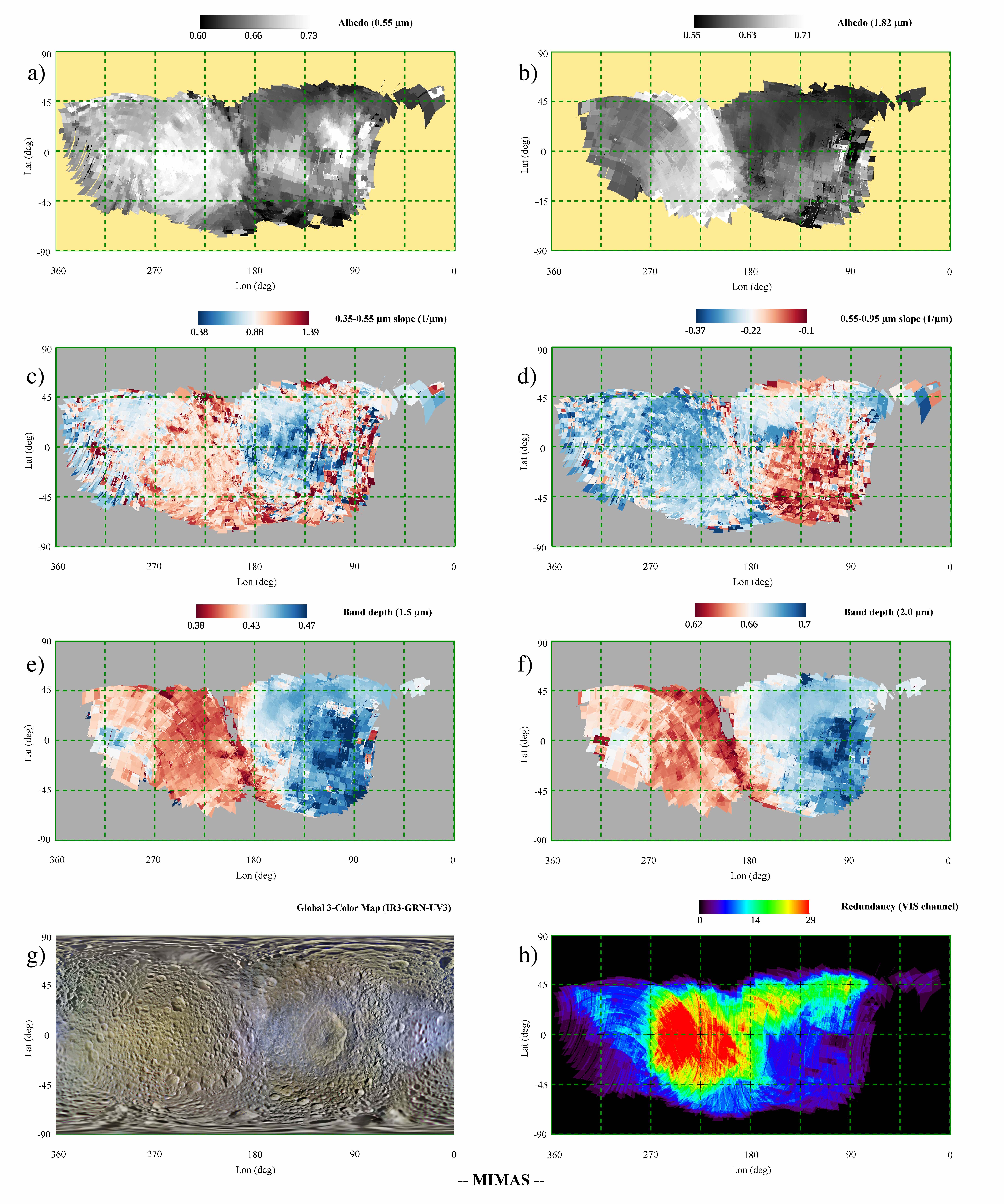}
    \caption{Mimas photometric-corrected maps rendered at 1.7 km/bin (at equator) in simple cylindrical projection.  Panel a): Equigonal albedo at 0.55 $\mu$m; b): Equigonal albedo at 1.82$\mu$m; c): 0.35-0.55 $\mu$m spectral slope; d): 0.55-0.95 $\mu$m spectral slope; e) 1.5 $\mu$m water ice band depth; f) 2.0 $\mu$m water ice band depth; g) ISS global color (IR, Green, UV) map at 200 m/px resolution (PIA18437, image credit P. Schenck, NASA/JPL-Caltech/Space Science Institute/Lunar and Planetary Institute); h): Visible channel data redundancy. On all maps: $0^\circ \leq lon \leq 180^\circ$ defines the leading hemisphere, $180^\circ \leq lon \leq 360^\circ$ is the trailing hemisphere, $lon=0^\circ$ is the saturnian meridian, $lon=180^\circ$ is the antisaturnian meridian.}
    \label{fig:10}
\end{figure}

Mimas albedo and spectral indicators maps are shown in Fig. \ref{fig:10}. An enhanced color map of Mimas' surface built from a mosaic of Cassini ISS images is shown in Fig. \ref{fig:10}-panel g), to identify morphological features and for comparison with VIMS maps. Albedo maps evidence a general brightening across the antisaturnian-trailing hemisphere ($180^\circ \le lon \le 270^\circ$) at both visible (panel a) and infrared (panel b) wavelengths. Across this quadrant the 0.55 and 1.82 $\mu$m equigonal albedos reach maximum values of 0.73 and 0.71, respectively. Conversely, the minimum values are 0.60 and 0.55, respectively, across the part of the leading hemisphere surface observed by VIMS. A similar albedo difference is in agreement with previous telescopic results by \cite{Verbiscer1992} and \cite{Buratti1998}. While the 0.55 $\mu$m albedo map is scarcely correlated with the median phase  angle associated with the observations (see Fig. \ref{fig:2} first row, third panel), not the same is happening on the 1.82 $\mu$m albedo where we observe lower values in correspondence of high phases conditions. This effect is causing the residual seams on the maps. In principle, those effects can be minimized by reducing the phase angle interval below the $90^\circ$ upper limit but at the cost of greatly reducing the overall spatial coverage and redundancy on a given map bin. For this reason, we prefer to maintain these criteria for all satellites since the weight of high phase observations is reduced on the rest of the satellites' datasets. The large impact crater Herschel on the leading hemisphere (centered at $lon=110^\circ$, $lat=0^\circ$) is the only morphological feature resolved. In general albedo images are strongly affected by local rough morphology: this happens because the geometries used to correct the single images are computed over the shape's ellipsoid and therefore larger uncertainties are found wherever local morphology deviates from it. Such uncertainties are expected on craters' rims, chasmata, and other rough morphologies. On Herschel's rims, the effect is particularly evident being the walls about 5 km high above the crater's floor while the map resolution is 1.7 km/bin. The "Thermal Anomaly Region" (TAR) lens \citep{Howett2011} located on the equatorial region of the leading hemisphere, and extending east and west of Herschel crater, appears brighter than the rest of the leading hemisphere on the 0.55 $\mu$m albedo map but it is not evident at 1.88 $\mu$m. The regions northward and southward the TAR are the darker units with an albedo of about 0.6 and 0.55 on the 0.55-1.82 $\mu$m albedo maps, respectively.
 \par 
 The 0.35-0.55 $\mu$m spectral slope (Fig. \ref{fig:10})-panel c) shows a similar contrast between the two hemispheres being redder (up to 1.39 1/$\mu$m) across the bright trailing side while across the TAR it appears much less red ($<$0.45 1/$\mu$m). Conversely, the 0.55-0.95 $\mu$m slope (panel d) is quite uniform and blue ($\le$ -0.25 1/$\mu$m) except for the Herschel crater and nearby southern terrains where the slope is almost neutral.    
The water ice band depths maps at 1.5 and 2.0 $\mu$m (panels e-f) evidence a well-defined contrast across the surface: the higher band depth values (respectively 0.47 and 0.7) are measured across the leading hemisphere, in particular above part of Herschel rim and southward areas while the bands are systematically fainter on the trailing hemisphere. We note that the band depth maps do not evidence the presence of the TAR lens like on the 0.55 $\mu$m albedo map. On the contrary, the 1.5 $\mu$m band map shows a local maximum on the equatorial region of the trailing hemisphere (at lon$\approx300^\circ$) which is much less evident on the 2.0 $\mu$m band map.

As reported in Table \ref{tbl:dataset} and shown in the redundancy map (computed at 0.55 $\mu$m, panel h), Mimas maps renderings are strongly affected by the scarcity of the available dataset. In particular, on the equatorial and south regions of the leading hemisphere and the trailing-Saturnian quadrant, only a few observations are at hand.  
This is a consequence of the Cassini spacecraft's trajectory which was preferentially flying outside Mimas' orbit resulting in limited possibilities to observe near the Saturnian meridian ($lon=0^\circ$).
\par

The spatial distribution of the spectral indicators we have presented here reflects both the surface's original state and successive endogenic and exogenic activities. By orbiting inside Enceladus' orbit, Mimas' trailing hemisphere surface is altered by three exogenic processes: 1) the accumulation of E-ring fine ice particles; 2) the bombardment of cold plasma particles and 3) the irradiation of low energy ($<$ 1 MeV) electrons from Saturn's magnetosphere. On the contrary, the leading side is affected only by high energy ($>$ 1 MeV) electrons flux \citep{Schenk2011, Nordheim2017, Howett2018} whose trajectories are focused within the TAR lens extending across the leading hemisphere equatorial region. The TAR is visible as a bluish oval on the ISS mosaic map (bottom left panel in Fig. \ref{fig:10}). The high energy electron bombardment's flux focusing on the TAR lens is altering not only optical response but also thermal behavior: the lens is characterized by high thermal inertia \citep{Howett2011} resulting in diurnal temperatures about 10 K colder than nearby regions outside it \citep{Filacchione2016b}.
 \par 
As a consequence of the localization of these phenomena, the high latitude regions of the leading hemisphere are the places less influenced by exogenic processes, apart from the falling of interplanetary dust particles \citep{Spahn2006, Poppe2016} and photolysis, which occur across the entire surface. The spectral differences visible on the VIMS maps we have previously described can be correlated with these phenomena. For example, the observed brightening and negative 0.55-0.95 $\mu$m slope of the trailing hemisphere is a consequence of the fine E-ring icy grains deposition on the surface. On the contrary, the minimum albedos are associated with the more pristine regions located at the high latitudes of the leading hemisphere. The high value of the 0.35-0.55 $\mu$m slope on the trailing hemisphere can be explained by the low energy electrons and cold plasma fluxes which can induce reddening in this spectral range. 
From a qualitative analysis based on comparison with pure water ice band reference values \citep{Filacchione2012}, the minima observed across the trailing hemisphere (0.38 and 0.62 for 1.5 and 2.0 $\mu$m bands, respectively) correspond to large cm-sized grain diameters while the maxima band depths (0.47 and 0.7) visible on the leading are compatible with much smaller (100 $\mu$m-0.5 cm) grains. We remark that water ice band depth intensities are anticorrelated with visible and infrared albedos: this is a shred of strong evidence that here we are not seeing only the effects of the water ice abundance but also of the regolith grain size (see i.e. Fig 15.27 in \cite{Cuzzi2009}). At the same time albedos are correlated with the 0.35-0.55 $\mu$m slope, meaning that brighter terrains are the ones with maximum reddening: this is the same behavior observed on A and B rings particles \citep{Nicholson2008, Cuzzi2009, Filacchione2014, Cuzzi2018a, Ciarniello2019} where the presence of a UV absorber material intimately mixed with the water ice particles together with a neutral absorber are used to reproduce a similar behavior. While the UV absorber causes the visible reddening without affecting the albedo, the neutral absorber develops a concurrent reduction of both slope and albedo.   
 \par 
 The fact that the TAR is not recognizable on the band depths maps is a remarkable difference with respect to the similar Tethys' TAR which we'll discuss in section \ref{sct:tethys}. Why Mimas lens is well-defined at visible wavelengths (Fig. \ref{fig:10}-panel a) while it disappears on the infrared albedo (panel b) and water ice bands maps (panels e-f)? A possible answer to this question is the presence of the 140 km-wide crater Herschel in the middle of the TAR area whose impact has caused a "reset" of the effects of the radiation accumulated before the event by mixing the surface material and altering the grain size distribution. Since visible photons are much less penetrating within the grains on the surface than the infrared ones, one could speculate that the alteration due to high energy electrons within the TAR is important only at very shallow skin depth ($\lesssim 1 \mu$m) while at infrared wavelengths, where the skin depth is larger ($\approx 10 \mu$m), the band depths are less responsive to the irradiation effects.     
 \par 
We note also that the 1.5 $\mu$m band depth growth (panel e) occurring around the trailing equator at lon$\approx300^\circ$ is compatible with the position of the second low energy electrons lens predicted by \cite{Nordheim2017} model but never observed on remote sensing visible and thermal infrared data. With respect to the leading lens here the energy of the electrons is much smaller resulting in a more shallow penetration depth. Apart from this, the band depth increase on the 1.5 $\mu$m map could be also related to alterations caused by the cold plasma flux or by E-ring grains deposition. 
\par
A more quantitative interpretation of the albedo spectra needs the application of radiative transfer codes \citep{Hapke1993a} in which several spectral parameters are inferred together, including the amounts and composition of the contaminants (amorphous carbon, tholins, silicates) mixed with water ice; different mixing modalities (areal, intimate, intraparticle, coating); and grain size distribution. This further effort will be the argument of a next paper in which we plan to derive composition maps of the midsized satellites by adopting a method similar to the one employed for disk-integrated observations \citep{Ciarniello2011, Filacchione2012}.

\clearpage
\subsection{Enceladus}
\label{sct:enceladus}
\begin{figure}[h!]
\centering
	\includegraphics[width=18cm]{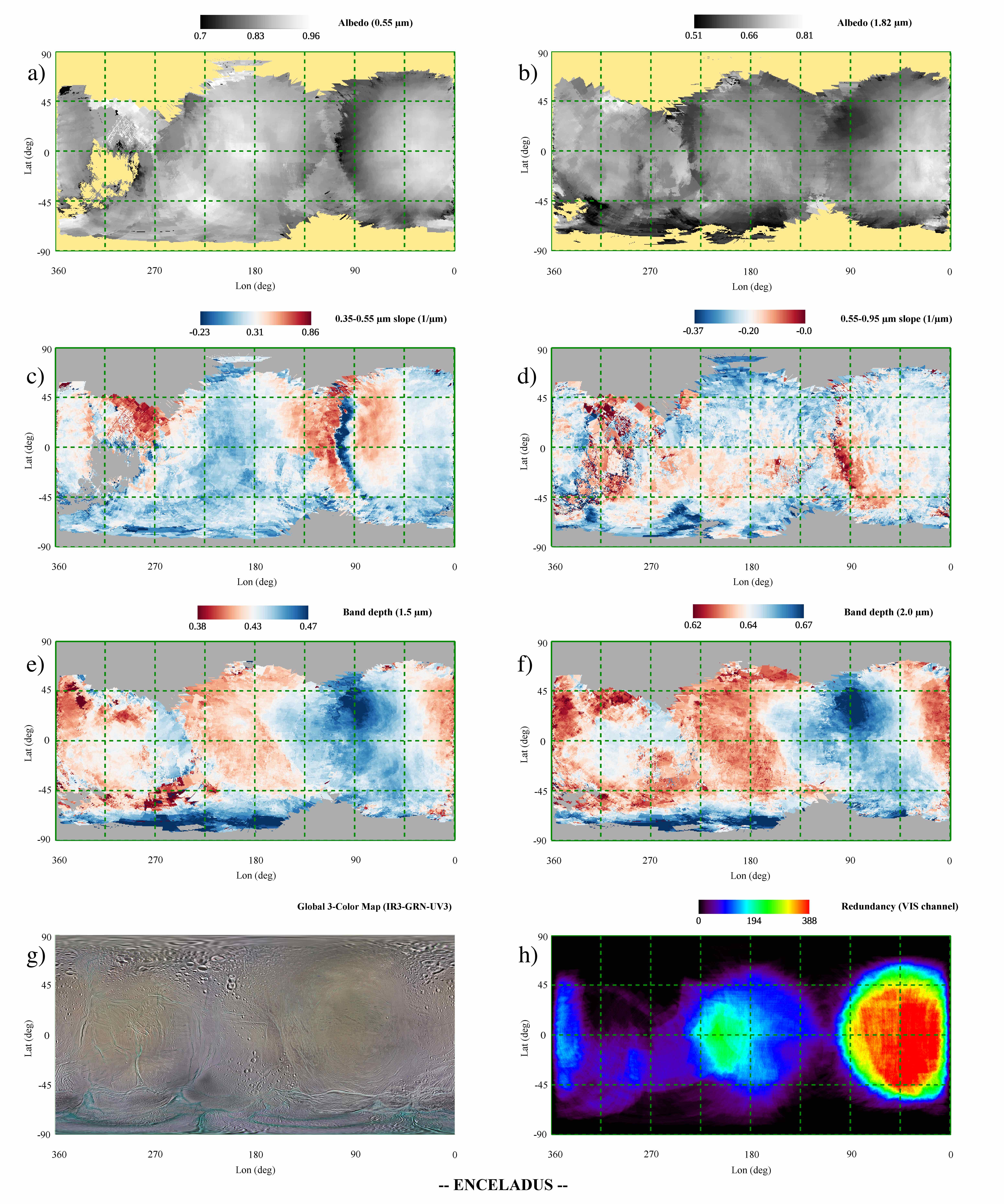}
    \caption{Enceladus photometric-corrected maps rendered at 2.2 km/bin (at equator) in simple cylindrical projection.  Panel a): Equigonal albedo at 0.55 $\mu$m; b): Equigonal albedo at 1.82$\mu$m; c): 0.35-0.55 $\mu$m spectral slope; d): 0.55-0.95 $\mu$m spectral slope; e) 1.5 $\mu$m water ice band depth; f) 2.0 $\mu$m water ice band depth; g) ISS global color (IR3, Green, UV3) map at 100 m/px resolution (PIA18435, image credit P. Schenck, NASA/JPL-Caltech/Space Science Institute/Lunar and Planetary Institute); h): Visible channel data redundancy.}
    \label{fig:11}
\end{figure}

Enceladus maps are shown in Fig. \ref{fig:11} following the same scheme adopted for Mimas. In this case, we are analyzing one of the brightest objects of the solar system with the equigonal albedo reaching a value of 0.96 at 0.55 $\mu$m above the antisaturnian hemisphere (panel a). The poor data redundancy across a great part of the trailing hemisphere and between $90^\circ \le lon \le 100^\circ$ on the leading one (Fig. \ref{fig:11}-panel h) together with the large excursion of the median phase angle across the dataset (Fig. \ref{fig:2} second row, third panel) are causing the occurrence of several residual seams which are particularly evident across the 0.55 $\mu$m albedo map (panel a) but less disturbing at 1.82 $\mu$m (panel b) and on the spectral indicators ones. 
\par
Noteworthy, at 1.82 $\mu$m the albedo systematically drops to low values (0.51) approaching the southern high latitudes: this trend appears independent from the phase angle distribution.
The 0.35-0.55 $\mu$m spectral slope is remarkably negative (colored in blue in Fig. \ref{fig:11}-panel c) across the antisaturnian hemisphere while it becomes positive, up to 0.86 1/$\mu$m across a wide area on the leading side between $45^\circ \le lon \le 135^\circ$ and on the north trailing side between $250^\circ \le lon \le 320^\circ$.
\par
As said before, an evident photometric residual seam is visible as a north-south feature with a low slope at $90^\circ \le lon \le 100^\circ$. The variability of the 0.55-0.95 $\mu$m slope (panel d) is much reduced across the map: the higher (null) values are reached in correspondence of photometric residuals (colored in red). The minimum slope values are measured on the southern hemisphere between $200^\circ \le lon \le 270^\circ$. Better renderings are achieved on the 1.5-2.0 $\mu$m band depth maps (panels e-f): here we observe the maximum values, up to 0.47 and 0.67 respectively for the 1.5 and 2.0 $\mu$m bands (color-coded in blue), across a wide area approximately centered on the leading side at $(lon, lat)=(90^\circ, 30^\circ)$. From here a region with high band depth values departs, spanning southward up to $lat\approx-60^\circ$ (coverage limit) across a wide region of the leading hemisphere. High band depth values are observed also across the south polar regions for latitudes above $-60^\circ$. On the equatorial part of the trailing hemisphere, another region of local enhancement of the 2.0 $\mu$m band depth is visible. 
\par
Despite some photometric residuals, VIMS maps allow the identification of some distinctive morphological features visible on the ISS IR3-Green-UV3 color mosaic map (panel g). The South Pole Terrains (SPT), where the active "Tiger Stripes" are located, are partially covered on the VIMS maps, including the two northernmost features, Damascus and Alexandria sulci, located at $lat\approx-60^\circ$ along the trailing and leading hemisphere, respectively. 
The comparison with ISS IR3-Green-UV3 color mosaic map displays higher water ice band depths distribution (panels e-f) correlated with the two yellowish units visible across the middle of the leading and trailing hemispheres. 
\par
The location of the wide feature centered at $(lon, lat)=(90^\circ, 30^\circ)$ characterized by low infrared albedo, high (positive) 0.35-0.55 $\mu$m slope, and maximum band depth values has been already reported in previous works based on VIMS data \citep{Scipioni2017, Combe2019, Robidel2020}. This spectral feature matches the morphological smooth unit located on the leading hemisphere reported in \cite{CrowWillard2015}. As noted by \cite{Robidel2020}, this position roughly corresponds to the origin of the 2.17 cm thermal emission measured by the Cassini RADAR radiometer \citep{Ries2015}. 
The emission could be induced by the local topography and ice shell thickness \citep{Cadek2016, Choblet2017}. These works conclude that the ice layer is thinner in this region, which could indicate the presence of hot spots similar to the Tiger Stripes. 
Different hypotheses about the origin of this emission are evaluated also by \cite{Ries2015}, concluding that a diapir buried at about 3-5 m below the surface is a more probable source than a thermal inertia anomaly caused by electron bombardment or a local micrometeoroid bombardment. From a spectral perspective, we note that this location shares remarkably similar albedo and band depth properties with the active tiger stripes area. A local alteration of the ice caused by thermal flux and upwelling of internal material could be the processes at the base of these shreds of evidence. In this respect, the infrared band depths maps could give useful information to constrain the dimensions of a putative diapir for future studies. 
\par
Moreover, we note that VIMS data do not show significant spectral variations in correspondence of the two broad wave-shaped plume redeposition areas placed along $lon=45^\circ$ and $205^\circ$ meridians foreseen by models \citep{Kempf2010, Southworth2019}. Finally, the data redundancy map (panel h) evidences how the VIMS dataset is preferentially covering the leading hemisphere between $0^\circ \le lon \le 90^\circ$ and the antisaturnian region while the trailing hemisphere has been scarcely observed. The residual seams on visible albedo and spectral slopes maps are correlated with the distribution of low redundancy data.

\clearpage
\subsection{Tethys}
\label{sct:tethys}
\begin{figure}[h!]
\centering
	\includegraphics[width=18cm]{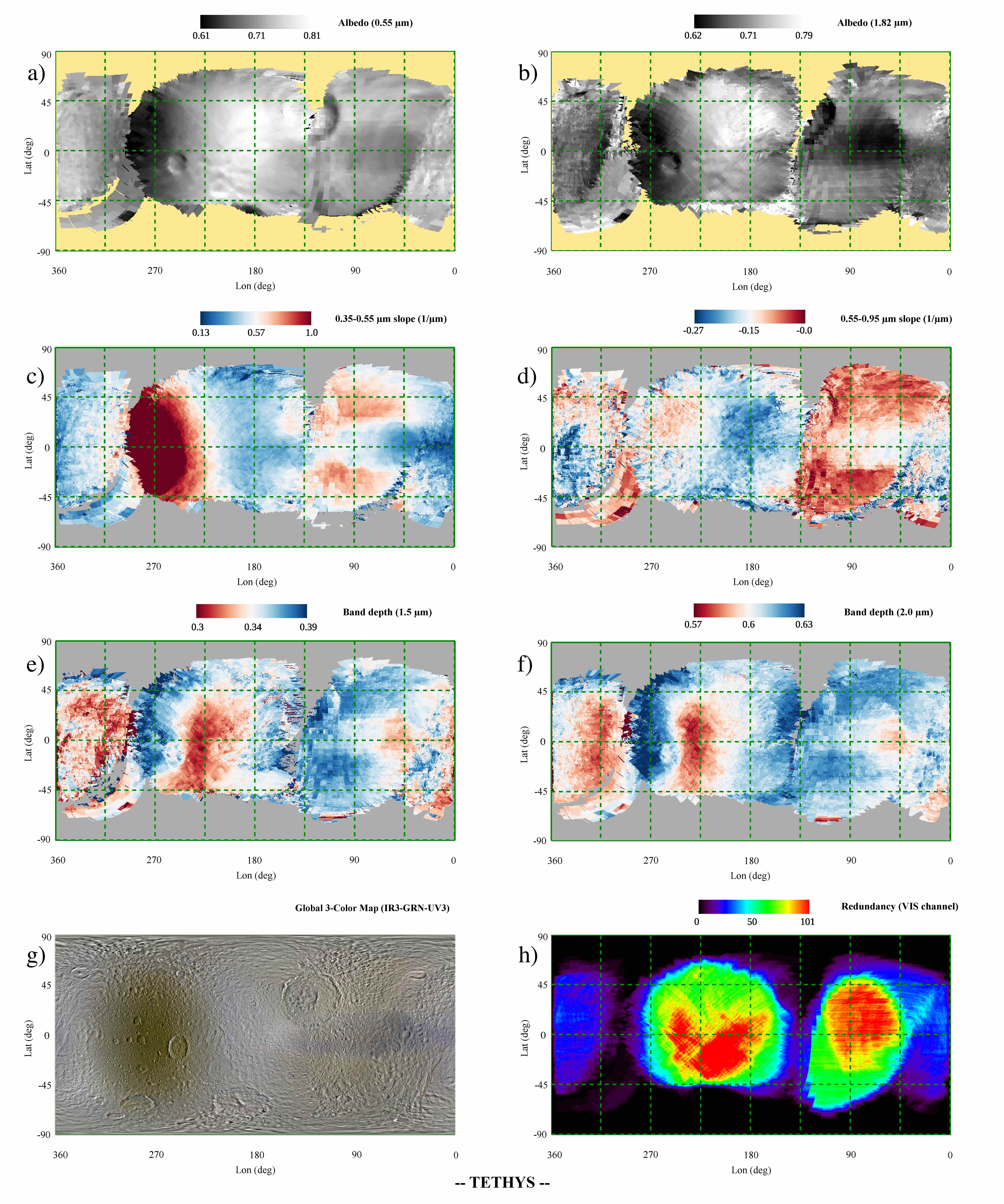}
        \caption{Tethys photometric-corrected maps rendered at 4.7 km/bin (at equator) in simple cylindrical projection.  Panel a): Equigonal albedo at 0.55 $\mu$m; b): Equigonal albedo at 1.82$\mu$m; c): 0.35-0.55 $\mu$m spectral slope; d): 0.55-0.95 $\mu$m spectral slope; e) 1.5 $\mu$m water ice band depth; f) 2.0 $\mu$m water ice band depth; g) ISS global color (IR, Green, UV) map at 250 m/px resolution (PIA18439, image credit P. Schenck, NASA/JPL-Caltech/Space Science Institute/Lunar and Planetary Institute); h): Visible channel data redundancy.}
            
    \label{fig:12}
\end{figure}

Several morphological features are visible on Tethys' maps including impact craters, the equatorial lens on the leading hemisphere, and a dark area on the trailing side (Fig. \ref{fig:12}): the largest craters resolved are Odysseus (445 km in diameter, placed at lon=128$^\circ$, lat=32$^\circ$), Penelope (207 km, 249$^\circ$, -10$^\circ$), Dolius (190 km, 210$^\circ$, -30$^\circ$) and Telemachus (92 km, 339$^\circ$, 54$^\circ$). Ithaca chasma, a thousand-km long, 100 km wide, and up to 5 km deep trough feature is identified across the subsaturnian meridian and running from about (lon, lat)=(35$^\circ$, -65$^\circ$) to (330$^\circ$, 65$^\circ$) and crossing Telemachus crater. The highest albedo values (Fig. \ref{fig:12}-panels a-b) are observed across the antisaturnian quadrant where values as high as $\approx$0.8 are measured on both 0.55 and 1.82 $\mu$m. The darkest areas of the surface correspond with thermal anomaly lens completely resolved across the equatorial region of the leading hemisphere and with the trailing hemisphere where albedo drops to about 0.6.
In general, Tethys maps show few seams due to photometric residuals in particular across the trailing  (lon=290$^\circ$) and leading (lon=135$^\circ$, visible in particular on the 0.55-0.95 $\mu$m slope map in panel d) hemispheres. By comparing with Fig. \ref{fig:2}, the first is caused by an abrupt change in emission and incidence angles, the latter by a set of high phase observations. Also, the not uniform redundancy distribution (Fig. \ref{fig:12}-panel h) is contributing to the presence of seams whenever there is any discontinuity in data availability. 
\par
Moreover, as a result of the application of a photometric correction computed on the average surface properties, the albedo maps are less accurate in correspondence of very bright and very dark terrains. Similar seams are visible also on the spectral indicators maps. The 0.35-0.55 $\mu$m spectral slope (panel c) is the parameter which allows the identification of the low albedo units previously described: while the dark terrains on the trailing side appear remarkably red (up to 1 $1/\mu$m), the TAR lens is on the contrary more neutral (0.13 $1/\mu$m) and similar to the bright terrains in the antisaturnian quadrant. The two halves of the "Pacman" feature are the reddest units visible north and southward of the TAR lens between $45^\circ \leq lon \leq 140^\circ$ on the leading side. The 0.55-0.95 $\mu$m slope map shows a distribution similar to 0.35-0.55 $\mu$m, at least for the leading side, with much less contrast and with more noise due to the photometric seams. The 1.5 and 2.0 $\mu$m band depth maps (panels e-f) show similar trends: across the leading hemisphere, the minimum band depths values are encountered within the TAR lens in the proximity of lon=45$^\circ$, a behavior similar to the one shown by the 0.35-0.55 $\mu$m spectral slopes. The band depths increase noticeably across the Pacman feature, on the floor of Odysseus crater, and above the southern part of Ithaca chasma. The distribution of the band depths on the trailing hemisphere is biased by photometric residuals in an area running along lon=$270^\circ$ meridian and recognizable on the map as a deep blue-colored area with high band depth values. Outside it, the bands' depth reaches the minimum values which are correlated with the spatial distribution of the dark and red units. 
Apart from the area between $360^\circ \le lon \le 320^\circ$ on the 1.5 $\mu$m map, we observe a strong correlation between the two water ice band depths maps and the dark units visible on the ISS color mosaic map (Fig. \ref{fig:12}-panel g).
\par   
The distribution of the albedo and spectral indicators across Tethys' surface is driven by exogenic processes similar to the ones already discussed in the Mimas section \ref{sct:mimas}. The maximum accumulation of E ring grains on the surface of Tethys is foreseen above the saturnian and antisaturnian hemispheres \cite{Kempf2018, Howett2018} where effectively VIMS measures the maximum albedo values (Fig. \ref{fig:12}-panels a-b). Instead, on Dione and Rhea that are orbiting farther from Enceladus, the flux of E ring particles at apoapsis is preferentially impacting on their leading hemispheres \citep{HamiltonBurns1994, Schenk2011, Howett2018}. On these satellites, VIMS observes an evident brightening across their respective leading hemispheres (see discussion on next sections \ref{sct:dione}-\ref{sct:rhea}). 
\par
The darkening visible across Tethys' trailing hemisphere can be traced back to the bombardment of cold plasma and implantation of small dust grains \citep{Schenk2011}: the high reddening and low band depths visible on the spectra are a result of these processes which can induce chemical alteration in the dominant water ice.    
Despite the formation of the TAR lenses on Mimas and Tethys are caused by the same process, e.g. the bombardment of the surface by high energy magnetospheric electrons able to penetrate within ice particles to 0.5-5 cm depths \citep{Zombeck1982}, their spectral properties appear strikingly different. On Tethys, the lens has minima 1.5-2.0 $\mu$m band depth values (respectively at 0.3 and 0.57) with respect to the rest of the surface, while on Mimas, on the contrary, they reach almost maximum values (0.45, 0.68). 
The comparison of such band depth values with reference water ice samples \citep{Filacchione2012} turns out that regolith's texture within Tethys' lens is dominated by tens of micron grain sizes while on Mimas by cm-sized particles. 
\par
Why this great difference in grain size distribution across the two lenses when they are generated by the same physical process? One possibility is that this difference is related to the different fragmentation of the grains by high-energy electrons bombardment. This hypothesis fails because the lower flux of electrons at Tethys' orbital distance, smaller by about 1/6 than at Mimas \citep{Paranicas2012}, implies a predominance of small grains where the flux is larger and vice versa. A similar trend is in disagreement with the VIMS data that show an opposite trend. To reconcile electrons bombardment energies with TARs' grain size distributions one needs to take into account the different cratering history of the two satellites: on Mimas, the lens region has been greatly processed by the impact of the giant crater Herschel which is located exactly on the middle of it. Conversely, on Tethys, the lens area does not show records of large impacts being the rim of the impact crater Odysseus placed immediately outside it.
This means that apart from electronic bombardment, the band depth properties within the two TARs are also influenced by the underlying grain size distribution driven by impacts. 
As reported in \cite{Howett2011, Howett2012}, the electron bombardment can mobilize water ice molecules and cause grains sintering resulting in an increase of both thermal inertia and conductivity. VIMS results indicate that this process is effective at both micron (on Tethys) and centimeter (on Mimas) grain sizes since thermal anomalies have been measured across the two lenses at infrared wavelengths \citep{Filacchione2016b} and they appear in good agreement with spatial distributions measured in the far-infrared by \cite{Howett2010}.  
\par
On the contrary, the electronic bombardment is influencing, in the same way, the visible spectral range where we observe similar trends of the albedo and 0.35-0.55 $\mu$m slope with both of them decreasing above Mimas' and Tethys' lenses. We interpret these results as a consequence of the fact that at visible wavelengths the optical skin depths at which photons are reflected/scattered is very shallow, much less than typical grain sizes. This explains why these indicators are much less sensitive to the grain size distribution with respect to band depth while being very responsive to the darkening and spectral slope changes occurring on the reflectance spectra. 
\par
As a final remark, we note that the low albedo - high reddening area visible on the center of Tethys's trailing hemisphere is originated from the accumulation of cold plasma and dust. A further discussion about Tethys's preliminary albedo maps properties is given in \cite{Filacchione2018b}.

\clearpage
\subsection{Dione}
\label{sct:dione}
\begin{figure}[h!]
\centering
	\includegraphics[width=18cm]{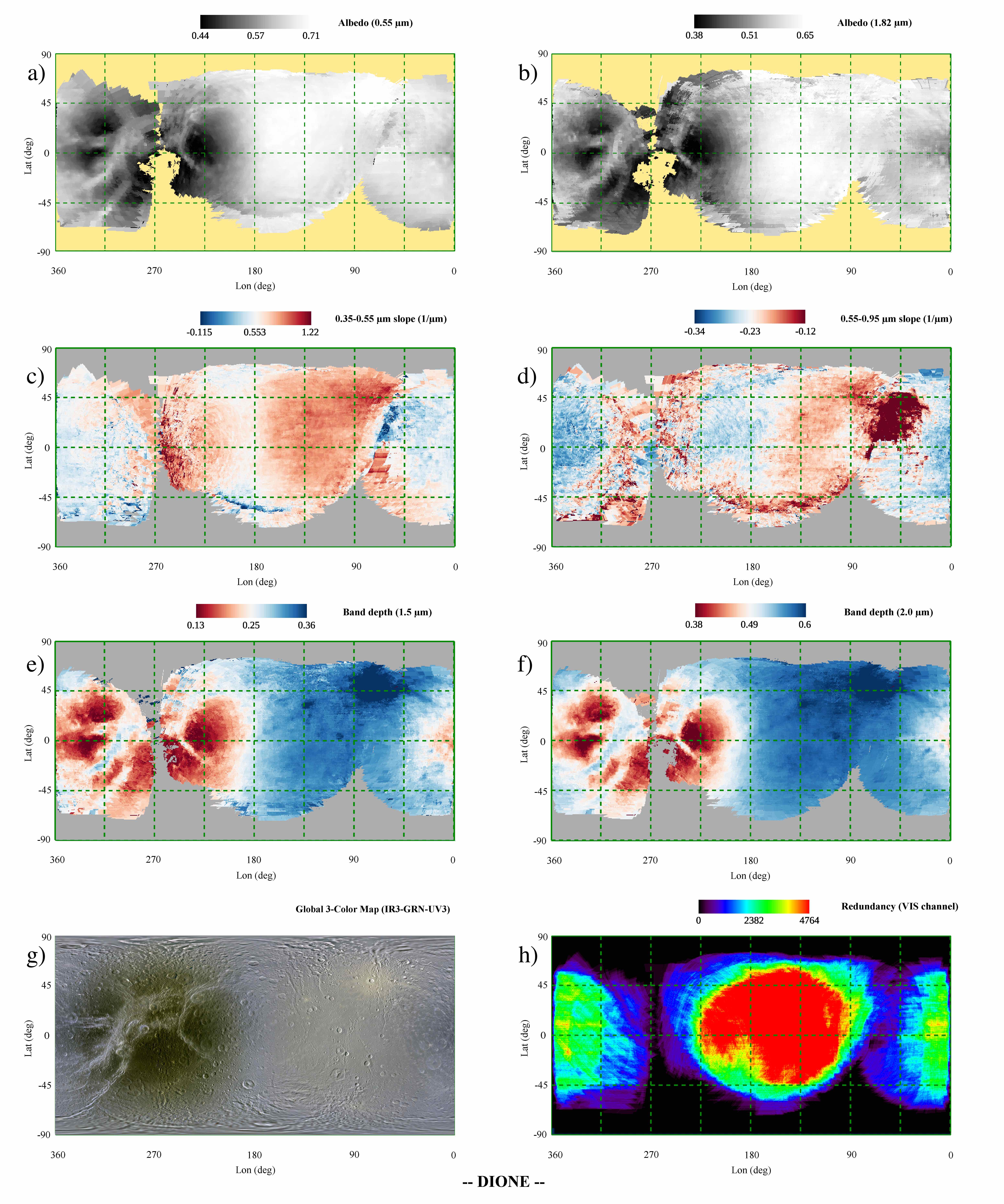}
            \caption{Dione photometric-corrected maps rendered at 4.5 km/bin (at equator) in simple cylindrical projection.  Panel a): Equigonal albedo at 0.55 $\mu$m; b): Equigonal albedo at 1.82$\mu$m; c): 0.35-0.55 $\mu$m spectral slope; d): 0.55-0.95 $\mu$m spectral slope; e) 1.5 $\mu$m water ice band depth; f) 2.0 $\mu$m water ice band depth; g) ISS global color (IR, Green, UV) map at 250 m/px resolution (PIA18434, image credit P. Schenck, NASA/JPL-Caltech/Space Science Institute/Lunar and Planetary Institute); h): Visible channel data redundancy.}
    \label{fig:13}
\end{figure}

The photometric correction method we have developed achieves the best rendering results on Dione and Rhea for which high redundancies, up to thousands of spectra, are available (Fig. \ref{fig:13}-\ref{fig:14}-panel h). Photometric seams and residuals are limited to very few areas on Dione's maps in correspondence of lower data redundancy conditions, like around lon=$50^\circ$, lat=$30^\circ$. Among midsized satellites, Dione shows the highest contrast between the bright leading and dark trailing hemispheres (0.71/0.44 at 0.55 $\mu$m and 0.65/0.38 at 1.82 $\mu$m, see Fig. \ref{fig:13}-panels a-b). The wispy terrains, or chasmata, across the trailing hemisphere are recognizable as a bright network developing within the dark area. Spectral slopes maps (panels c-d) are evidencing also the differences between the two hemispheres being the high albedo terrains the ones showing more red colors, like across the leading side and above the wispy terrains, whereas the low albedo units are associated with negative slopes. The two water ice band depth maps (panels e-f) evidence very similar spatial distributions: maximum values of 0.36 and 0.6 (in deep blue color on maps) are measured above the north region of the leading hemisphere roughly centered on Creusa crater (lon=$75^\circ$, lat=$45^\circ$, easily recognizable on the ISS color mosaic image in Fig. \ref{fig:13}-panel g) including the ejecta rays and the nearby terrains roughly extending up to Tibur Chasmata and Arpi Fossa; a similar behavior is observed on impact craters Lucagus-Liger (lon=$135^\circ$, lat=$25^\circ$) and Sagaris (lon=$105^\circ$, lat=$5^\circ$) too. All these craters have a similar response at visible wavelengths being much redder with respect to surrounding terrains on the spectral slopes maps (see panels c-d). The minimum band depths values (0.13 and 0.38 for the 1.5 and 2.0 $\mu$m bands, respectively) across Dione's surface are measured above the dark terrain units of the trailing hemisphere. In comparison, the more fresh ice within the wispy terrains shows higher band depths.
The distribution and properties of the spectral indicators are in agreement with the expected brightening caused by the E-ring particles across the leading hemisphere and darkening due to the cold plasma and dust flux above the trailing side. These two ongoing phenomena are locally biased by the resurfacing of fresh material due to tectonism (wispy terrains) or recent impact craters (Creusa, Lucagus-Liger, and Sagaris). Dione's preliminary albedo maps properties are discussed in \cite{Filacchione2018b}.

\clearpage
\subsection{Rhea}
\label{sct:rhea}
\begin{figure}[h!]
\centering
	\includegraphics[width=18cm]{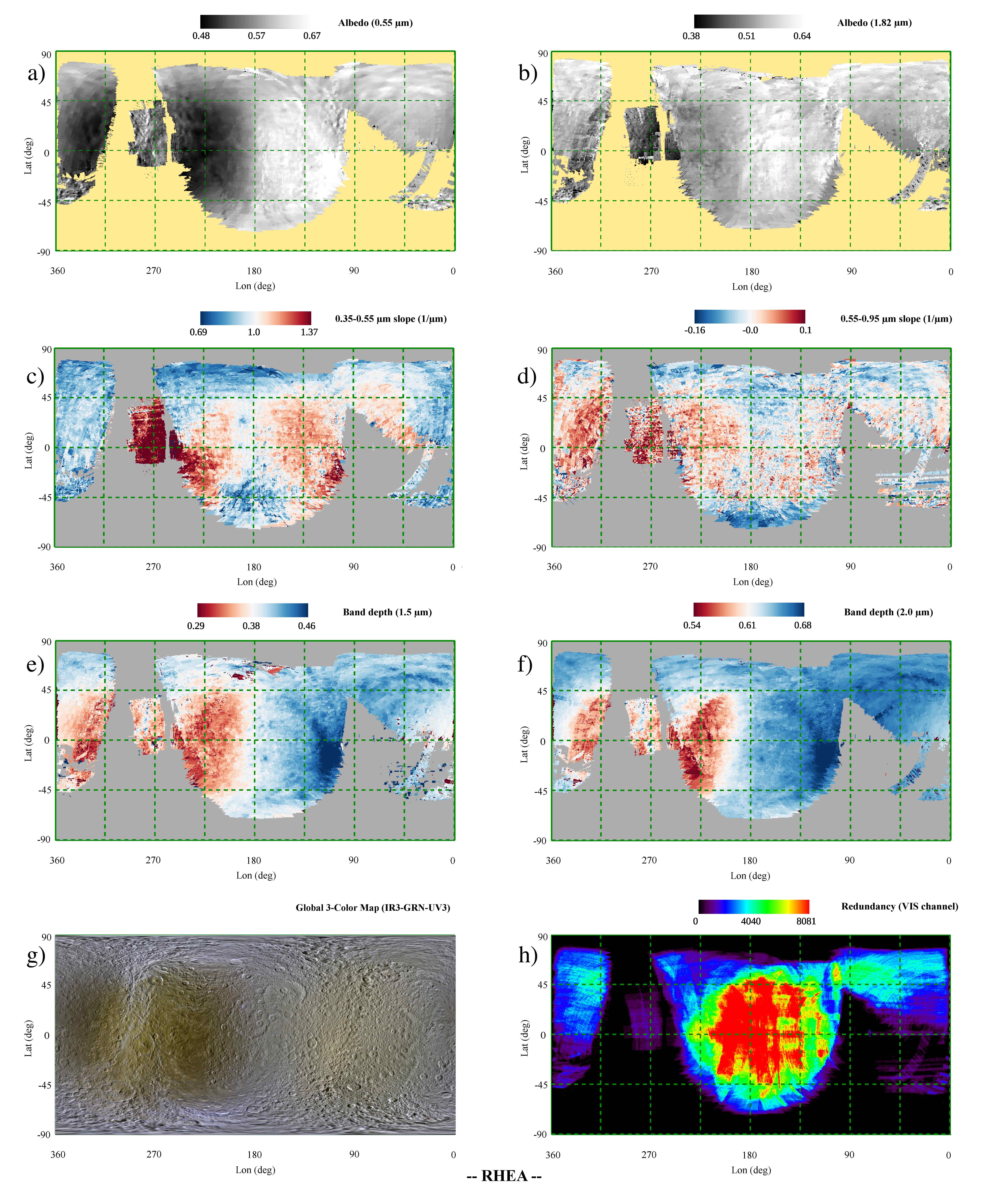}
    \caption{Rhea photometric-corrected maps rendered at 6.3 km/bin (at equator) in simple cylindrical projection.  Panel a): Equigonal albedo at 0.55 $\mu$m; b): Equigonal albedo at 1.82$\mu$m; c): 0.35-0.55 $\mu$m spectral slope; d): 0.55-0.95 $\mu$m spectral slope; e) 1.5 $\mu$m water ice band depth; f) 2.0 $\mu$m water ice band depth; g) ISS global color (IR, Green, UV) map at 400 m/px resolution (PIA18438, image credit P. Schenck, NASA/JPL-Caltech/Space Science Institute/Lunar and Planetary Institute); h): Visible channel data redundancy.}
    \label{fig:14}
\end{figure}

Rhea's maps offer limited coverage across the southern hemisphere and in the middle of the trailing hemisphere (Fig. \ref{fig:14}-panel h). Similarly with Dione, also on Rhea the albedo contrast (panels a-b) between the leading and trailing hemispheres is remarkable (0.67/0.48 at 0.55 $\mu$m and 0.64/0.38 at 1.82 $\mu$m). Above the leading hemisphere, the spectral slopes show higher reddening correlated with the high albedo and high water ice band depths as a consequence of the layering of E ring fine grains and electronic bombardment. These trends do not show significant variations across ancient and large impact craters, like Tirawa (lon=151$^{\circ}$, lat=34$^{\circ}$) and Mamaldi (lon=184$^{\circ}$, lat=14$^{\circ}$) suggesting that the exogenous alteration of the surface is the dominant process across a wide part of the leading hemisphere. 
\par  
On the contrary, the spectral indicators are peaked above the recent Inktomi crater (lon=112$^\circ$, lat=-14$^\circ$) and ejecta. A similar distribution is compatible with the exposure of more pristine water ice from the subsurface caused by the impact. The trailing side shows an extended dark terrain distribution characterized by fainter water ice bands but high reddening caused by the bombardment of cold plasma and dust particles. This distribution is broken apart around the 270$^\circ$ meridian where the bright Avaiki Chasmata \citep{Smith1981, Thomas1988, Stephan2012} runs along the north-south direction. The Chasmata, caused by recent tectonics and consequently resurfacing of more pristine material, is evident on the VIMS visible albedo (Fig. \ref{fig:14}-panel a) and ISS maps (panel g). The chasmata region appears remarkably dark (albedo 0.38) with respect to the rest of the satellite's surface on the 1.82 $\mu$m map (panel b), very red (1.37 $\mu$m$^{-1}$) on the 0.35-0.55 $\mu$m spectral slope, panel c) and with shallow water ice band depths (panels e-f). Similar behavior is very peculiar and its interpretation needs more efforts in spectral modeling. 
The fact that Avaiki Chasmata, which has a rough morphology with a relative elevation up to about 2.5 km \citep{Beddingfield2015}, was imaged by VIMS 1) at high spatial resolution (up to 5.9 km/px) with respect to the rest of the dataset; 2) with poor data redundancy (panel h); 3) with uniform illumination geometry (Fig \ref{fig:2)} implies that some photometric residuals could affect the results due to the poor statistics and geometric parameters (i, e, g) incertitudes. Conversely, given these conditions, the map shows sharp contrast due to the rendering of a single illumination condition imaged at high spatial resolution.
\par
Even though the cold plasma particles are similarly hitting the trailing sides of Dione and Rhea, the dark material on those two moons has a different response being blue (negative slopes) on Dione and red (positive) on Rhea. This difference could be explained by the presence at Dione of a flux of submicron dust particles embedded in the cold plasma which are very efficient to scatter blue light in the Rayleigh regime \citep{Clark2008}. Comparing Rhea and Dione albedo maps it is evident that the two moons have very similar albedos values: on the leading hemisphere of Dione the 0.55 $\mu$m albedo is 0.71 compared to 0.67 on Rhea; on the trailing side we measure 0.44 on Dione and 0.48 on Rhea. In the infrared, these differences become even smaller, with a maximum value of 0.65 for the 1.82 $\mu$m albedo across the leading hemisphere of Dione and 0.64 on Rhea, and a minimum of 0.38 on both trailing sides. These findings are in contradiction with previous spectral analyses of VIMS data: based on spectroscopic classification, \cite{Scipioni2014} was reporting higher infrared brightness on Rhea with respect to Dione. Instead, a similar difference in brightness is observed on Radar's data on Cassini at 2.2 cm and Arecibo at 13 cm and it is interpreted in terms of volume scattering of the regoliths of the two moons at cm-scale \citep{Ostro2010, LeGall2019}.  A more detailed discussion about correlations between spectral indicators and morphological structures on preliminary photometric-corrected VIMS data of Rhea is given in \cite{Filacchione2020}.

\clearpage

\section{A comparative analysis of Saturn's icy satellites average equigonal albedo and spectral indicators}
\label{sct:comparative_analysis}

\begin{figure}[h!]
\centering
	\includegraphics[width=18cm]{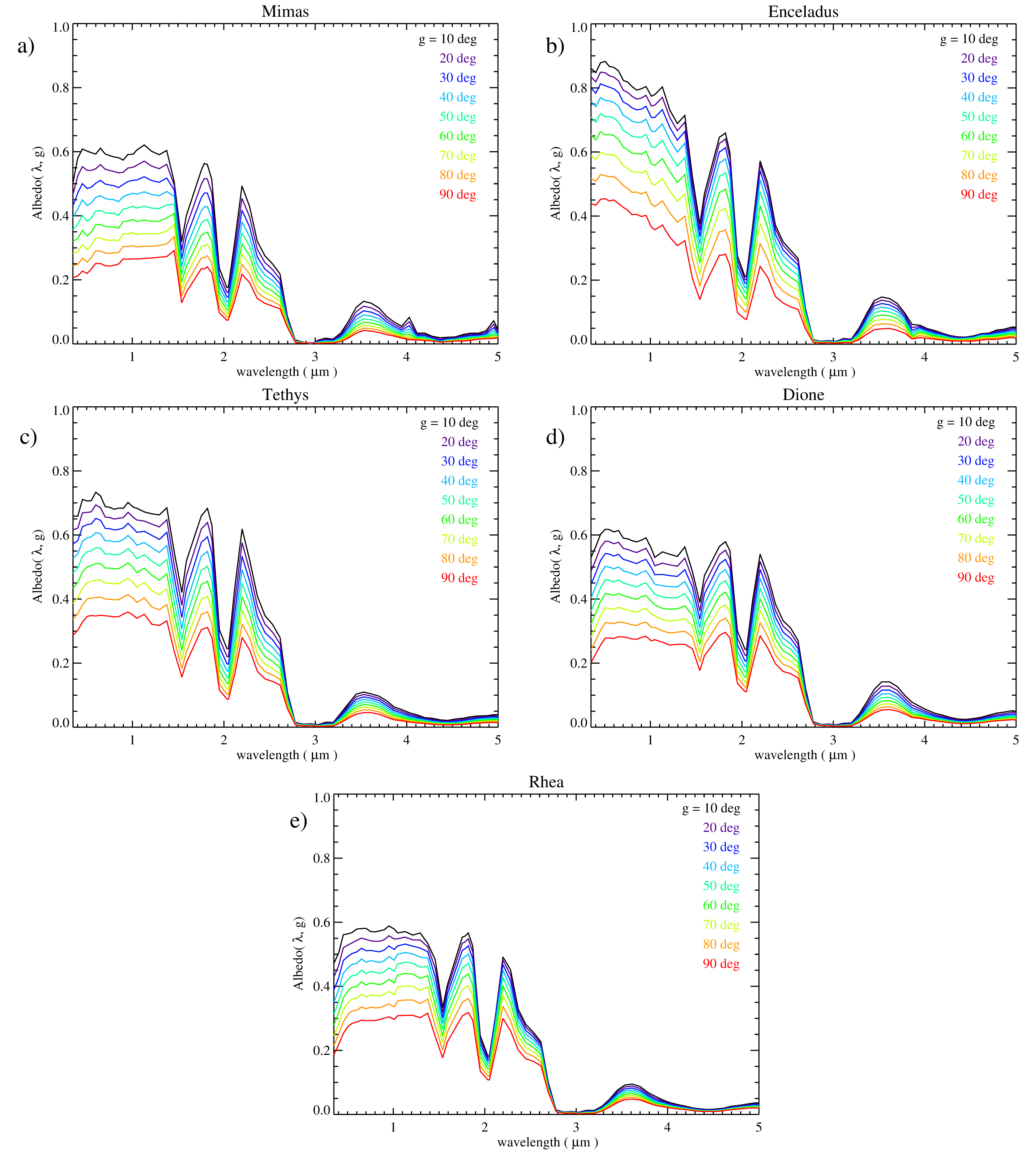}
    \caption{Icy satellites average spectral albedos as a function of phase angle $10^\circ \le g \le 90^\circ$. Panel a): Mimas; b) Enceladus; c) Tethys; d) Dione; e) Rhea.}
    \label{fig:15}
\end{figure}

Taking advantage of the computation of the photometric parameters (Tables \ref{tbl:mimas_parameters}-\ref{tbl:rhea_parameters}), one can derive the average spectral albedos of the five midsized satellites in the phase range $10^\circ \leq g \leq 90^\circ$ employing eq. \ref{eq:5}. The resulting spectral albedos shown in Fig. \ref{fig:15} are suitable to perform a comparative analysis among satellites scaled at the same illumination conditions. We note that these spectra are affected by a few calibration residuals, like the recurrent faint peaks in the visible range and the merging between VIMS visible and infrared channels at about 1 $\mu$m which is particularly evident on Dione's low phase spectra (Fig. \ref{fig:15}-panel d). Some artifacts are also recognizable on Mimas data (panel a) at 4.03 and 4.95 $\mu$m and on Enceladus at 3.85 $\mu$m (panel b). All visible albedos show a change of slope at 0.55 $\mu$m which has been exploited to define the range of 0.35-0.55 and 0.55-0.95 $\mu$m spectral slopes. This wavelength is also used to normalize the albedo spectra before the computation of the spectral slope value allowing to disentangle color variations from albedo differences.   
The water ice bands at 1.5, 2.05, and 3.0 $\mu$m are the dominating features across the average infrared spectra of all satellites at any phase angle. The fainter water ice signatures at 1.05 and 1.25 $\mu$m are resolved only on Enceladus' spectra (panel b). For comparison, the average I/F spectra of the satellites derived from a set of disk-integrated observations at full VIMS spectral resolution are discussed in \cite{Filacchione2012}.

\begin{figure}[h!]
\centering
	\includegraphics[width=14cm]{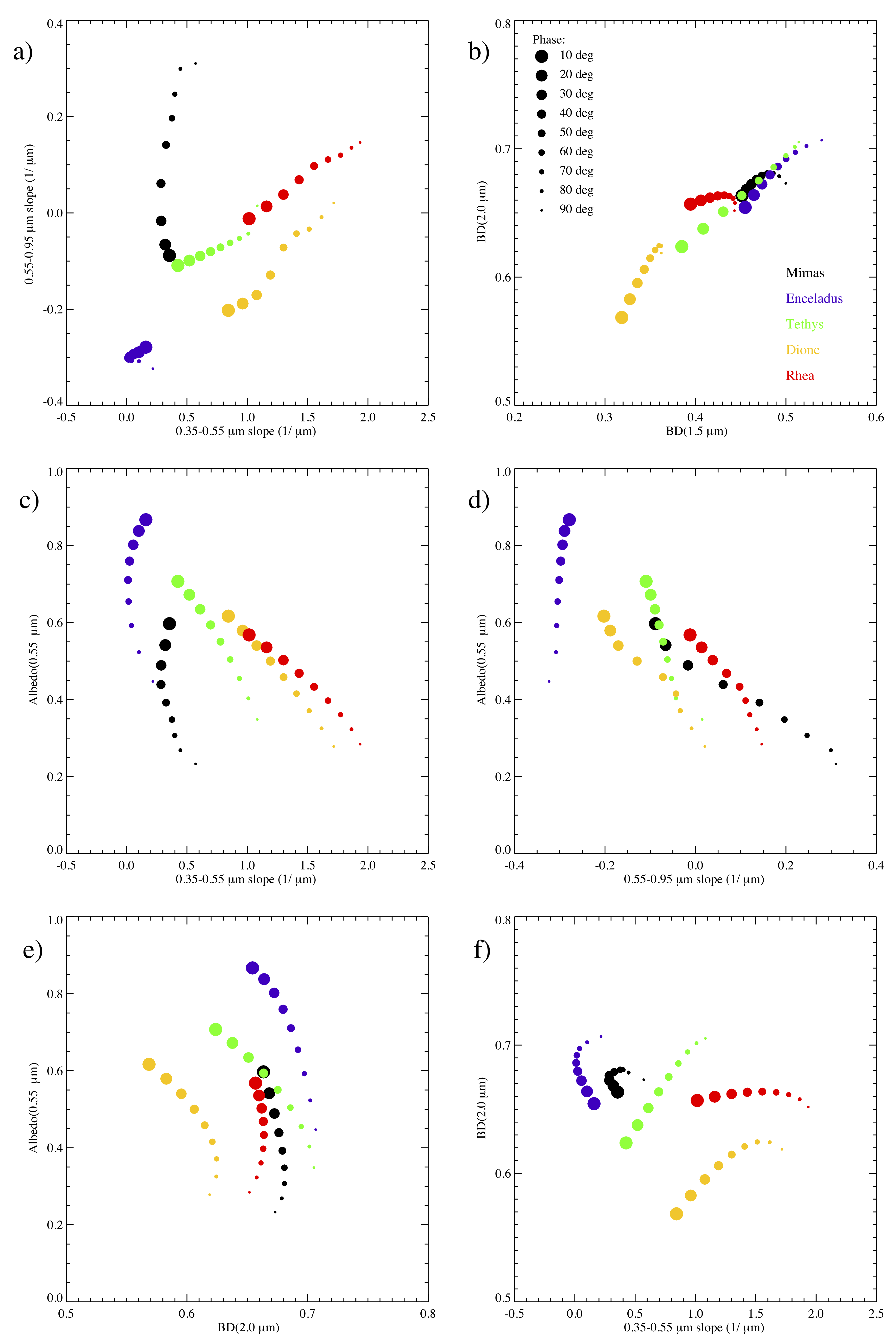}
    \caption{Scatterplots showing the variability of visible spectral slopes (top left panel), water ice band depths (top right panel), spectral slopes vs. 0.55 $\mu$m albedo (middle row panels), 2.0 $\mu$m band depth vs. 0.55 $\mu$m albedo (bottom left) and 0.35-0.55 $\mu$m slope vs. 2.0 $\mu$m band depth (bottom right) as a function of the phase angle within the range $10^\circ \le g \le 90^\circ$. These data are derived from average spectral albedo curves shown in Fig. \ref{fig:15}.}
    \label{fig:16}
\end{figure}

The variability of the spectral parameters computed from the albedo spectra as a function of phase angle is shown in six scatterplots in Fig. \ref{fig:16}. For all satellites but Enceladus (and partially Mimas), visible slopes increase with phase angle from g=$10^\circ$ to g=$90^\circ$ (Fig. \ref{fig:16}-panel a).  The maximum 0.35-0.55 $\mu$m spectral slope change is measured on Rhea and Dione, followed by Tethys, whereas the minimum is seen on Enceladus and Mimas. Mimas shows the maximum variability of the 0.55-0.95 $\mu$m slope which runs from -0.1 to 0.32 1/$\mu$m while the 0.35-0.55 $\mu$m spectral slope evidences a little bluening at intermediate phases. Conversely, Enceladus' slopes are the bluer ones among satellites and show a peculiar behavior because they steadily decrease from g=$10^\circ$ to $70^\circ$, resulting in a remarkable blueing effect on visible albedo spectra (Fig. \ref{fig:15}-panel b).  
Rhea and Dione do not have only the higher average 0.35-0.55 $\mu$m spectral slopes but also the highest reddening above their leading hemispheres where E ring grains preferentially accumulate. This result is in agreement with the scenario proposed by \cite{Hendrix2018} in which radiolytic processing of organic matter embedded in E ring particles by magnetospheric particles is the main cause of color alteration. The reddening appears correlated with the time of flight of the grains within the E ring in such a way that they accumulate higher doses, and become redder, during the long transfer times necessary to reach the farthest satellites from Enceladus. The reddening gradient observed on satellites orbiting within the E ring as a function of distance from Enceladus reported for the first time by \cite{Filacchione2013} can be explained through this mechanism. The correlations between the 0.55 $\mu$m albedo and spectral slopes (Fig. \ref{fig:16}-panels c-d) show how the satellites' albedo progressively decreases with orbital distance from Enceladus while the reddening increases at the same time. However, since the density of E ring particles at Rhea is smaller than at Dione \citep{Verbiscer2007}, the reddening process occurring on the trailing hemispheres must take into account also the much lower flux of particles reaching the moons orbiting at farther distances from Enceladus.  
  
Spectral analyses of the icy satellites and Saturn's rings particles (which show many similarities with satellites) by different authors \citep{Cuzzi1998, Poulet2002, Clark2008, Ciarniello2011, Filacchione2012, Cuzzi2018b, Ciarniello2019} indicate that the reddening is driven by the presence of a still-debated UV absorber, made of minor amounts of tholins, nanophase iron, hematite (or their combinations) mixed within water ice particles. In addition, the presence of neutral absorbers, like amorphous carbon and silicates, results in the reduction of albedo and spectral contrast across water ice bands.  
All these materials are called chromophores because they can alter the visible colorless spectral response of water ice which is the dominant endmember of the saturnian satellites. Radiative transfer models show that the 0.35-0.55 $\mu$m spectral slope is very sensitive to intraparticle mixing of those chromophores embedded in water ice while the 0.55-0.95 $\mu$m slope is driven by intimate and areal mixing modes \citep{Filacchione2012, Ciarniello2019}.

The water ice 1.5-2.0 $\mu$m band depths scatterplot (Fig. \ref{fig:16}-panel b) displays how all satellites data are distributed along a diagonal branch which runs from low phase observations of Dione on the lower end to high phase observations of Enceladus and Tethys on the higher end of the branch. For all satellites, we observe a general increase of the band depth values with phase angle. However, on Mimas, Dione and Rhea this monotonic trend is interrupted at about g=70$^\circ$ where a sudden inversion is observed. The overall trend of the two band depths scatterplot is driven by both water ice purity and grain size distribution. Uncontaminated ice implies high albedo and high band depth occurring at the same time, a condition met on Enceladus (Fig. \ref{fig:16}-panel e) but not on the remaining satellites which despite having high band depths show much darker albedos. This is a direct consequence of the mixing of the chromophores within water ice.     

Finally, the dispersion of the 0.35-0.55 $\mu$m slope as a function of 2.0 $\mu$m band depth (panel f) evidences three different responses: 1) Tethys' points are characterized by a linear dispersion resulting in a steady increase of the band depth and reddening with phase angle; 2) Enceladus and Mimas show a quadratic dispersion caused by the bluening effect at intermediate phases; 3) Rhea and Dione are the reddest objects and their reddening increases with phase angle. These differences are induced by the relative efficiency of the single and multiple scattering within the medium. The multiple scattering among grains is more efficient at intermediate phases while the single scattering dominates at low and high phases. The observed bluening on Enceladus and Mimas can be explained through the dominance of the multiple scattering which is advantaged by high albedo at shorter wavelengths. On the contrary, the phase reddening observed on Dione, Tethys and Rhea occur thanks to the lower albedo and intrinsic redder color of these surfaces where single scattering prevails.  

\clearpage

\section{Summary and conclusions}
The whole Cassini/VIMS dataset has been mined and processed to select all pixels suitable to compute photometric parameters and maps of Mimas, Enceladus, Tethys, Dione, and Rhea satellites. By applying the widely-tested photometric correction method developed by \cite{Shkuratov2011} we have systematically derived spectral albedo maps of Saturn's midsized satellites on 65 visible and infrared wavelengths out of the 352 measured by VIMS. While in principle the method can be applied to all VIMS wavelengths, a similar reduction in spectral bands processing has been necessary to keep the long computation, optimization, and verification times within a reasonable duration. Spectral indicators, including spectral slopes and water ice band depths, have been derived and their average values have been compared among satellites as a function of phase angle. Apart from deriving photometric parameters through a second-order fit (Tables \ref{tbl:mimas_parameters}-\ref{tbl:rhea_parameters}) and equigonal albedo and spectral indicators maps (Fig. \ref{fig:10}-\ref{fig:14}) released in digital format as supplementary material, the major findings of this work can be summarized as follows:
\begin{itemize}
\item Satellites' phase curves can be modeled through a quadratic law which, according to \cite{Schroeder2014}, it is typical behavior of smooth surfaces dominated by forward-scattering particles.  
\item the hemispheric dichotomies between leading and trailing hemispheres common to all midsized satellites but Enceladus have been resolved and explored through quantitive measurements of visible-infrared albedos and spectral indicators. 
\item the role played by the redeposition of fine E ring grains, cold plasma particles, high and low energy electronic bombardment, and meteoritic flux has been investigated by studying the spatial distribution of the spectral indicators across the expected places of exogenous alteration. VIMS albedo and spectral indicators maps allow to resolve: 1) the brightening of Mimas' trailing hemisphere, Tethys' antisaturnian hemisphere, and Dione's and Rhea's trailing hemispheres which are caused by the accumulation of icy E ring particles following forecasts provided by dynamical models of Enceladus' plumes  \citep{Kempf2018, Howett2018}; 2) the dark and red-colored material mapped across the trailing hemispheres of Tethys, Dione, and Rhea which is associated with the implantation of cold plasma particles; 3) the thermal anomaly lenses located on the leading equatorial regions of Mimas and Tethys which are the consequence of the impacts with high energy magnetospheric electrons; the distribution of the water ice band depths on Mimas' lens appears biased by the presence of the large Herschel impact crater possibly pointing to a different regolith size distribution with respect to Tethys' lens.
\item the local changes of albedo and spectral indicators in correspondence of recent impact craters (Inktomi on Rhea, Creusa on Dione) and on Dione's wispy terrains are compatible with the exposure of pristine water ice.
\item Enceladus' tiger stripes, the active cryovolcanic sources of plumes in the southern polar region, are partially resolved on VIMS maps and characterized by exceedingly high band depths in comparison with the rest of the surface \citep{Scipioni2017}.
\item Enceladus' smooth terrains located on the leading hemisphere around $(lon,lat)=(90^\circ, 30^\circ)$ show peculiar properties (low infrared albedo, positive 0.35-0.55 $\mu$m slope and maximum band depth) possibly associated with the presence of a buried diapir in this area \citep{Ries2015} or to local topography and ice shell thickness variations \citep{Cadek2016, Choblet2017}.
\item VIMS data do not show spectral evidence of the presence of the two wave-shaped E ring grains redeposition areas around meridians at $lon=45^\circ, 205^\circ$ on Enceladus as predicted by \cite{Kempf2010, Southworth2019}.
\item the analysis of albedos, spectral slopes, and band depths show how these parameters are phase-dependent through the different efficiency of single and multiple scattering. In particular, we observe that Enceladus' spectra are characterized by a decrease (bluening) of the spectral slopes with phase angle,
Mimas' evidence a little bluening at intermediate phases, while Dione's, Tethys' and Rhea's surfaces respond with a steady phase reddening. In the former case, the bluening is caused by the dominance of the multiple scattering in bright media while in the latter by the single scattering prevailing in more dark and contaminated surfaces.   
\end{itemize}

Our next effort for paper VI will be to perform a quantitative spectral analysis of the photometric-corrected data discussed here by applying a radiative transfer model on each $0.5^\circ \times 0.5^\circ$ bin of the satellites maps. This approach will allow us to derive composition endmembers percentages, mixing modalities, grain sizes distribution, and regolith's parameters across the surfaces of each satellite and to correlate the distribution of these quantities with geomorphological features.

\section*{Acknowledgements}

The authors acknowledge the financial support from Italian Space Agency (ASI) and from Istituto Nazionale di Astrofisica-IAPS. This research has made use of NASA's Astrophysics Data System.
Cassini-VIMS data are publicly available through NASA PDS node at
\\
 \texttt{https://pds-imaging.jpl.nasa.gov/volumes/vims.html}. 
\\
 The analysis discussed in this paper has been made by means of proprietary software developed in IDL-ENVI language by the authors.

\label{lastpage}
\section*{References}

\bibliography{Filacchione_Icarus_5_arxiv.bib}

\bibliographystyle{plainnat}

\appendix

\begin{table}[h!]
\caption{Mimas photometric parameters.}
\begin{center}
\begin{scriptsize}
\begin{tabular}{|c|c|c|c|c|}
\hline
wavelength (nm) & $a_0$ (adim.) & $a_1$ (deg$^{-1}$) & $a_2$ (deg$^{-2}$) &  $\chi^2$ \\
\hline
350 & 0.534362 $\pm$ 0.0372414 & -0.00353327 $\pm$ 0.00281849 & -1.68918e-06 $\pm$ 3.48485e-05 & 9.40740 \\
402 & 0.632356 $\pm$ 0.0205091 & -0.00597678 $\pm$ 0.000594014 & 1.41209e-05 $\pm$ 4.04266e-06 & 47.6230 \\
453 & 0.659615 $\pm$ 0.0189060 & -0.00579098 $\pm$ 0.000525200 & 1.04772e-05 $\pm$ 3.54492e-06 & 53.7091 \\
506 & 0.658717 $\pm$ 0.0225513 & -0.00648206 $\pm$ 0.000657704 & 1.74923e-05 $\pm$ 4.38304e-06 & 13.8010 \\
549 & 0.649537 $\pm$ 0.0413843 & -0.00595072 $\pm$ 0.00227855 & 1.42732e-05 $\pm$ 3.12055e-05 & 2.50143 \\
599 & 0.634896 $\pm$ 0.0177909 & -0.00473968 $\pm$ 0.000509178 & 4.79886e-06 $\pm$ 3.49446e-06 & 68.8080 \\
651 & 0.658532 $\pm$ 0.0182773 & -0.00535511 $\pm$ 0.000529654 & 8.73124e-06 $\pm$ 3.56242e-06 & 42.0049 \\
702 & 0.636657 $\pm$ 0.0183432 & -0.00489470 $\pm$ 0.000527319 & 5.43284e-06 $\pm$ 3.52723e-06 & 58.5770 \\
753 & 0.630439 $\pm$ 0.0169427 & -0.00478404 $\pm$ 0.000504747 & 4.82377e-06 $\pm$ 3.42187e-06 & 50.0659 \\
797 & 0.625471 $\pm$ 0.0167822 & -0.00466339 $\pm$ 0.000504637 & 4.40943e-06 $\pm$ 3.47352e-06 & 45.7230 \\
849 & 0.625474 $\pm$ 0.0154952 & -0.00498036 $\pm$ 0.000503628 & 8.46396e-06 $\pm$ 3.64802e-06 & 47.0736 \\
900 & 0.607891 $\pm$ 0.0169955 & -0.00377677 $\pm$ 0.000553670 & -8.32545e-07 $\pm$ 3.94117e-06 & 29.0028 \\
951 & 0.629705 $\pm$ 0.0159384 & -0.00429357 $\pm$ 0.000485369 & 2.13996e-06 $\pm$ 3.43160e-06 & 60.5974 \\
1010 & 0.628934 $\pm$ 0.0175826 & -0.00406435 $\pm$ 0.000574155 & -1.95211e-07 $\pm$ 4.18139e-06 & 23.9909 \\
1047 & 0.648290 $\pm$ 0.0279353 & -0.00490798 $\pm$ 0.000823728 & 6.74681e-06 $\pm$ 5.47472e-06 & 24.9387 \\
1130 & 0.667918 $\pm$ 0.0267349 & -0.00529284 $\pm$ 0.000799633 & 8.85330e-06 $\pm$ 5.38731e-06 & 27.7722 \\
1212 & 0.643991 $\pm$ 0.0257534 & -0.00481959 $\pm$ 0.000767494 & 6.70785e-06 $\pm$ 5.15442e-06 & 28.2999 \\
1294 & 0.630714 $\pm$ 0.0250366 & -0.00449884 $\pm$ 0.000766312 & 5.09373e-06 $\pm$ 5.22046e-06 & 18.9758 \\
1376 & 0.643200 $\pm$ 0.0230410 & -0.00439576 $\pm$ 0.000707592 & 2.61376e-06 $\pm$ 4.86124e-06 & 38.9980 \\
1458 & 0.510982 $\pm$ 0.172917 & -0.000379365 $\pm$ 0.00527351 & -2.34316e-05 $\pm$ 3.28624e-05 & 1.34710 \\
1540 & 0.344405 $\pm$ 0.0108894 & -0.00283071 $\pm$ 0.000304982 & 4.52791e-06 $\pm$ 2.05966e-06 & 155.902 \\
1590 & 0.431843 $\pm$ 0.0121689 & -0.00362396 $\pm$ 0.000348611 & 6.60127e-06 $\pm$ 2.44673e-06 & 116.403 \\
1754 & 0.582456 $\pm$ 0.0170482 & -0.00456588 $\pm$ 0.000475409 & 7.41514e-06 $\pm$ 3.23568e-06 & 123.373 \\
1787 & 0.607270 $\pm$ 0.0182339 & -0.00495652 $\pm$ 0.000505499 & 8.61669e-06 $\pm$ 3.43861e-06 & 119.200 \\
1820 & 0.602263 $\pm$ 0.0177354 & -0.00474123 $\pm$ 0.000493148 & 7.60120e-06 $\pm$ 3.37458e-06 & 128.345 \\
1869 & 0.551903 $\pm$ 0.0153743 & -0.00436645 $\pm$ 0.000436775 & 7.38235e-06 $\pm$ 3.05260e-06 & 130.353 \\
1951 & 0.255196 $\pm$ 0.00936507 & -0.00206276 $\pm$ 0.000263065 & 3.27578e-06 $\pm$ 1.79062e-06 & 159.242 \\
2034 & 0.190336 $\pm$ 0.00797768 & -0.00159391 $\pm$ 0.000224547 & 3.19683e-06 $\pm$ 1.52152e-06 & 103.436 \\
2050 & 0.192872 $\pm$ 0.00779414 & -0.00178265 $\pm$ 0.000218974 & 5.13364e-06 $\pm$ 1.48478e-06 & 130.105 \\
2116 & 0.336612 $\pm$ 0.0110215 & -0.00282511 $\pm$ 0.000312700 & 5.75720e-06 $\pm$ 2.14703e-06 & 102.884 \\
2199 & 0.528483 $\pm$ 0.0147410 & -0.00399688 $\pm$ 0.000424445 & 5.65258e-06 $\pm$ 2.99674e-06 & 116.994 \\
2282 & 0.459437 $\pm$ 0.0126247 & -0.00332996 $\pm$ 0.000367992 & 3.63187e-06 $\pm$ 2.59977e-06 & 140.382 \\
2367 & 0.343513 $\pm$ 0.0108672 & -0.00259688 $\pm$ 0.000309768 & 3.69458e-06 $\pm$ 2.14698e-06 & 122.269 \\
2450 & 0.294315 $\pm$ 0.00958281 & -0.00217879 $\pm$ 0.000279630 & 3.32287e-06 $\pm$ 1.97278e-06 & 132.862 \\
2532 & 0.266797 $\pm$ 0.00906125 & -0.00183678 $\pm$ 0.000267396 & 1.91189e-06 $\pm$ 1.89908e-06 & 133.713 \\
2615 & 0.233088 $\pm$ 0.00842628 & -0.00165540 $\pm$ 0.000253127 & 2.97592e-06 $\pm$ 1.81733e-06 & 120.226 \\
2696 & 0.0918798 $\pm$ 0.00529498 & -0.000557875 $\pm$ 0.000152757 & 1.30293e-06 $\pm$ 1.05704e-06 & 161.798 \\
2781 & 0.0138359 $\pm$ 0.00194283 & -0.000153183 $\pm$ 5.43307e-05 & 6.28789e-07 $\pm$ 3.51206e-07 & 23.8527 \\
2866 & 0.00652901 $\pm$ 0.00198743 & -8.54227e-06 $\pm$ 5.39238e-05 & -3.97605e-07 $\pm$ 3.43464e-07 & 8.89869 \\
2947 & 0.00343152 $\pm$ 0.00109927 & 0.000173704 $\pm$ 3.82792e-05 & -1.05645e-06 $\pm$ 2.64912e-07 & 106.513 \\
3029 & 0.0137495 $\pm$ 0.00222218 & -0.000255182 $\pm$ 6.20136e-05 & 1.35433e-06 $\pm$ 4.42996e-07 & 13.8962 \\
3112 & 0.0211174 $\pm$ 0.00264720 & -0.000361673 $\pm$ 6.51497e-05 & 1.72615e-06 $\pm$ 3.79101e-07 & 7.33515 \\
3197 & 0.0176703 $\pm$ 0.00255559 & -0.000293663 $\pm$ 6.82316e-05 & 1.43123e-06 $\pm$ 4.28572e-07 & 8.57555 \\
3282 & 0.0394597 $\pm$ 0.00359761 & -0.000567709 $\pm$ 9.63415e-05 & 2.35786e-06 $\pm$ 5.89325e-07 & 30.9476 \\
3365 & 0.0801127 $\pm$ 0.00489257 & -0.00103694 $\pm$ 0.000135987 & 3.95542e-06 $\pm$ 8.78699e-07 & 47.5509 \\
3448 & 0.126794 $\pm$ 0.00575553 & -0.00148030 $\pm$ 0.000158902 & 4.84195e-06 $\pm$ 1.06284e-06 & 31.4724 \\
3530 & 0.148055 $\pm$ 0.00587300 & -0.00170588 $\pm$ 0.000170254 & 5.80321e-06 $\pm$ 1.18482e-06 & 29.4872 \\
3613 & 0.141083 $\pm$ 0.00671708 & -0.00158916 $\pm$ 0.000182581 & 5.10235e-06 $\pm$ 1.19879e-06 & 64.7781 \\
3699 & 0.123169 $\pm$ 0.00760139 & -0.00126001 $\pm$ 0.000209593 & 3.10670e-06 $\pm$ 1.37910e-06 & 39.3947 \\
3784 & 0.0931164 $\pm$ 0.00597939 & -0.000904607 $\pm$ 0.000167527 & 2.06402e-06 $\pm$ 1.12474e-06 & 19.6275 \\
3861 & 0.0761850 $\pm$ 0.0236173 & -0.000800021 $\pm$ 0.000655858 & 2.46969e-06 $\pm$ 4.24100e-06 & 4.96848 \\
3947 & 0.0612005 $\pm$ 0.00627396 & -0.000684625 $\pm$ 0.000163335 & 2.29204e-06 $\pm$ 1.01952e-06 & 9.05016 \\
4029 & 0.0964293 $\pm$ 0.00333442 & -0.00155819 $\pm$ 0.000125804 & 6.98882e-06 $\pm$ 9.52558e-07 & 79.4474 \\
4114 & 0.0379139 $\pm$ 0.00566453 & -0.000437703 $\pm$ 0.000149761 & 1.71570e-06 $\pm$ 9.31809e-07 & 10.5843 \\
4198 & 0.0381774 $\pm$ 0.00615595 & -0.000460087 $\pm$ 0.000156887 & 1.71586e-06 $\pm$ 9.57374e-07 & 14.8454 \\
4278 & 0.0272350 $\pm$ 0.00808886 & -0.000370636 $\pm$ 0.000213204 & 1.92435e-06 $\pm$ 1.36903e-06 & 4.72155 \\
4366 & 0.0240649 $\pm$ 0.00595989 & -0.000559407 $\pm$ 0.000168651 & 4.10491e-06 $\pm$ 1.15761e-06 & 38.1641 \\
4447 & 0.0235719 $\pm$ 0.00911423 & -0.000310498 $\pm$ 0.000233584 & 1.32716e-06 $\pm$ 1.40138e-06 & 4.85009 \\
4533 & 0.0243647 $\pm$ 0.0125463 & -0.000281013 $\pm$ 0.000389846 & 1.26277e-06 $\pm$ 2.64588e-06 & 4.31743 \\
4620 & 0.0331182 $\pm$ 0.0125372 & -0.000413056 $\pm$ 0.000312765 & 1.57933e-06 $\pm$ 1.86678e-06 & 5.86964 \\
4702 & 0.0374945 $\pm$ 0.0105603 & -0.000428169 $\pm$ 0.000285462 & 1.51558e-06 $\pm$ 1.95259e-06 & 7.10978 \\
4786 & 0.0368585 $\pm$ 0.0127449 & -0.000386270 $\pm$ 0.000336186 & 1.56369e-06 $\pm$ 2.20541e-06 & 9.58456 \\
4869 & 0.0458377 $\pm$ 0.0127171 & -0.000555946 $\pm$ 0.000355640 & 2.51469e-06 $\pm$ 2.41784e-06 & 4.60153 \\
4953 & 0.0838533 $\pm$ 0.0147530 & -0.00132716 $\pm$ 0.000421372 & 6.50462e-06 $\pm$ 3.08108e-06 & 40.8348 \\
5040 & 0.0213654 $\pm$ 0.00878035 & -5.19945e-05 $\pm$ 0.000329477 & 4.96310e-07 $\pm$ 2.47628e-06 & 5.49659 \\
\hline
\end{tabular}
\end{scriptsize}
\end{center}
\label{tbl:mimas_parameters}
\end{table}

\begin{table}[h!]
\caption{Enceladus photometric parameters.}
\begin{center}
\begin{scriptsize}
\begin{tabular}{|c|c|c|c|c|}
\hline
wavelength (nm) & $a_0$ (adim.) & $a_1$ (deg$^{-1}$) & $a_2$ (deg$^{-2}$) &  $\chi^2$ \\
\hline
350 & 0.878931 $\pm$ 0.0291541 & -0.00166205 $\pm$ 0.000905813 & -3.68755e-05 $\pm$ 6.81845e-06 & 67.7050 \\
402 & 0.868410 $\pm$ 0.0263120 & -0.00198719 $\pm$ 0.000804571 & -3.25403e-05 $\pm$ 6.03920e-06 & 108.573 \\
453 & 0.899159 $\pm$ 0.0278268 & -0.00199714 $\pm$ 0.000832638 & -3.39609e-05 $\pm$ 6.16610e-06 & 113.365 \\
506 & 0.909894 $\pm$ 0.0266856 & -0.00280542 $\pm$ 0.000803174 & -2.60275e-05 $\pm$ 6.00060e-06 & 114.733 \\
549 & 0.887595 $\pm$ 0.0275526 & -0.00197692 $\pm$ 0.000854473 & -3.33943e-05 $\pm$ 6.51879e-06 & 55.6607 \\
599 & 0.886070 $\pm$ 0.0283470 & -0.00217728 $\pm$ 0.000847358 & -3.05459e-05 $\pm$ 6.37132e-06 & 54.3968 \\
651 & 0.875807 $\pm$ 0.0250636 & -0.00225747 $\pm$ 0.000782616 & -2.97717e-05 $\pm$ 6.00309e-06 & 77.7935 \\
702 & 0.847247 $\pm$ 0.0235811 & -0.00237559 $\pm$ 0.000717996 & -2.63313e-05 $\pm$ 5.42955e-06 & 90.1901 \\
753 & 0.842410 $\pm$ 0.0250520 & -0.00251155 $\pm$ 0.000755645 & -2.53524e-05 $\pm$ 5.56716e-06 & 66.6901 \\
797 & 0.827022 $\pm$ 0.0251166 & -0.00225416 $\pm$ 0.000761997 & -2.80193e-05 $\pm$ 5.64594e-06 & 79.3096 \\
849 & 0.813137 $\pm$ 0.0239556 & -0.00196588 $\pm$ 0.000718744 & -2.93569e-05 $\pm$ 5.34235e-06 & 74.7544 \\
900 & 0.818532 $\pm$ 0.0239915 & -0.00207616 $\pm$ 0.000701065 & -2.96857e-05 $\pm$ 5.13469e-06 & 82.6514 \\
951 & 0.823289 $\pm$ 0.0228267 & -0.00182740 $\pm$ 0.000660323 & -3.26240e-05 $\pm$ 4.89192e-06 & 88.0443 \\
1010 & 0.796628 $\pm$ 0.0287063 & -0.00222771 $\pm$ 0.000813523 & -2.94267e-05 $\pm$ 5.77691e-06 & 86.1866 \\
1047 & 0.803527 $\pm$ 0.0306524 & -0.00247774 $\pm$ 0.00102556 & -2.85169e-05 $\pm$ 7.49484e-06 & 112.693 \\
1130 & 0.826933 $\pm$ 0.0323137 & -0.00237435 $\pm$ 0.00107063 & -3.07435e-05 $\pm$ 7.79846e-06 & 111.029 \\
1212 & 0.745118 $\pm$ 0.0328385 & -0.00128386 $\pm$ 0.00105685 & -3.74700e-05 $\pm$ 7.52683e-06 & 85.0249 \\
1294 & 0.702882 $\pm$ 0.0320546 & -0.00124340 $\pm$ 0.00101108 & -3.59051e-05 $\pm$ 7.11413e-06 & 77.3462 \\
1376 & 0.726244 $\pm$ 0.0329749 & -0.000987567 $\pm$ 0.00104161 & -3.97860e-05 $\pm$ 7.34290e-06 & 78.0487 \\
1458 & 0.537241 $\pm$ 0.0284273 & -0.00118961 $\pm$ 0.000849419 & -2.85016e-05 $\pm$ 5.73351e-06 & 95.7250 \\
1540 & 0.386892 $\pm$ 0.0211993 & -0.00124494 $\pm$ 0.000624858 & -1.72778e-05 $\pm$ 4.16280e-06 & 109.358 \\
1590 & 0.483202 $\pm$ 0.0263920 & -0.00125279 $\pm$ 0.000788623 & -2.39913e-05 $\pm$ 5.29958e-06 & 117.098 \\
1754 & 0.657102 $\pm$ 0.0334150 & -0.000835428 $\pm$ 0.00103315 & -3.84669e-05 $\pm$ 7.13399e-06 & 89.0237 \\
1787 & 0.662681 $\pm$ 0.0339311 & -0.000813612 $\pm$ 0.00105071 & -3.94264e-05 $\pm$ 7.25181e-06 & 96.8306 \\
1820 & 0.669113 $\pm$ 0.0340544 & -0.000730216 $\pm$ 0.00105104 & -4.07270e-05 $\pm$ 7.25150e-06 & 96.1297 \\
1869 & 0.606893 $\pm$ 0.0318272 & -0.000686592 $\pm$ 0.000978746 & -3.79015e-05 $\pm$ 6.71887e-06 & 100.225 \\
1951 & 0.285615 $\pm$ 0.0160506 & -0.000939465 $\pm$ 0.000477447 & -1.30552e-05 $\pm$ 3.18879e-06 & 94.7952 \\
2034 & 0.217333 $\pm$ 0.0125851 & -0.000625850 $\pm$ 0.000382774 & -1.04442e-05 $\pm$ 2.59991e-06 & 84.0748 \\
2050 & 0.221016 $\pm$ 0.0123626 & -0.000835674 $\pm$ 0.000371543 & -8.80034e-06 $\pm$ 2.49604e-06 & 100.571 \\
2116 & 0.372775 $\pm$ 0.0212201 & -0.000671465 $\pm$ 0.000643482 & -2.09130e-05 $\pm$ 4.35112e-06 & 101.503 \\
2199 & 0.573896 $\pm$ 0.0300574 & 0.000169286 $\pm$ 0.000921618 & -4.36705e-05 $\pm$ 6.27791e-06 & 114.565 \\
2282 & 0.511676 $\pm$ 0.0257108 & -0.000727349 $\pm$ 0.000785989 & -3.00593e-05 $\pm$ 5.37696e-06 & 91.2256 \\
2367 & 0.378581 $\pm$ 0.0197919 & -0.000553214 $\pm$ 0.000603639 & -2.27985e-05 $\pm$ 4.10886e-06 & 98.8147 \\
2450 & 0.326319 $\pm$ 0.0167679 & -0.000564080 $\pm$ 0.000513774 & -1.86136e-05 $\pm$ 3.50707e-06 & 87.8024 \\
2532 & 0.300071 $\pm$ 0.0154617 & -0.000544286 $\pm$ 0.000477239 & -1.70515e-05 $\pm$ 3.27825e-06 & 101.967 \\
2615 & 0.272141 $\pm$ 0.0136445 & -0.000353908 $\pm$ 0.000423143 & -1.66767e-05 $\pm$ 2.90285e-06 & 92.3812 \\
2696 & 0.112191 $\pm$ 0.00653588 & -2.23718e-05 $\pm$ 0.000200254 & -7.94001e-06 $\pm$ 1.36872e-06 & 94.4326 \\
2781 & 0.0146191 $\pm$ 0.00209419 & -3.95264e-05 $\pm$ 5.93977e-05 & -6.74377e-07 $\pm$ 3.93031e-07 & 31.0751 \\
2866 & 0.00958769 $\pm$ 0.00115591 & -5.55944e-05 $\pm$ 4.17765e-05 & -1.41074e-07 $\pm$ 3.20939e-07 & 8.25357 \\
2947 & 0.0125162 $\pm$ 0.00130964 & -0.000106928 $\pm$ 4.37991e-05 & 2.13841e-07 $\pm$ 3.33691e-07 & 8.45815 \\
3029 & 0.00968945 $\pm$ 0.00127630 & -7.31844e-05 $\pm$ 4.35208e-05 & -3.44312e-08 $\pm$ 3.31298e-07 & 14.0367 \\
3112 & 0.0193915 $\pm$ 0.00206476 & -0.000183765 $\pm$ 6.39898e-05 & 2.27687e-07 $\pm$ 4.55443e-07 & 21.9621 \\
3197 & 0.0171737 $\pm$ 0.00177083 & -0.000158888 $\pm$ 5.60614e-05 & 1.75221e-07 $\pm$ 3.95627e-07 & 12.8205 \\
3282 & 0.0557142 $\pm$ 0.00319722 & -0.000616978 $\pm$ 9.59164e-05 & 1.69249e-06 $\pm$ 6.49433e-07 & 43.4713 \\
3365 & 0.100871 $\pm$ 0.00601119 & -0.000548578 $\pm$ 0.000182091 & -2.67537e-06 $\pm$ 1.22687e-06 & 62.8609 \\
3448 & 0.142726 $\pm$ 0.00781532 & -0.000671296 $\pm$ 0.000235637 & -4.72317e-06 $\pm$ 1.59731e-06 & 64.0727 \\
3530 & 0.155371 $\pm$ 0.00776950 & -0.000863553 $\pm$ 0.000238139 & -3.87520e-06 $\pm$ 1.62728e-06 & 62.4043 \\
3613 & 0.150878 $\pm$ 0.00766498 & -0.000882150 $\pm$ 0.000232878 & -2.87625e-06 $\pm$ 1.58513e-06 & 50.5081 \\
3699 & 0.134250 $\pm$ 0.00763070 & -0.000711239 $\pm$ 0.000231053 & -3.49666e-06 $\pm$ 1.58253e-06 & 51.8372 \\
3784 & 0.101522 $\pm$ 0.00620469 & -0.000530528 $\pm$ 0.000191576 & -2.56094e-06 $\pm$ 1.34759e-06 & 38.3701 \\
3861 & 0.0713970 $\pm$ 0.0160138 & -0.00112335 $\pm$ 0.000459231 & 5.71469e-06 $\pm$ 3.03514e-06 & 24.3550 \\
3947 & 0.0618074 $\pm$ 0.00514304 & -0.000268979 $\pm$ 0.000163968 & -2.18926e-06 $\pm$ 1.16390e-06 & 22.5793 \\
4029 & 0.0506756 $\pm$ 0.00548291 & -0.000223100 $\pm$ 0.000161746 & -1.62270e-06 $\pm$ 1.08398e-06 & 22.0285 \\
4114 & 0.0457759 $\pm$ 0.00420847 & -0.000366585 $\pm$ 0.000129818 & 3.15942e-07 $\pm$ 9.16791e-07 & 10.7268 \\
4198 & 0.0356618 $\pm$ 0.00447185 & -0.000196039 $\pm$ 0.000137530 & -6.83020e-07 $\pm$ 9.56205e-07 & 9.60116 \\
4278 & 0.0288412 $\pm$ 0.00501954 & -0.000209802 $\pm$ 0.000153372 & 5.95396e-08 $\pm$ 1.08050e-06 & 9.61164 \\
4366 & 0.0243086 $\pm$ 0.00569822 & -0.000178919 $\pm$ 0.000167396 & 1.67740e-07 $\pm$ 1.13806e-06 & 7.22713 \\
4447 & 0.0233647 $\pm$ 0.00533797 & -0.000103055 $\pm$ 0.000169608 & -5.27759e-07 $\pm$ 1.22592e-06 & 6.85509 \\
4533 & 0.0273640 $\pm$ 0.0139897 & -0.000135732 $\pm$ 0.000434061 & -6.61722e-07 $\pm$ 2.90733e-06 & 3.61203 \\
4620 & 0.0340532 $\pm$ 0.00780880 & -7.93589e-05 $\pm$ 0.000222297 & -1.77309e-06 $\pm$ 1.43073e-06 & 13.7323 \\
4702 & 0.0360671 $\pm$ 0.0123696 & -0.000157308 $\pm$ 0.000345880 & -9.06800e-07 $\pm$ 2.32468e-06 & 6.78682 \\
4786 & 0.0472307 $\pm$ 0.00917058 & -0.000353081 $\pm$ 0.000280820 & -1.13216e-07 $\pm$ 1.99390e-06 & 5.56612 \\
4869 & 0.0500730 $\pm$ 0.00870447 & -0.000364740 $\pm$ 0.000273189 & 9.72967e-08 $\pm$ 1.98938e-06 & 5.87916 \\
4953 & 0.0575194 $\pm$ 0.0133943 & -0.000443517 $\pm$ 0.000440582 & 3.68452e-07 $\pm$ 3.38492e-06 & 4.32100 \\
5040 & 0.0585442 $\pm$ 0.0154851 & -0.000635180 $\pm$ 0.000486952 & 1.78965e-06 $\pm$ 3.58073e-06 & 2.81925 \\
\hline
\end{tabular}
\end{scriptsize}
\end{center}
\label{tbl:enceladus_parameters}
\end{table}

\begin{table}[h!]
\caption{Tethys photometric parameters.}
\begin{center}
\begin{scriptsize}
\begin{tabular}{|c|c|c|c|c|}
\hline
wavelength (nm) & $a_0$ (adim.) & $a_1$ (deg$^{-1}$) & $a_2$ (deg$^{-2}$) &  $\chi^2$ \\
\hline
350 & 0.696788 $\pm$ 0.0367416 & -0.00415584 $\pm$ 0.00101166 & -5.01087e-06 $\pm$ 6.41702e-06 & 23.1663 \\
402 & 0.690351 $\pm$ 0.0345137 & -0.00342041 $\pm$ 0.000968016 & -1.04861e-05 $\pm$ 6.17025e-06 & 22.7638 \\
453 & 0.747063 $\pm$ 0.0330661 & -0.00393178 $\pm$ 0.000945299 & -8.67318e-06 $\pm$ 6.09980e-06 & 17.1289 \\
506 & 0.735376 $\pm$ 0.0303450 & -0.00315262 $\pm$ 0.000898946 & -1.37236e-05 $\pm$ 5.91567e-06 & 25.6141 \\
549 & 0.736281 $\pm$ 0.0308483 & -0.00310144 $\pm$ 0.000910277 & -1.41087e-05 $\pm$ 5.96401e-06 & 20.9418 \\
599 & 0.766672 $\pm$ 0.0319447 & -0.00358972 $\pm$ 0.000929919 & -1.26177e-05 $\pm$ 6.05260e-06 & 6.04056 \\
651 & 0.751931 $\pm$ 0.0301970 & -0.00325114 $\pm$ 0.000898288 & -1.42701e-05 $\pm$ 5.92709e-06 & 8.72324 \\
702 & 0.715489 $\pm$ 0.0281358 & -0.00287179 $\pm$ 0.000857745 & -1.41549e-05 $\pm$ 5.76695e-06 & 23.5707 \\
753 & 0.712496 $\pm$ 0.0289675 & -0.00276283 $\pm$ 0.000868584 & -1.51645e-05 $\pm$ 5.73769e-06 & 17.5099 \\
797 & 0.704906 $\pm$ 0.0283116 & -0.00259876 $\pm$ 0.000850092 & -1.63151e-05 $\pm$ 5.60985e-06 & 12.8655 \\
849 & 0.710181 $\pm$ 0.0309207 & -0.00276305 $\pm$ 0.000914560 & -1.50440e-05 $\pm$ 5.99047e-06 & 13.8507 \\
900 & 0.706356 $\pm$ 0.0303812 & -0.00241077 $\pm$ 0.000908759 & -1.75184e-05 $\pm$ 5.98367e-06 & 17.4280 \\
951 & 0.726834 $\pm$ 0.0298228 & -0.00263666 $\pm$ 0.000895596 & -1.67451e-05 $\pm$ 5.90945e-06 & 15.7379 \\
1010 & 0.712688 $\pm$ 0.0334241 & -0.00276393 $\pm$ 0.000987792 & -1.48602e-05 $\pm$ 6.51848e-06 & 12.5640 \\
1047 & 0.715061 $\pm$ 0.0313833 & -0.00342283 $\pm$ 0.00114784 & -8.88043e-06 $\pm$ 8.75626e-06 & 14.2859 \\
1130 & 0.694860 $\pm$ 0.0291354 & -0.00240454 $\pm$ 0.00112965 & -1.62080e-05 $\pm$ 8.80068e-06 & 21.9267 \\
1212 & 0.692184 $\pm$ 0.0257540 & -0.00276478 $\pm$ 0.000884290 & -1.57757e-05 $\pm$ 6.10067e-06 & 16.0820 \\
1294 & 0.694908 $\pm$ 0.0254327 & -0.00351943 $\pm$ 0.000810562 & -8.07145e-06 $\pm$ 5.31530e-06 & 59.0347 \\
1376 & 0.716679 $\pm$ 0.0292815 & -0.00344716 $\pm$ 0.000894249 & -9.83072e-06 $\pm$ 5.75299e-06 & 55.2487 \\
1458 & 0.580022 $\pm$ 0.0251992 & -0.00393366 $\pm$ 0.000715315 & 7.13336e-07 $\pm$ 4.42188e-06 & 68.4204 \\
1540 & 0.458733 $\pm$ 0.0193721 & -0.00430962 $\pm$ 0.000528432 & 1.02468e-05 $\pm$ 3.18422e-06 & 74.5365 \\
1590 & 0.561942 $\pm$ 0.0241582 & -0.00483288 $\pm$ 0.000672738 & 8.86887e-06 $\pm$ 4.12890e-06 & 66.6030 \\
1754 & 0.696297 $\pm$ 0.0297546 & -0.00394344 $\pm$ 0.000896781 & -5.09509e-06 $\pm$ 5.71387e-06 & 54.3112 \\
1787 & 0.708552 $\pm$ 0.0304631 & -0.00413825 $\pm$ 0.000913473 & -4.13446e-06 $\pm$ 5.80283e-06 & 55.6067 \\
1820 & 0.722356 $\pm$ 0.0312046 & -0.00431520 $\pm$ 0.000930350 & -3.37205e-06 $\pm$ 5.88764e-06 & 55.5818 \\
1869 & 0.669557 $\pm$ 0.0291147 & -0.00439225 $\pm$ 0.000856972 & 1.25474e-07 $\pm$ 5.41827e-06 & 57.9708 \\
1951 & 0.329679 $\pm$ 0.0148678 & -0.00289601 $\pm$ 0.000411667 & 5.52221e-06 $\pm$ 2.50613e-06 & 56.7764 \\
2034 & 0.268282 $\pm$ 0.0121947 & -0.00263569 $\pm$ 0.000331178 & 6.84875e-06 $\pm$ 2.00162e-06 & 51.2673 \\
2050 & 0.268394 $\pm$ 0.0121236 & -0.00267324 $\pm$ 0.000325609 & 7.16340e-06 $\pm$ 1.94661e-06 & 48.8637 \\
2116 & 0.466660 $\pm$ 0.0195273 & -0.00431920 $\pm$ 0.000545481 & 1.01641e-05 $\pm$ 3.33740e-06 & 48.6566 \\
2199 & 0.656896 $\pm$ 0.0240751 & -0.00424452 $\pm$ 0.000702241 & 1.23646e-07 $\pm$ 4.43604e-06 & 40.2259 \\
2282 & 0.550887 $\pm$ 0.0232737 & -0.00324417 $\pm$ 0.000689793 & -2.65057e-06 $\pm$ 4.34297e-06 & 33.2639 \\
2367 & 0.437617 $\pm$ 0.0181402 & -0.00327727 $\pm$ 0.000524173 & 3.56037e-06 $\pm$ 3.26575e-06 & 28.8401 \\
2450 & 0.372489 $\pm$ 0.0158142 & -0.00275966 $\pm$ 0.000455457 & 3.02371e-06 $\pm$ 2.82620e-06 & 25.6855 \\
2532 & 0.343304 $\pm$ 0.0140890 & -0.00251120 $\pm$ 0.000410193 & 2.92708e-06 $\pm$ 2.55476e-06 & 23.4109 \\
2615 & 0.297348 $\pm$ 0.0126408 & -0.00199625 $\pm$ 0.000374803 & 1.51588e-06 $\pm$ 2.35649e-06 & 20.8802 \\
2696 & 0.114462 $\pm$ 0.00635488 & -0.000659794 $\pm$ 0.000190458 & 2.28231e-07 $\pm$ 1.21678e-06 & 22.2746 \\
2781 & 0.0165311 $\pm$ 0.00111513 & -0.000140307 $\pm$ 3.43761e-05 & 3.91964e-07 $\pm$ 2.26691e-07 & 19.2484 \\
2866 & 0.0101379 $\pm$ 0.000839111 & -8.88387e-05 $\pm$ 2.60361e-05 & 2.03711e-07 $\pm$ 1.70379e-07 & 5.21170 \\
2947 & 0.0119934 $\pm$ 0.000875564 & -0.000110278 $\pm$ 2.54385e-05 & 3.10525e-07 $\pm$ 1.55310e-07 & 6.48191 \\
3029 & 0.00880047 $\pm$ 0.00105424 & -0.000121731 $\pm$ 3.27372e-05 & 5.63706e-07 $\pm$ 2.12267e-07 & 7.71997 \\
3112 & 0.0180165 $\pm$ 0.00126783 & -0.000290810 $\pm$ 3.54707e-05 & 1.40763e-06 $\pm$ 2.13142e-07 & 20.8907 \\
3197 & 0.0174447 $\pm$ 0.00108889 & -0.000252004 $\pm$ 3.27630e-05 & 1.07764e-06 $\pm$ 2.10738e-07 & 9.15868 \\
3282 & 0.0496506 $\pm$ 0.00246093 & -0.000550971 $\pm$ 6.97718e-05 & 1.77610e-06 $\pm$ 4.42525e-07 & 16.4943 \\
3365 & 0.0805788 $\pm$ 0.00460558 & -0.000630550 $\pm$ 0.000118700 & 5.73859e-07 $\pm$ 6.86131e-07 & 24.3957 \\
3448 & 0.111169 $\pm$ 0.00656156 & -0.000843633 $\pm$ 0.000175959 & 5.97566e-07 $\pm$ 1.07101e-06 & 27.8259 \\
3530 & 0.116072 $\pm$ 0.00703646 & -0.000707699 $\pm$ 0.000182759 & -9.56788e-07 $\pm$ 1.06137e-06 & 34.8098 \\
3613 & 0.109853 $\pm$ 0.00702651 & -0.000653272 $\pm$ 0.000192707 & -7.85011e-07 $\pm$ 1.17829e-06 & 36.0331 \\
3699 & 0.102701 $\pm$ 0.00630851 & -0.000657249 $\pm$ 0.000168638 & -4.98683e-07 $\pm$ 9.90343e-07 & 29.0501 \\
3784 & 0.0885122 $\pm$ 0.00502491 & -0.000764176 $\pm$ 0.000139615 & 1.32258e-06 $\pm$ 8.54061e-07 & 19.9329 \\
3861 & 0.0691256 $\pm$ 0.00961470 & -0.000624076 $\pm$ 0.000313185 & 1.20349e-06 $\pm$ 2.08842e-06 & 5.76973 \\
3947 & 0.0601500 $\pm$ 0.00363437 & -0.000586674 $\pm$ 0.000104562 & 1.50817e-06 $\pm$ 6.58374e-07 & 19.6413 \\
4029 & 0.0494466 $\pm$ 0.00335559 & -0.000501130 $\pm$ 9.94580e-05 & 1.43227e-06 $\pm$ 6.37691e-07 & 18.1151 \\
4114 & 0.0426249 $\pm$ 0.00309645 & -0.000464351 $\pm$ 9.33105e-05 & 1.52716e-06 $\pm$ 6.02480e-07 & 16.9582 \\
4198 & 0.0322586 $\pm$ 0.00285395 & -0.000271070 $\pm$ 8.36418e-05 & 4.83177e-07 $\pm$ 5.13445e-07 & 20.0542 \\
4278 & 0.0314696 $\pm$ 0.00328079 & -0.000399880 $\pm$ 0.000102860 & 1.64963e-06 $\pm$ 6.83836e-07 & 4.80272 \\
4366 & 0.0255910 $\pm$ 0.00308873 & -0.000306602 $\pm$ 9.81237e-05 & 1.19373e-06 $\pm$ 6.44967e-07 & 6.58957 \\
4447 & 0.0235604 $\pm$ 0.00335721 & -0.000263049 $\pm$ 0.000104612 & 8.50027e-07 $\pm$ 6.85657e-07 & 5.35084 \\
4533 & 0.0296437 $\pm$ 0.0106761 & -0.000472764 $\pm$ 0.000370358 & 2.58106e-06 $\pm$ 2.86288e-06 & 2.39558 \\
4620 & 0.0319490 $\pm$ 0.00435908 & -0.000433200 $\pm$ 0.000128242 & 1.96191e-06 $\pm$ 7.88154e-07 & 5.84142 \\
4702 & 0.0367381 $\pm$ 0.00510127 & -0.000452422 $\pm$ 0.000158963 & 1.87743e-06 $\pm$ 1.03714e-06 & 4.44055 \\
4786 & 0.0384862 $\pm$ 0.00492297 & -0.000366217 $\pm$ 0.000157048 & 8.67445e-07 $\pm$ 1.02949e-06 & 10.5770 \\
4869 & 0.0382403 $\pm$ 0.00536156 & -0.000280522 $\pm$ 0.000165646 & 1.18051e-07 $\pm$ 1.08376e-06 & 11.8732 \\
4953 & 0.0423836 $\pm$ 0.00757406 & -0.000360865 $\pm$ 0.000229238 & 5.80995e-07 $\pm$ 1.49145e-06 & 9.13096 \\
5040 & 0.0430713 $\pm$ 0.00732141 & -0.000463358 $\pm$ 0.000230657 & 1.61463e-06 $\pm$ 1.53485e-06 & 3.73760 \\
\hline
\end{tabular}
\end{scriptsize}
\end{center}
\label{tbl:tethys_parameters}
\end{table}

\begin{table}[h!]
\caption{Dione photometric parameters.}
\begin{center}
\begin{scriptsize}
\begin{tabular}{|c|c|c|c|c|}
\hline
wavelength (nm) & $a_0$ (adim.) & $a_1$ (deg$^{-1}$) & $a_2$ (deg$^{-2}$) &  $\chi^2$ \\
\hline
350 & 0.569054 $\pm$ 0.0241859 & -0.00433361 $\pm$ 0.000712115 & 2.32692e-06 $\pm$ 4.85007e-06 & 69.0012 \\
402 & 0.600564 $\pm$ 0.0239839 & -0.00406881 $\pm$ 0.000720175 & -1.21272e-06 $\pm$ 4.94044e-06 & 73.7408 \\
453 & 0.637653 $\pm$ 0.0260714 & -0.00391785 $\pm$ 0.000779200 & -4.13066e-06 $\pm$ 5.34139e-06 & 79.6003 \\
506 & 0.650044 $\pm$ 0.0253119 & -0.00350285 $\pm$ 0.000785638 & -7.54402e-06 $\pm$ 5.45884e-06 & 87.9191 \\
549 & 0.649476 $\pm$ 0.0258560 & -0.00356622 $\pm$ 0.000799496 & -6.79378e-06 $\pm$ 5.56309e-06 & 83.3035 \\
599 & 0.641847 $\pm$ 0.0296400 & -0.00352333 $\pm$ 0.000880623 & -6.48497e-06 $\pm$ 5.99377e-06 & 80.9544 \\
651 & 0.645660 $\pm$ 0.0284870 & -0.00346212 $\pm$ 0.000867839 & -6.93321e-06 $\pm$ 5.96896e-06 & 82.4667 \\
702 & 0.618009 $\pm$ 0.0271362 & -0.00299024 $\pm$ 0.000834173 & -8.84922e-06 $\pm$ 5.76356e-06 & 85.0305 \\
753 & 0.616170 $\pm$ 0.0272745 & -0.00308794 $\pm$ 0.000833289 & -7.84143e-06 $\pm$ 5.74508e-06 & 80.6644 \\
797 & 0.610286 $\pm$ 0.0272360 & -0.00303872 $\pm$ 0.000831578 & -8.10680e-06 $\pm$ 5.71635e-06 & 78.9526 \\
849 & 0.607019 $\pm$ 0.0276168 & -0.00311541 $\pm$ 0.000838350 & -6.96454e-06 $\pm$ 5.75903e-06 & 78.4437 \\
900 & 0.610816 $\pm$ 0.0280369 & -0.00293258 $\pm$ 0.000855161 & -8.77834e-06 $\pm$ 5.88242e-06 & 79.1544 \\
951 & 0.619403 $\pm$ 0.0286504 & -0.00305565 $\pm$ 0.000877456 & -8.02641e-06 $\pm$ 6.05745e-06 & 77.8382 \\
1010 & 0.582345 $\pm$ 0.0295187 & -0.00263027 $\pm$ 0.000881137 & -9.95835e-06 $\pm$ 6.06727e-06 & 70.3872 \\
1047 & 0.551960 $\pm$ 0.0367672 & -0.00237672 $\pm$ 0.00120914 & -8.49538e-06 $\pm$ 8.54308e-06 & 20.0698 \\
1130 & 0.579053 $\pm$ 0.0220572 & -0.00352670 $\pm$ 0.000599442 & 2.24195e-08 $\pm$ 4.30934e-06 & 13.3715 \\
1212 & 0.568676 $\pm$ 0.0338122 & -0.00319082 $\pm$ 0.00112631 & -3.59173e-06 $\pm$ 8.05878e-06 & 35.9243 \\
1294 & 0.559587 $\pm$ 0.0323566 & -0.00291242 $\pm$ 0.00106822 & -5.23963e-06 $\pm$ 7.58657e-06 & 40.6596 \\
1376 & 0.593850 $\pm$ 0.0344379 & -0.00341356 $\pm$ 0.00114639 & -3.96602e-06 $\pm$ 8.21976e-06 & 25.8107 \\
1458 & 0.506004 $\pm$ 0.0248322 & -0.00207809 $\pm$ 0.000789625 & -9.55438e-06 $\pm$ 5.45114e-06 & 44.7206 \\
1540 & 0.412173 $\pm$ 0.0162736 & -0.00256203 $\pm$ 0.000511469 & -9.12718e-07 $\pm$ 3.54266e-06 & 74.8676 \\
1590 & 0.498095 $\pm$ 0.0220240 & -0.00272574 $\pm$ 0.000691637 & -4.11852e-06 $\pm$ 4.80165e-06 & 86.7547 \\
1754 & 0.585954 $\pm$ 0.0318192 & -0.00229295 $\pm$ 0.00100612 & -1.20175e-05 $\pm$ 7.03671e-06 & 89.8573 \\
1787 & 0.593281 $\pm$ 0.0324892 & -0.00227675 $\pm$ 0.00102625 & -1.25008e-05 $\pm$ 7.17675e-06 & 91.1046 \\
1820 & 0.600577 $\pm$ 0.0329484 & -0.00228333 $\pm$ 0.00104135 & -1.27655e-05 $\pm$ 7.28380e-06 & 91.1653 \\
1869 & 0.573316 $\pm$ 0.0301976 & -0.00239051 $\pm$ 0.000952473 & -1.03566e-05 $\pm$ 6.65778e-06 & 89.7025 \\
1951 & 0.320622 $\pm$ 0.0122829 & -0.00225942 $\pm$ 0.000383317 & 2.19166e-06 $\pm$ 2.66749e-06 & 57.5766 \\
2034 & 0.260975 $\pm$ 0.00900118 & -0.00193702 $\pm$ 0.000282417 & 2.90007e-06 $\pm$ 1.97795e-06 & 54.9908 \\
2050 & 0.258220 $\pm$ 0.00888691 & -0.00189214 $\pm$ 0.000279497 & 2.66009e-06 $\pm$ 1.96009e-06 & 55.8296 \\
2116 & 0.412938 $\pm$ 0.0178824 & -0.00215678 $\pm$ 0.000564647 & -3.38834e-06 $\pm$ 3.98078e-06 & 87.3135 \\
2199 & 0.557964 $\pm$ 0.0289449 & -0.00187027 $\pm$ 0.000915380 & -1.33822e-05 $\pm$ 6.47814e-06 & 95.9348 \\
2282 & 0.494094 $\pm$ 0.0233164 & -0.00193849 $\pm$ 0.000748346 & -9.33715e-06 $\pm$ 5.34277e-06 & 84.2399 \\
2367 & 0.401933 $\pm$ 0.0160691 & -0.00187883 $\pm$ 0.000526181 & -4.60978e-06 $\pm$ 3.81225e-06 & 80.9138 \\
2450 & 0.347701 $\pm$ 0.0129177 & -0.00164444 $\pm$ 0.000430817 & -3.43785e-06 $\pm$ 3.16830e-06 & 77.6972 \\
2532 & 0.323388 $\pm$ 0.0112093 & -0.00147727 $\pm$ 0.000382337 & -3.11844e-06 $\pm$ 2.86358e-06 & 76.6808 \\
2615 & 0.280972 $\pm$ 0.00966723 & -0.00114097 $\pm$ 0.000335673 & -3.29924e-06 $\pm$ 2.55978e-06 & 74.6416 \\
2696 & 0.131379 $\pm$ 0.00340836 & -0.000989951 $\pm$ 0.000122657 & 3.55864e-06 $\pm$ 9.71442e-07 & 40.3731 \\
2781 & 0.0188240 $\pm$ 0.00118498 & -0.000188498 $\pm$ 3.83483e-05 & 1.08451e-06 $\pm$ 2.69249e-07 & 49.5532 \\
2866 & 0.00929229 $\pm$ 0.000749389 & -7.06288e-05 $\pm$ 2.44861e-05 & 2.13658e-07 $\pm$ 1.79962e-07 & 19.2307 \\
2947 & 0.0108443 $\pm$ 0.000567713 & -9.84170e-05 $\pm$ 1.91030e-05 & 3.30497e-07 $\pm$ 1.41268e-07 & 9.48759 \\
3029 & 0.00836773 $\pm$ 0.000610162 & -0.000122735 $\pm$ 2.02307e-05 & 6.68186e-07 $\pm$ 1.49372e-07 & 13.5315 \\
3112 & 0.0141117 $\pm$ 0.000640664 & -0.000211690 $\pm$ 2.17200e-05 & 1.04220e-06 $\pm$ 1.60668e-07 & 23.3300 \\
3197 & 0.0150830 $\pm$ 0.000690692 & -0.000218972 $\pm$ 2.33927e-05 & 1.07958e-06 $\pm$ 1.72934e-07 & 21.2153 \\
3282 & 0.0450745 $\pm$ 0.00130244 & -0.000522123 $\pm$ 4.13881e-05 & 2.04072e-06 $\pm$ 2.95998e-07 & 30.6952 \\
3365 & 0.0864166 $\pm$ 0.00296179 & -0.000883941 $\pm$ 9.50339e-05 & 2.72714e-06 $\pm$ 6.87324e-07 & 57.0459 \\
3448 & 0.129110 $\pm$ 0.00500415 & -0.00124034 $\pm$ 0.000158558 & 3.25531e-06 $\pm$ 1.13845e-06 & 70.5392 \\
3530 & 0.153852 $\pm$ 0.00574320 & -0.00141499 $\pm$ 0.000184677 & 3.25256e-06 $\pm$ 1.32474e-06 & 79.9196 \\
3613 & 0.153867 $\pm$ 0.00568420 & -0.00139937 $\pm$ 0.000182977 & 3.29216e-06 $\pm$ 1.31546e-06 & 73.1956 \\
3699 & 0.138335 $\pm$ 0.00475493 & -0.00128727 $\pm$ 0.000156146 & 3.39666e-06 $\pm$ 1.13947e-06 & 57.9483 \\
3784 & 0.107828 $\pm$ 0.00329340 & -0.00104084 $\pm$ 0.000110206 & 3.28207e-06 $\pm$ 8.21609e-07 & 36.3792 \\
3861 & 0.0871152 $\pm$ 0.00704379 & -0.00103846 $\pm$ 0.000233419 & 4.51125e-06 $\pm$ 1.68677e-06 & 5.38206 \\
3947 & 0.0714928 $\pm$ 0.00191227 & -0.000778889 $\pm$ 6.50121e-05 & 3.23564e-06 $\pm$ 4.93767e-07 & 17.7761 \\
4029 & 0.0589616 $\pm$ 0.00195421 & -0.000666217 $\pm$ 6.47889e-05 & 2.93508e-06 $\pm$ 4.79756e-07 & 15.1808 \\
4114 & 0.0481161 $\pm$ 0.00187207 & -0.000539870 $\pm$ 6.33053e-05 & 2.48402e-06 $\pm$ 4.70914e-07 & 16.0241 \\
4198 & 0.0417028 $\pm$ 0.00180547 & -0.000485115 $\pm$ 6.08335e-05 & 2.32908e-06 $\pm$ 4.47998e-07 & 17.4127 \\
4278 & 0.0337870 $\pm$ 0.00190491 & -0.000391102 $\pm$ 6.17161e-05 & 1.88919e-06 $\pm$ 4.49897e-07 & 11.5079 \\
4366 & 0.0287593 $\pm$ 0.00209433 & -0.000353199 $\pm$ 7.15163e-05 & 1.80446e-06 $\pm$ 5.36449e-07 & 10.7259 \\
4447 & 0.0273988 $\pm$ 0.00214976 & -0.000352276 $\pm$ 7.14895e-05 & 1.87276e-06 $\pm$ 5.25599e-07 & 6.51833 \\
4533 & 0.0302125 $\pm$ 0.00638157 & -0.000376986 $\pm$ 0.000294432 & 2.01997e-06 $\pm$ 2.66905e-06 & 2.55228 \\
4620 & 0.0346577 $\pm$ 0.00261570 & -0.000408855 $\pm$ 8.60599e-05 & 1.97648e-06 $\pm$ 6.23494e-07 & 6.79367 \\
4702 & 0.0415696 $\pm$ 0.00309834 & -0.000447450 $\pm$ 0.000101931 & 1.98572e-06 $\pm$ 7.40410e-07 & 4.94594 \\
4786 & 0.0489468 $\pm$ 0.00300470 & -0.000566404 $\pm$ 0.000101619 & 2.71458e-06 $\pm$ 7.64309e-07 & 5.89297 \\
4869 & 0.0519387 $\pm$ 0.00308774 & -0.000549558 $\pm$ 0.000104003 & 2.34438e-06 $\pm$ 7.66572e-07 & 6.30304 \\
4953 & 0.0567377 $\pm$ 0.00415756 & -0.000618546 $\pm$ 0.000148302 & 2.73839e-06 $\pm$ 1.12780e-06 & 3.30309 \\
5040 & 0.0541138 $\pm$ 0.00412723 & -0.000601125 $\pm$ 0.000137560 & 2.65198e-06 $\pm$ 1.02490e-06 & 5.15585 \\
\hline
\end{tabular}
\end{scriptsize}
\end{center}
\label{tbl:dione_parameters}
\end{table}

\begin{table}[h!]
\caption{Rhea photometric parameters.}
\begin{center}
\begin{scriptsize}
\begin{tabular}{|c|c|c|c|c|}
\hline
wavelength (nm) & $a_0$ (adim.) & $a_1$ (deg$^{-1}$) & $a_2$ (deg$^{-2}$) &  $\chi^2$ \\
\hline
350 & 0.512044 $\pm$ 0.0351535 & -0.00452830 $\pm$ 0.00105159 & 9.80795e-06 $\pm$ 7.30227e-06 & 13.5940 \\
402 & 0.540246 $\pm$ 0.0400253 & -0.00409084 $\pm$ 0.00105835 & 5.30443e-06 $\pm$ 6.71828e-06 & 50.7906 \\
453 & 0.598151 $\pm$ 0.0445684 & -0.00413351 $\pm$ 0.00118516 & 2.85153e-06 $\pm$ 7.54935e-06 & 49.5443 \\
506 & 0.594842 $\pm$ 0.0467737 & -0.00342633 $\pm$ 0.00125313 & -1.83094e-06 $\pm$ 8.01952e-06 & 58.2542 \\
549 & 0.596226 $\pm$ 0.0470739 & -0.00310465 $\pm$ 0.00127070 & -4.49371e-06 $\pm$ 8.18182e-06 & 50.0908 \\
599 & 0.610461 $\pm$ 0.0730245 & -0.00352956 $\pm$ 0.00195804 & -1.00710e-06 $\pm$ 1.28327e-05 & 3.96116 \\
651 & 0.606040 $\pm$ 0.0475247 & -0.00277289 $\pm$ 0.00130070 & -8.20553e-06 $\pm$ 8.45803e-06 & 21.7050 \\
702 & 0.611581 $\pm$ 0.0479941 & -0.00341137 $\pm$ 0.00128989 & -1.90779e-06 $\pm$ 8.28031e-06 & 52.1318 \\
753 & 0.594847 $\pm$ 0.0464900 & -0.00260065 $\pm$ 0.00127480 & -8.79556e-06 $\pm$ 8.30919e-06 & 20.3477 \\
797 & 0.594794 $\pm$ 0.0462435 & -0.00269886 $\pm$ 0.00126816 & -7.81879e-06 $\pm$ 8.26061e-06 & 20.3786 \\
849 & 0.601416 $\pm$ 0.0468344 & -0.00313457 $\pm$ 0.00126596 & -3.31275e-06 $\pm$ 8.16883e-06 & 48.1283 \\
900 & 0.603978 $\pm$ 0.0467907 & -0.00300556 $\pm$ 0.00127201 & -4.49503e-06 $\pm$ 8.23376e-06 & 45.6636 \\
951 & 0.614548 $\pm$ 0.0484755 & -0.00284924 $\pm$ 0.00133187 & -7.12253e-06 $\pm$ 8.69244e-06 & 20.0922 \\
1010 & 0.606242 $\pm$ 0.0456210 & -0.00297826 $\pm$ 0.00123586 & -5.60249e-06 $\pm$ 7.97442e-06 & 19.5761 \\
1047 & 0.579522 $\pm$ 0.0540204 & -0.00126032 $\pm$ 0.00147737 & -2.00917e-05 $\pm$ 9.34156e-06 & 15.2937 \\
1130 & 0.582691 $\pm$ 0.0554518 & -0.00112043 $\pm$ 0.00150808 & -2.19215e-05 $\pm$ 9.58870e-06 & 13.7469 \\
1212 & 0.569393 $\pm$ 0.0542274 & -0.000846618 $\pm$ 0.00147971 & -2.32961e-05 $\pm$ 9.40832e-06 & 15.3346 \\
1294 & 0.583598 $\pm$ 0.0509265 & -0.00175623 $\pm$ 0.00139233 & -1.58158e-05 $\pm$ 8.81160e-06 & 15.7311 \\
1376 & 0.539758 $\pm$ 0.0485621 & -0.000418759 $\pm$ 0.00134165 & -2.35435e-05 $\pm$ 8.54960e-06 & 32.3123 \\
1458 & 0.505369 $\pm$ 0.0370631 & -0.00244601 $\pm$ 0.00100968 & -5.55828e-06 $\pm$ 6.45837e-06 & 17.5507 \\
1540 & 0.345474 $\pm$ 0.0242264 & -0.00130635 $\pm$ 0.000664389 & -6.63880e-06 $\pm$ 4.28329e-06 & 17.5500 \\
1590 & 0.411901 $\pm$ 0.0305545 & -0.000905003 $\pm$ 0.000839096 & -1.33432e-05 $\pm$ 5.39592e-06 & 18.0486 \\
1754 & 0.567597 $\pm$ 0.0466196 & -0.00142986 $\pm$ 0.00128131 & -1.67574e-05 $\pm$ 8.17340e-06 & 17.2433 \\
1787 & 0.570110 $\pm$ 0.0470016 & -0.00128065 $\pm$ 0.00129072 & -1.80802e-05 $\pm$ 8.22716e-06 & 17.0876 \\
1820 & 0.579828 $\pm$ 0.0475787 & -0.00130797 $\pm$ 0.00130503 & -1.83167e-05 $\pm$ 8.31190e-06 & 17.4500 \\
1869 & 0.539053 $\pm$ 0.0430928 & -0.00139015 $\pm$ 0.00118504 & -1.52864e-05 $\pm$ 7.58274e-06 & 16.7201 \\
1951 & 0.256180 $\pm$ 0.0181021 & -0.00110528 $\pm$ 0.000511204 & -2.89585e-06 $\pm$ 3.37322e-06 & 16.3728 \\
2034 & 0.188211 $\pm$ 0.0139837 & -0.000531389 $\pm$ 0.000395087 & -4.12826e-06 $\pm$ 2.60339e-06 & 17.8852 \\
2050 & 0.186490 $\pm$ 0.0136991 & -0.000511729 $\pm$ 0.000387694 & -4.20759e-06 $\pm$ 2.55428e-06 & 17.2864 \\
2116 & 0.306109 $\pm$ 0.0226653 & 4.54105e-05 $\pm$ 0.000645386 & -1.48279e-05 $\pm$ 4.23827e-06 & 16.9406 \\
2199 & 0.496728 $\pm$ 0.0379226 & -0.000399802 $\pm$ 0.00107546 & -2.03726e-05 $\pm$ 7.00984e-06 & 14.8050 \\
2282 & 0.464609 $\pm$ 0.0333652 & -0.00134956 $\pm$ 0.000945830 & -1.05332e-05 $\pm$ 6.18333e-06 & 13.5791 \\
2367 & 0.335619 $\pm$ 0.0221905 & -0.000697431 $\pm$ 0.000645952 & -9.21792e-06 $\pm$ 4.32361e-06 & 13.4562 \\
2450 & 0.287433 $\pm$ 0.0183897 & -0.000688607 $\pm$ 0.000544002 & -6.59859e-06 $\pm$ 3.68793e-06 & 13.2691 \\
2532 & 0.261896 $\pm$ 0.0160561 & -0.000527364 $\pm$ 0.000485183 & -6.20683e-06 $\pm$ 3.34513e-06 & 14.2777 \\
2615 & 0.231030 $\pm$ 0.0141255 & -0.000472200 $\pm$ 0.000436122 & -4.70354e-06 $\pm$ 3.05769e-06 & 13.6898 \\
2696 & 0.104691 $\pm$ 0.00632273 & -0.000653025 $\pm$ 0.000198189 & 2.76112e-06 $\pm$ 1.43883e-06 & 17.4842 \\
2781 & 0.0156109 $\pm$ 0.00134050 & -0.000157822 $\pm$ 3.90099e-05 & 8.75587e-07 $\pm$ 2.74452e-07 & 31.8919 \\
2866 & 0.00829627 $\pm$ 0.000786979 & -4.91275e-05 $\pm$ 2.28570e-05 & 1.38511e-08 $\pm$ 1.54693e-07 & 12.0355 \\
2947 & 0.0102901 $\pm$ 0.000784749 & -8.48649e-05 $\pm$ 2.18176e-05 & 2.08733e-07 $\pm$ 1.42388e-07 & 12.2502 \\
3029 & 0.00763761 $\pm$ 0.000830174 & -9.75132e-05 $\pm$ 2.37442e-05 & 4.11285e-07 $\pm$ 1.59300e-07 & 9.78033 \\
3112 & 0.0150120 $\pm$ 0.00132555 & -0.000216421 $\pm$ 3.53110e-05 & 9.18027e-07 $\pm$ 2.23997e-07 & 12.8433 \\
3197 & 0.0141761 $\pm$ 0.00112444 & -0.000186595 $\pm$ 3.05026e-05 & 7.23240e-07 $\pm$ 1.95769e-07 & 13.9914 \\
3282 & 0.0279771 $\pm$ 0.00176146 & -0.000262621 $\pm$ 4.97251e-05 & 7.44350e-07 $\pm$ 3.25815e-07 & 24.2698 \\
3365 & 0.0506577 $\pm$ 0.00385778 & -0.000367856 $\pm$ 0.000112209 & 6.06234e-07 $\pm$ 7.47135e-07 & 33.6537 \\
3448 & 0.0797381 $\pm$ 0.00650617 & -0.000528572 $\pm$ 0.000186151 & 5.25032e-07 $\pm$ 1.22170e-06 & 26.6507 \\
3530 & 0.0970469 $\pm$ 0.00826045 & -0.000601976 $\pm$ 0.000233504 & 2.93135e-07 $\pm$ 1.52343e-06 & 26.2412 \\
3613 & 0.100410 $\pm$ 0.00835900 & -0.000625241 $\pm$ 0.000238862 & 4.04884e-07 $\pm$ 1.57292e-06 & 22.3221 \\
3699 & 0.0926928 $\pm$ 0.00741652 & -0.000624547 $\pm$ 0.000215626 & 1.01241e-06 $\pm$ 1.44179e-06 & 23.4661 \\
3784 & 0.0735067 $\pm$ 0.00543037 & -0.000525727 $\pm$ 0.000160949 & 1.23522e-06 $\pm$ 1.09013e-06 & 20.7677 \\
3861 & 0.0631079 $\pm$ 0.00954394 & -0.000663086 $\pm$ 0.000294388 & 3.12195e-06 $\pm$ 2.08014e-06 & 8.25823 \\
3947 & 0.0491730 $\pm$ 0.00344771 & -0.000416629 $\pm$ 0.000101074 & 1.56042e-06 $\pm$ 6.87297e-07 & 11.9181 \\
4029 & 0.0396739 $\pm$ 0.00322359 & -0.000347729 $\pm$ 9.64173e-05 & 1.46851e-06 $\pm$ 6.74349e-07 & 16.6258 \\
4114 & 0.0333062 $\pm$ 0.00281417 & -0.000304830 $\pm$ 8.14473e-05 & 1.37167e-06 $\pm$ 5.49546e-07 & 12.1752 \\
4198 & 0.0281470 $\pm$ 0.00272554 & -0.000265902 $\pm$ 8.12075e-05 & 1.31353e-06 $\pm$ 5.69660e-07 & 19.6792 \\
4278 & 0.0219414 $\pm$ 0.00250828 & -0.000196959 $\pm$ 7.29764e-05 & 9.31217e-07 $\pm$ 5.01312e-07 & 6.85060 \\
4366 & 0.0190158 $\pm$ 0.00260720 & -0.000197020 $\pm$ 8.06152e-05 & 9.97716e-07 $\pm$ 5.62996e-07 & 5.70432 \\
4447 & 0.0168486 $\pm$ 0.00253435 & -0.000148841 $\pm$ 7.36772e-05 & 7.09562e-07 $\pm$ 5.01819e-07 & 8.91186 \\
4533 & 0.0189400 $\pm$ 0.00482484 & -0.000206499 $\pm$ 0.000202532 & 1.06559e-06 $\pm$ 1.54969e-06 & 3.60822 \\
4620 & 0.0217010 $\pm$ 0.00336745 & -0.000197166 $\pm$ 0.000100743 & 9.85540e-07 $\pm$ 6.99623e-07 & 5.71905 \\
4702 & 0.0281346 $\pm$ 0.00413315 & -0.000260679 $\pm$ 0.000124325 & 1.25366e-06 $\pm$ 8.68075e-07 & 6.21439 \\
4786 & 0.0304752 $\pm$ 0.00455458 & -0.000232103 $\pm$ 0.000140287 & 9.55733e-07 $\pm$ 9.87834e-07 & 9.17557 \\
4869 & 0.0349120 $\pm$ 0.00463767 & -0.000284493 $\pm$ 0.000135346 & 1.24908e-06 $\pm$ 9.15293e-07 & 14.4720 \\
4953 & 0.0391132 $\pm$ 0.00566360 & -0.000351654 $\pm$ 0.000178402 & 1.60497e-06 $\pm$ 1.24658e-06 & 4.88742 \\
5040 & 0.0401744 $\pm$ 0.00608593 & -0.000438818 $\pm$ 0.000185325 & 2.28852e-06 $\pm$ 1.30121e-06 & 5.65671 \\

\hline
\end{tabular}
\end{scriptsize}
\end{center}
\label{tbl:rhea_parameters}
\end{table}


\end{document}